\documentclass[journal]{IEEEtran}
\usepackage[colorlinks,bookmarksopen,bookmarksnumbered,citecolor=red,urlcolor=red]{hyperref}
\usepackage{mathtools,cuted,epsfig,float,amsmath,amssymb,amsmath,amsthm,mathrsfs,color,subfigure,graphicx,anyfontsize,t1enc,cite,epstopdf}
\usepackage{dblfloatfix}
\usepackage{multirow}
\usepackage[normalem]{ulem}
\useunder{\uline}{\ul}{}
\usepackage{tabularx,booktabs,caption}
\usepackage{algorithm,algpseudocode,authblk}
\usepackage{textcomp}
\algnewcommand{\Inputs}[1]{%
	\State \textbf{Inputs:}
	\Statex \hspace*{\algorithmicindent}\parbox[t]{.8\linewidth}{\raggedright #1}
}
\algnewcommand{\Initialize}[1]{%
	\State \textbf{Initialize:}
	\Statex \hspace*{\algorithmicindent}\parbox[t]{.8\linewidth}{\raggedright #1}
}
\ifCLASSINFOpdf
\else
\fi

\begin{document}

	\title{Hybrid FSO/THz-based Backhaul Network for mmWave Terrestrial Communication}
	\author{
		\IEEEauthorblockN{Praveen Kumar Singya$^{1}$, \IEEEmembership{Member, IEEE}, Behrooz Makki$^{2}$, \IEEEmembership{Senior Member, IEEE}, Antonio D'Errico$^{3}$, \IEEEmembership{Senior Member, OPTICA}, and Mohamed-Slim Alouini$^{1}$, \IEEEmembership{Fellow, IEEE}}
		\thanks{$^{1}$P. K. Singya and M.-S. Alouini are with the Computer, Electrical, and Mathematical Science and Engineering (CEMSE) Division, King Abdullah University	of Science and Technology (KAUST), Thuwal 23955-6900, Saudi Arabia (e-mail:praveen.singya@kaust.edu.sa, slim.alouini@kaust.edu.sa)}
		\thanks{$^{2}$B. Makki is with Ericsson Research, Ericsson, 41756 Göteborg, Sweden
			(e-mail:behrooz.makki@ericsson.com)}
		\thanks{$^{3}$D'Errico is with Ericsson Research, Ericsson, 56124 Pisa, Italy (e-mail:antonio.d.errico@ericsson.com)}
	\vspace{-1em}}
	
	\maketitle
	
\begin{abstract}
	In this work, a hybrid free-space optics (FSO)/ teraHertz (THz) based backhaul network is considered to provide high-data-rate reliable communication to the terrestrial mobile users (MUs) operating at millimeter-wave (mmWave) bands. The FSO link is affected by atmospheric turbulence and pointing error impairments. At the FSO receiver, both intensity-modulated direct detection and heterodyne detection techniques are considered.
	The multi-antenna THz link suffers from high path-loss, small-scale fading, and misalignment error. To minimize the effect of back-and-forth switching, soft switching method is introduced at the access point (AP) to select the signal coming through the hybrid FSO/THz link, and a comparison with hard switching method is presented. 
	Selective decode-and-forward relaying is considered at the AP.	In this context, we derive closed-form expressions of the individual link's outage probability, end-to-end (E2E) outage probability, asymptotic outage probability, ergodic capacity, and generalized average bit-error-rate. Finally, we study the effect of different parameters such as atmospheric turbulence, pointing/misalignment errors, link distance, atmospheric attenuation/path-loss, fading parameters of the THz and access links, and number of antennas on the network performance. Our results indicate that, with a proper switching method, the joint implementation of FSO/THz links improves the rate/reliability of the backhaul links with limited switching overhead.
	
\end{abstract}

\section{Introduction} 
Recently, teraHertz (THz) wireless communication has gained increased research attention and is considered to be a potential solution for wireless backhauling in 6G.
THz communication exists  between the
millimeter-wave (mmWave) and infrared
light band from 100 GHz to 10 THz frequency band (including the sub THz band) and has various advantages such as high
directional gain, highly secured communication,
and large available bandwidths \cite{elayan2019terahertz,yi2021design,rajatheva2020scoring,li2021performance}.	
However, THz communication is highly susceptible to blockage in a dense environment. Further, it has huge path-loss and sensitivity to misalignment error, which reduces its propagation range and  makes it less suitable for the fronthaul communication scenario.

From a different perspective, free-space optics (FSO) has been proven to be a promising cost-effective alternative to fiber optical cables for short-range backhauling.  FSO provides high speed secured line-of-sight (LoS) communication, is easy to deploy, and is cost-effective due to its operation in license-free band \cite{trichili2020roadmap}. However, pointing error, atmospheric turbulence, and weather issues are problematic  in FSO communications, which limits the FSO link's performance. The FSO link is severely affected by fog and atmospheric turbulence, whereas the THz link is severely affected by the rain.	  
Hence, a hybrid paradigm which combines the benefits of both the FSO and THz links is required to guarantee reliable and robust high data-rate backhaul applications.
For fronthaul terrestrial mobile communication, the FSO and the THz are not suitable due to their LoS operation, blockage, and users mobility. In such situations, more robust and reliable communication can be established through the radio-frequency (RF) links to the terrestrial mobile users (MUs) operating at mmWave bands.
Therefore, a cooperative dual-hop communication system which has  hybrid FSO/THz link for the backhaul communication and multiple mmWave RF links for the fronthaul communication to the terrestrial MUs is expected to guarantee proper end-to-end (E2E) performance.

Considering mmWave bands for the RF link, the performance of hybrid FSO/RF links has been studied in different works which can be divided into two parts. 
In,  \cite{singya2021performance,singya2020performance,bhatnagar2013performance,zedini2016performance,makki2017performance,makki2017performance1,zedini2020performance,lee2020throughput,xu2021performance,xu2020performance,balti2019tractable,balti2021joint,yang2018performance,michailidis2018outage,singya2022Haps,hassan2019hybrid,zhang20203d} and reference therein, dual-hop or multi-hop networks are considered, where each hop can be based on either FSO or mmWave RF. 
In \cite{douik2016hybrid,Nock2016,sharma2019switching,swaminathan2021haps,nath2019impact,gupta2019hard,althunibat2020secure,shah2021adaptive,rakia2015outage,makki2016performance,he2009bit,zhang2009soft,abdulhussein2010rateless,trichili2021retrofitting,Moradi2010,usman2014practical,bag2018performance}, on the other hand, at-least one link is equipped with both the FSO and RF transceiver and can, possibly, switch between them or use them simultaneously. 
\cite{Nock2016} proposes an adaptive algorithm for a hybrid FSO/RF communication system.
In \cite{sharma2019switching,swaminathan2021haps}, the performance of hybrid FSO/RF decode-and-forward (DF)
relay systems is analyzed. Then, \cite{nath2019impact} proposes a hybrid FSO/RF cognitive system, where RF link provides backup to two FSO links in a cognitive manner. In \cite{gupta2019hard}, the performance of a hybrid RF/visible-light communication system is analyzed. Also, \cite{althunibat2020secure} proposes an index modulation based switching for a hybrid FSO/RF communication system.
Note that in \cite{Nock2016,sharma2019switching,swaminathan2021haps,nath2019impact,gupta2019hard,althunibat2020secure}, hard switching method is considered, where the communication  switches between the FSO and the RF links, as soon as the signal-to-noise ratio (SNR) in one of the links exceeds that of the other one. 
In \cite{shah2021adaptive,rakia2015outage},  switching-based hybrid FSO/RF communication systems are studied, where both the FSO and RF links are working simultaneously if the FSO link's SNR falls below threshold.  
Finally, \cite{makki2016performance}  studies a simultaneous data transmission method in parallel FSO/RF links with and without hybrid automatic repeat request.

One of the problems of the hard switching method is frequent back-and-forth switching between the RF and the FSO links, which affects the links performance and increases the implementation complexity/cost. To minimize the effect of frequent back-and-forth switching, a soft switching method can be preferred.

Different works on the soft switching-based hybrid FSO/RF communication system are available in the literature \cite{he2009bit,zhang2009soft,abdulhussein2010rateless,trichili2021retrofitting,Moradi2010,usman2014practical,bag2018performance}.
\cite{he2009bit,zhang2009soft,abdulhussein2010rateless} proposed the soft switching based hybrid FSO/RF systems, where the soft switching between the FSO and RF links is achieved through the joint design of coding schemes.
In \cite{he2009bit}, the performance of a  hybrid FSO/RF communication system with bit-interleaved coded modulation is investigated. 
In \cite{zhang2009soft}, a comparison of soft and hard switching-based hybrid FSO/RF communication system is presented, where the soft switching is performed by using soft-length raptor codes. Also, for hybrid FSO/RF systems, \cite{abdulhussein2010rateless} proposes a rateless coded automatic repeat request.
Further, \cite{Moradi2010} discusses the capacity performance  of a hybrid FSO/RF communication system with dual FSO threshold-based switching.
In  \cite{usman2014practical},  a comparison of hard and soft switching-based hybrid FSO/RF communication system is presented in the cases with intensity modulation and direct detection (IM/DD) in the FSO link and mmWave communication at the RF link. 
In   \cite{bag2018performance}, a dual-hop FSO and hybrid FSO/RF based communication system is proposed by considering both the hard and soft switching operations.
Note that none of the works \cite{singya2021performance,singya2020performance,bhatnagar2013performance,zedini2016performance,makki2017performance,makki2017performance1,zedini2020performance,lee2020throughput,xu2021performance,xu2020performance,balti2019tractable,balti2021joint,yang2018performance,michailidis2018outage,singya2022Haps,hassan2019hybrid,zhang20203d,douik2016hybrid,Nock2016,sharma2019switching,swaminathan2021haps,nath2019impact,gupta2019hard,althunibat2020secure,shah2021adaptive,rakia2015outage,makki2016performance,he2009bit,zhang2009soft,abdulhussein2010rateless,trichili2021retrofitting,Moradi2010,usman2014practical,bag2018performance} consider THz bands and its corresponding channel properties. Also, without FSO, \cite{li2021performance,boulogeorgos2019error,bhardwaj2021performance}  study the performance of dual-hop RF networks, where one of the hops operates in THz and other one in either THz or mmWave band. Further, \cite{li2021performancecon,li2022mixed} study the performance of AF based dual-hop FSO/THz network, where one hop is FSO and other hop operates in THz band.
Note that, although \cite{balti2019tractable,boulogeorgos2019error,bhardwaj2021performance,makki2017performance,zedini2016performance,zedini2020performance,makki2017performance1,lee2020throughput,xu2021performance,xu2020performance,li2021performancecon,li2022mixed} do not explicitly mention the backhauling as the main usecase of their channel model/analysis, the presence of multi-hop networks with stationary nodes using FSO and/or THz links fits best to cases with backhauling. Further, most of the hybrid FSO/RF systems with switching operations are suitable for the backhaul applications.  \par

\subsection{Motivation and Contributions}The major motivation is that in 6G high-rate reliable backhaul links are required to support multiple high-rate MUs. Then, FSO is a promising option. However, pointing error, atmospheric turbulence, and weather issues are problematic, which limit the FSO link's performance. This directs us to consider backup links for the FSO. The mmWave RF links can not support the same rate as the FSO, instead THz links with multi-antenna can support the same order of rate. This is the reason that we consider FSO and THz links in parallel. Further, both of these links are sensitive to atmospheric turbulence/fading and various imperfections especially the pointing/misalignment error, which must be modeled carefully. Thus, the parallel implementation of the FSO and the THz links is expected to improve the diversity and guarantee a stable performance. However, one needs to be careful with the frequent switching between these links, based on the temporal channel variations. Thus, there is a need to design proper switching methods with low complexity and proper link performance.

\begin{figure*}[]
	\centering
	\includegraphics[width=6in,height=2.7in]{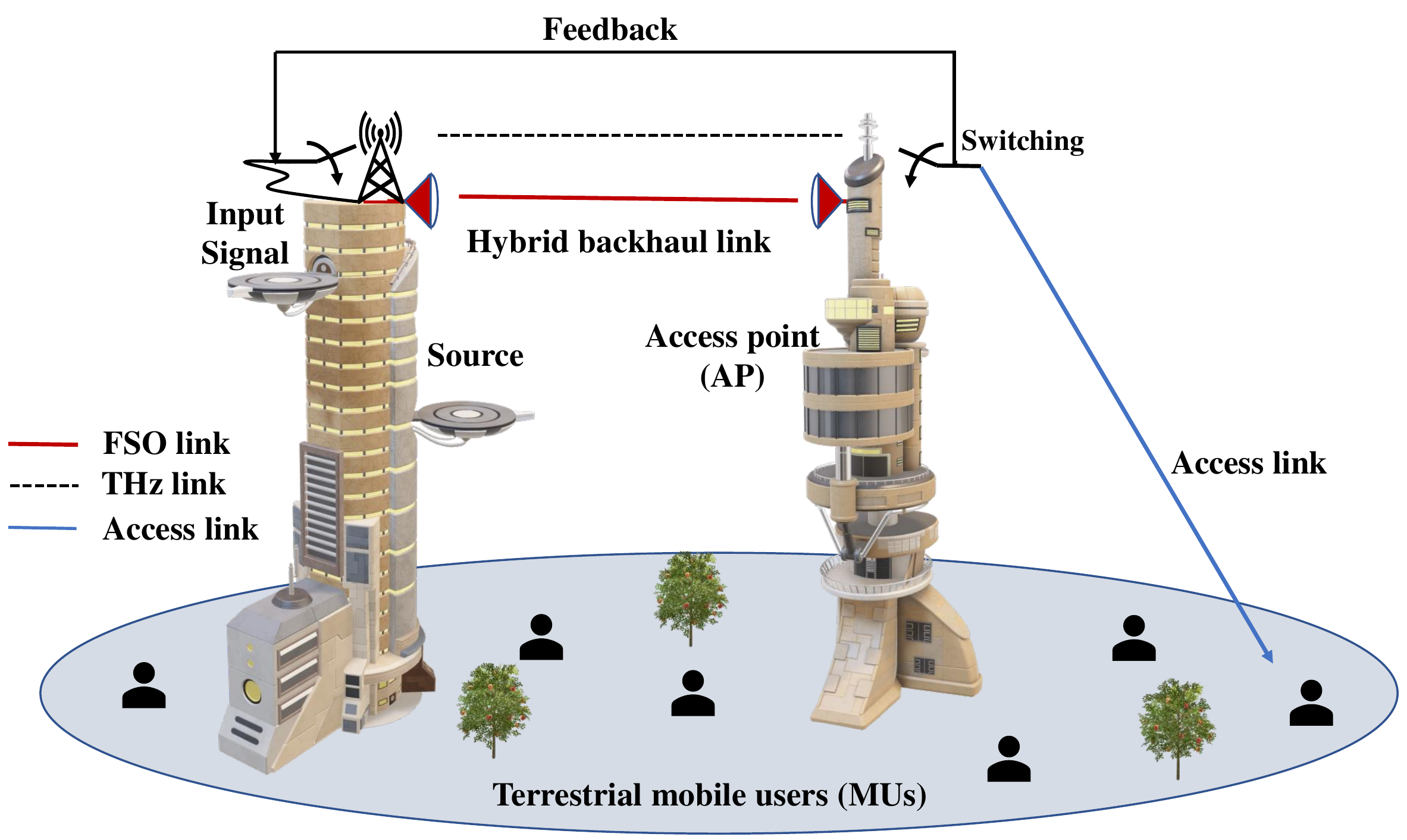}
	\caption{{Considered switching-based system model.}}
	\label{systmod}
	\hrulefill
\end{figure*}

Motivated with the above, in this work, we study the performance of hybrid FSO/THz backhaul system with different imperfections and switching methods. The fronthaul communication to the terrestrial MUs is completed by mmWave access links (see Fig. \ref{systmod}).  The FSO link suffers from both atmospheric turbulence and pointing error which are statistically modeled by Gamma-Gamma and Rayleigh distributions, respectively. 
The THz link is characterized with small-scale fading and misalignment error which are statistically modeled by $\alpha-\mu$ and Rayleigh distributions, respectively. To improve the THz link's performance against path-loss, we consider multiple antennas at the THz receiver to achieve an equivalent performance as in the FSO link. At the access point (AP), we apply a soft switching method to select the signal coming either from the FSO link or from the THz link  to minimize the back-and-forth switching, and we compare the results with those obtained with hard switching. 
Further, we apply a selective DF relaying at the AP to forward the successfully decoded high-data rate backhaul  signals to the terrestrial MUs. Here, while we consider a dual-hop network, consisting of an FSO/THz backhaul link and an mmWave access link, the main focus of the theoretical analysis is on the backhaul link, where the presence of the parallel THz and FSO links with different switching methods and pointing errors makes the analysis challenging.	
The major contributions of this work are as follows:
\begin{itemize}
	\item We analyze the performance of soft switching method in the  backhaul hybrid FSO/THz link, to minimize the frequent back-and-forth switching in practical applications, and compare the results with those obtained by hard switching.
	
	\item We derive the outage probability of the individual links as well as the E2E outage probability of the considered communication system for both  hard and soft switching cases. Then, using high-SNR approximations and for different switching methods, we derive the diversity order in each hop as well as in the E2E system.
	
	\item We derive the analytical ergodic capacity expressions for the individual links. By using these expressions, we obtain the ergodic capacity of the hybrid FSO/THz link for both  hard and soft switching cases, and study the impact of backhaul and access links' performance on the E2E ergodic capacity.
	
	\item We derive the average bit-error-rate (ABER) expressions of the individual links and the E2E system for various modulation schemes  and switching methods. 
	
	\item Finally,  we illustrate the impact of atmospheric turbulence, pointing/misalignment error, atmospheric attenuation/path-loss, fading parameters of the THz and access links, link distance, and number of RF antennas on the outage probability, ergodic capacity, and ABER performance.
\end{itemize}

Our analytical and simulation results show that the soft switching provides marginal gain, in terms of outage probability, over the hard 
switching, and reduces the frequent back-and-forth switching. Thus, from a practical point of view, soft switching between the links may be of interest. Observing the impact of pointing/misalignment
error on the hybrid FSO/THz
link, it is concluded that with the increased jitter standard
deviation, the FSO, the THz, and consequentially the
hybrid FSO/THz links reach towards the outage. Further, for a given transmit power, the optimum outage performance is obtained with a specific value of beamwidth for each jitter standard deviation. Moreover, the atmospheric turbulence and the
pointing error (resp. the fading and the misalignment error) directly affect the diversity order of the FSO (resp. the THz) link. Finally, with typical parameter settings and proper switching method, the hybrid FSO/THz links provide high rate reliable backhaul, while the achievable rate of the mmWave access links is the bottleneck of the E2E performance.

Note that as opposed to  \cite{singya2021performance,li2021performance,bhardwaj2021performance,singya2020performance,bhatnagar2013performance,zedini2016performance,makki2017performance,makki2017performance1,zedini2020performance,lee2020throughput,xu2021performance,xu2020performance,balti2019tractable,balti2021joint,yang2018performance,michailidis2018outage,singya2022Haps,boulogeorgos2019error,li2021performancecon,li2022mixed,hassan2019hybrid,zhang20203d} which consider serial FSO and THz hops where each link is only based on one of these technologies, we consider parallel FSO/THz links with different switching methods. Moreover, different from \cite{douik2016hybrid,Nock2016,sharma2019switching,swaminathan2021haps,nath2019impact,gupta2019hard,althunibat2020secure,shah2021adaptive,rakia2015outage,makki2016performance,he2009bit,zhang2009soft,abdulhussein2010rateless,trichili2021retrofitting,Moradi2010,usman2014practical,bag2018performance}, we consider the THz bands and its corresponding channel properties/pointing error. Finally, our analytical and simulation results on the performance of soft/hard switching methods in FSO/THz links and the effect of different parameters on the network performance have not been studied before. The differences in the system model and communication setup makes our paper completely different from the state-of-the-art.

\begin{table*}[]
	\caption{{Various parameters considered throughout the paper.}}
	\begin{tabular}{|l|l|l|l|}
		\cline{1-4}
		Symbol & Definition & Symbol & Definition \\ \cline{1-4}
		$\text{i}\in \{\text{F},\text{T},\text{R}\}$ & Parameter to select the FSO, THz, and mmWave access links &  $\tau$ &  Selection parameter for the FSO detector type    \\
		$P_{\text{i}}$& Transmit power at the source i & $L_{\text{i}}$ &  Length of the  i link \\ 
		$N_{\text{r}}$& Number of antennas at the THz receiver & $N_{\text{t}}$ & Number of antennas at the AP transmitter   \\ 
		$\alpha_\text{F}$,$\beta_\text{F}$& Atmospheric turbulence parameter of the FSO link & $\alpha$,$\mu$ & Fading parameters of $\alpha-\mu$ distributed THz link   \\
		$G_{\text{t}}^{\text{i}}$ & Transmit antenna gain at node i   & $G_{\text{r}}^{\text{i}}$& Receive antenna gain at node i \\ 
		$\xi_\text{i}$,$A_{0,\text{i}}$ & Pointing error parameters for the $\text{i}$ link &   $f_\text{i}$   & Operating frequency of the i link \\ 
		$a_\text{i}$,$\omega_{\text{L,i}}$,$\sigma_{\text{j,i}}$ &  Receiver radius, beamwidth, and jitter standard deviation of $\text{i}$ link & $\lambda_\text{i}$   & Wavelength of the i link  \\ 
		$I_\text{l}$,$h_{\text{l}}$,$p_{\text{l}}$ & Path-loss associated with the FSO, THz, and mmWave access links  & m &  Fading parameter of mmWave access link  \\ 
		$\eta$& Optical-to-electrical conversion coefficient & c &  Speed of light \\ 
		$a_{\text{oxy}}$      & Oxygen absorption in mmWave access link  &  $a_{\text{rain}}$& Rain attenuation in mmWave access link  \\ 
		$\gamma_{\text{th}}$ & Threshold SNR at the FSO and THz receiver for hard switching  & 	$\gamma_{\text{i}}$ &   Instantaneous received SNR at node i \\ 
		$\gamma^{\text{F}}_{\text{th,L}}$, $\gamma^{\text{F}}_{\text{th,U}}$,$\gamma^{\text{T}}_{\text{th}}$ & Lower and upper FSO and THz thresholds for soft switching &$\gamma^{\text{R}}_{\text{th}}$ & Threshold SNR at the mmWave link's receiver \\
		\cline{1-4}
	\end{tabular}
	\label{parameter}
\end{table*}
We organize the rest of the paper as follows. System and channel models are discussed in Section II. Sections III and IV consist of the analytical results of the outage probability, the diversity order, the ergodic capacity, and the ABER for hard and soft switching methods, respectively. In Section V, the numerical results for the derived analytical expressions are obtained and are validated through the simulation results. Finally, conclusions are made in Section VI. 
\vspace{-1em}
\section{System and Channel Models}
In Table \ref{parameter}, we mention the parameters considered throughout the paper. We consider a hybrid FSO/THz-based backhaul network providing high data-rates to the mmWave terrestrial MUs, as shown in Fig. \ref{systmod}. The FSO and the THz transmitters and receivers are deployed at the source (building 1) and at the AP (building 2), respectively. For implementation simplicity, the data decoding is performed in one of the FSO or the THz receivers. 
By default, FSO link is given high priority over the THz link (if it has acceptable link performance) due to its improved capability to provide high data-rate secured communication.
Hence, the AP checks the instantaneous received SNR of both links, and decides which link is to be selected for the transmission. Further, a selective DF relaying is applied at the AP to forward the signal coming through the hybrid FSO/THz link to the terrestrial MUs, if they are correctly decoded. Finally, note that, while we present the discussions for downlink communication, as explained in the simulation results, the same discussions hold for the cases with uplink transmission.

	\subsection{FSO Link}
	During the first phase of transmission, the information signal received at the AP through the FSO link  can be expressed as
	\begin{align}
		y_{\text{F}}\left(t\right)=\left({P_\text{F}}\eta I_{\text{l}} I\right)^{\frac{\tau}{2}}s(t)+n_{\text{F}}\left(t\right),
	\end{align}
	where $P_\text{F}$ is the transmit power at the source, $I=I_\text{a} I_\text{p}$ is the FSO link's fading coefficient, wherein $I_\text{a}$ is the atmospheric turbulence induced fading and $I_\text{p}$ is the pointing error impairment during the transmission through the FSO link. Further, $I_\text{l}$ represents the atmospheric attenuation associated with the FSO link.
	Here, $\eta$ represents the coefficient of the optical-to-electrical conversion and $\tau$ is a selection parameter of FSO detector's type, where $\tau=1$ and $\tau=2$ represent respectively the heterodyne detection and IM/DD.  
	Also, $n_{\text{F}}(t)\sim \mathcal{N}(0, \sigma_{\text{o}}^2)$ is the  zero mean and $\sigma_{\text{o}}^2$ variance additive white Gaussian noise (AWGN) associated with the FSO link.  	
	The FSO link's instantaneous received SNR at the AP will be
	\begin{align}
		\gamma_{\text{F}}=\frac{\left({P_{\text{F}}}\eta I_{{\text{l}}} I\right)^\tau}{\sigma^2_{\text{o}}}.
	\end{align}
	
	\subsubsection{Atmospheric Attenuation} Beer-Lambert Law \cite{henniger2010introduction} is used to model the atmospheric attenuation as
	\begin{align}
		I_{\text{l}}=\text{exp}\left(-\sigma_{\text{l}}L_{\text{F}}\right).
	\end{align}
	Here, $L_{\text{F}}$ is the FSO  link distance and $\sigma_{\text{l}}$ is the attenuation coefficient which can be modeled as
	\begin{align}
		\sigma_{\text{l}}=\frac{3.912}{\text{Vi}}\left(\frac{\lambda_{\text{F}}}{550}\right)^{\text{q}\left(\text{Vi}\right)},
	\end{align}
	where $\text{Vi}$ represents the visibility and $\text{q}\left(\text{Vi}\right)$ is the visibility
	coefficient of  atmospheric attenuation which is given as
	\begin{align}
		\text{q}(\text{Vi})=
		\begin{cases}
			1.6,  & \text{for } \text{Vi} > 50 \,\text{km} \\
			1.3,  & \text{for } 6 < \text{Vi} < 50 \,\text{km}\\
			0.585\text{Vi}^{1/3},  & \text{for }  \text{Vi} < 6 \,\text{km}.
		\end{cases}
	\end{align}
	\subsubsection{Pointing/Misalignment Error}
	We assume that an $\omega_{\text{L,i}}$ beamwidth Gaussian beam is propagated from the transmitter to the receiver of $a_\text{i}$ aperture radius which is ${L_\text{i}}$ distance apart. Hence, the approximate received optical power is  given by \cite{farid2007outage}
	\begin{align}\label{pnt}
		I_{\text{p,i}}\left(r_{\text{d,i}};L_\text{i}\right)\approx A_{\text{0,i}}~ \text{exp}\bigg(-\frac{2r_{\text{d,i}}^2}{\omega^2_{\text{L}_{\text{eq}},\text{i}}}\bigg),
	\end{align}
	where $r_{\text{d,i}}$ represents the radial displacement between the beam center and the detector, $A_{0,\text{i}}$ is the fraction of the collected power at $r_{\text{d,i}}=0$, and $\omega^2_{\text{L}_{\text{eq}},\text{i}}=\frac{\sqrt{A_{0,\text{i}}\pi}}{2v_{0,\text{i}}~\text{exp}(-v_{0,\text{i}}^2)}\omega_{\text{L,i}}^2$ is the equivalent beamwidth.
	Here, subscript $\text{i} \in [\text{T,F}]$ represents the parameters for the THz or FSO links, respectively. Note that the same pointing/misalignment error modeling is considered  for both FSO and THz links, however, the selection parameters may differ for the links.
	Further, $A_{0,\text{i}}=\left[\text{erf}(v_{0,\text{i}})\right]^2$, where $v_{\text{0,i}}=\sqrt{\frac{a_\text{i}^2\pi}{2\omega_{\text{L,i}}^2}}$ and $\text{erf}\left(x\right)=\frac{2}{\sqrt{\pi}}\int_{0}^{x}\exp\left(-t^2\right)\text{d}t$ is the error function.
	Approximation (\ref{pnt}) satisfies only if $\omega_{\text{L,i}} > 6a_\text{i}$.\par
	
\begin{figure}[]
		\centering
		\includegraphics[width=2.8in,height=2.1in]{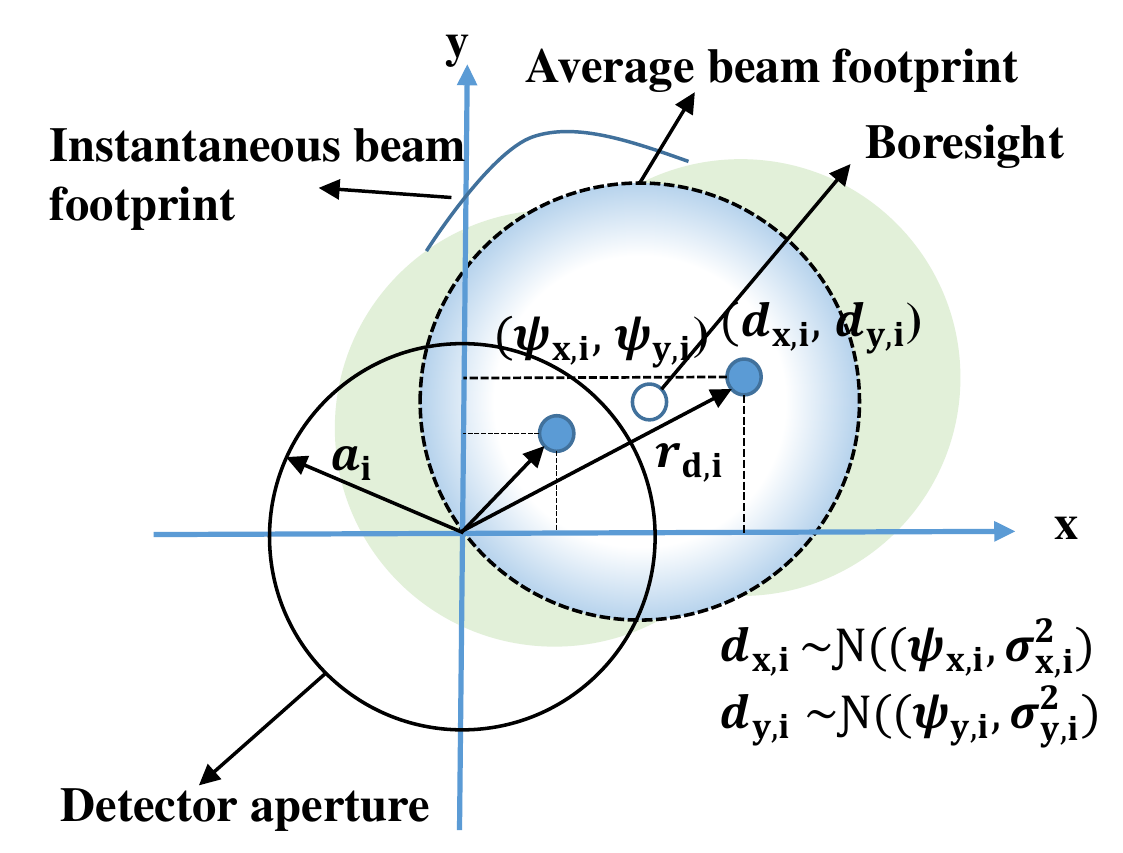}
		\caption{An illustration of the pointing/misalignment error.}
		\label{pntfig}
	\end{figure}

	A graphical representation of pointing errors is shown in Fig. \ref{pntfig}.	Let us consider that the beam displacements in the horizontal and vertical directions are respectively $d_{\text{x,i}}$ and $d_{\text{y,i}}$ in the detector plane. Then, the radial displacement  vector is $r_{\text{d,i}}=\left[d_{\text{x,i}},d_{\text{y,i}}\right]^\text{T}$, as shown in Fig. \ref{pntfig}.
	Let's assume $d_{\text{x,i}} \approx \mathcal{N}\left(\psi_{\text{x,i}},\sigma_{\text{x,i}}^2\right)$ and $d_{\text{y,i}} \approx \mathcal{N}\left(\psi_{\text{y,i}},\sigma_{\text{y,i}}^2\right)$ are independent and Gaussian distributed, then the radial vector's amplitude $|r_{\text{d,i}}|=\sqrt{d_{\text{x,i}}^2+d_{\text{y,i}}^2}$  is Beckmann distributed. Therefore, with various jitter and boresight values, pointing error can be modeled in various ways \cite{jung2020unified}.
	If no boresight error and identical jitters are considered, the mean values of the  radial vectors $d_{\text{x,i}}$ and $d_{\text{y,i}}$ are zero (i.e., $\psi_{\text{x,i}}$ = $\psi_{\text{y,i}}$ =0) with identical variance (i.e., $\sigma_{\text{x,i}}^2=\sigma_{\text{y,i}}^2=\sigma_{\text{j,i}}^2$). For such a case, radial displacement can be modeled by Rayleigh distribution and the probability density function (PDF) of the pointing error can be represented as
	\begin{align}\label{Ray_PDF}
		f_{\text{I}_{\text{p,i}}}(x)=\frac{\xi_\text{i}^2}{A_{0,\text{i}}^{\xi_\text{i}^2}}x^{\xi_\text{i}^2-1},~~~~ \text{for } 0\leq x \leq A_{0,\text{i}}
	\end{align}
	where $\xi_\text{i}=\frac{\omega_{\text{L}_{\text{eq}},\text{i}}}{2\sigma_{\text{j,i}}}$ represents the ratio of the equivalent beam radius and  jitter standard deviation at the receiver. 	
	
	The atmospheric turbulence $I_{\text{a}}$ is modeled with Gamma-Gamma distribution, hence, its PDF is given as
	\begin{align}\label{GG}
		f_{I_{\text{a}}}(x)&=\frac{2\left(\alpha_{\text{F}}\beta_{\text{F}}\right)^{\frac{\alpha_{\text{F}}+\beta_{\text{F}}}{2}}}{\Gamma(\alpha_{\text{F}})\Gamma(\beta_{\text{F}})}\nonumber\\&
		\times
		x^{\frac{\alpha_{\text{F}}+\beta_{\text{F}}}{2}-1}\text{K}_{\alpha_{\text{F}}-\beta_{\text{F}}}\left(2\sqrt{\alpha_{\text{F}}\beta_{\text{F}} x}\right),
	\end{align}
	where $\text{K}_{{v}}(x)=\frac{\Gamma\left(v+\frac{1}{2}\right)\left(2x\right)^v}{\sqrt{\pi}}\int_{0}^{\infty}\frac{\text{cos}(t)}{\left(x^2+t^2\right)^{v+\frac{1}{2}}}\text{d}t$ represents the ${v}^{\text{th}}$ order modified Bessel function of second kind \cite[(9.6.25)]{abramowitz1988handbook} and $\Gamma(x)=\int_{0}^{\infty}t^{x-1}\exp\left(-t\right)\text{d}t$ is the complete Gamma function \cite[(6.1.1)]{abramowitz1988handbook}. Considering the plane wave propagation, the fading coefficients of the atmospheric turbulence are given as
	\begin{align}\label{Atm_tur}
		\alpha_{\text{F}}=\left[\text{exp}\left(\frac{0.49\sigma_{\text{RV}}^2}{\left(1+1.11\sigma_{\text{RV}}^{12/5}\right)^{7/6}}\right)-1\right]^{-1},\nonumber\\
		\beta_{\text{F}}=\left[\text{exp}\left(\frac{0.51\sigma_{\text{RV}}^2}{\left(1+0.69\sigma_{\text{RV}}^{12/5}\right)^{5/6}}\right)-1\right]^{-1},
	\end{align}
	where $\sigma_{\text{RV}}^2=1.23C_{{n}}^2k_{{0}}L_{\text{F}}^{11/6}$ is the Rytov variance corresponding to the turbulence strength metric. Here, $k_{\text{0}}=2\pi/\lambda_{\text{F}}$ is the wave number, $L_{\text{F}}$ is the FSO link length, and $C_{{n}}^2$ is the refractive index structure parameter related to the atmospheric turbulence strength.
	The PDF of the FSO channel fading coefficient ($I$), having the combined impact of the atmospheric turbulence and pointing error, is given as
	\begin{align}\label{main}
		f_{\text{I}}(I)=\int f_{I|I_{\text{a}}}(I|I_{\text{a}})~f_{I_{\text{a}}}(I_{\text{a}})\text{d}I_{\text{a}}.
	\end{align}

	Considering the  identical jitters and zero boresight error case,  the PDF of the FSO link is given by 
	\begin{align}\label{combined1}
		f_{\text{I}}(I)=\frac{\xi^2_{\text{F}}\alpha_{\text{F}}\beta_{\text{F}}}{A_{0,\text{F}}I_{\text{l}}\Gamma(\alpha_{\text{F}})\Gamma(\beta_{\text{F}})}\text{G}^{3,0}_{1,3}\left[\frac{\alpha_{\text{F}}\beta_{\text{F}}}{A_{0,\text{F}}I_\text{l}}I\Big{|}^{~~~~~~~\xi^2_{\text{F}}}_{\xi^2_{\text{F}}-1,\alpha_{\text{F}}-1, \beta_{\text{F}}-1}\right],	
	\end{align}
	where $\text{G}^{m,n}_{p,q}\left[z\big{|}^{a_1,...,a_n,a_{n+1},...,a_p}_{b_1,...,b_m,b_{m+1},...,b_q}\right]$ represents the Meijer-G function.
	The instantaneous SNR of the FSO link is calculated as $\gamma_\text{F}=\delta_{\tau} I^{{\tau}}$, where  $\delta_{\tau}$ is the average received power of the FSO link. Hence, applying the change of variable $I=\left(\frac{\gamma_\text{F}}{\delta_{\tau}}\right)^{{\frac{1}{\tau}}}$ in (\ref{combined1}), the PDF of the FSO link's SNR is given by
	\begin{align}\label{Gamma_PDF}
		f_{\gamma_{\text{F}}}(\gamma_{\text{F}})&=\frac{\xi^2_{\text{F}}}{\tau\Gamma(\alpha_{\text{F}})\Gamma(\beta_{\text{F}})\gamma_{\text{F}}}\text{G}^{3,0}_{1,3}\left[\frac{\alpha_{\text{F}} \beta_{\text{F}}}{A_{0,{\text{F}}}}\left(\frac{\gamma_{\text{F}}}{\delta_{{\tau}}}\right)^{\frac{1}{\tau}}\Big{|}^{\xi^2_{\text{F}}+1}_{\xi^2_{\text{F}},\alpha_{\text{F}},\beta_{\text{F}}}\right].
	\end{align}
	Substituting (\ref{Gamma_PDF}) in $F_{{\gamma_{\text{F}}}}(x)=\int_{0}^{x}f_{\gamma_{\text{F}}}(\gamma)\text{d}\gamma$ and applying \cite[(07.34.21.0084.01)]{wolframe} with some mathematical computations, the cumulative distribution function (CDF) of the  FSO link's SNR for both  detection techniques is given as 
	\begin{align}\label{Gamma_CDF}
		F_{\gamma_{\text{F}}}(\gamma_{\text{F}})=\mathbb{D}_1\text{G}^{3\tau,1}_{\tau+1,3\tau+1}\left[\frac{\mathbb{D}_2\gamma_{\text{F}}}{\left(A_{0,{\text{F}}}\right)^{{\tau}}\delta_{{\tau}}}\Big{|}^{1, \varrho_1}_{\varrho_2, 0}\right],
	\end{align}
	where 
	\begin{align} \varrho_1&=\left[\frac{\xi_{\text{F}}^2+1}{\tau},...,\frac{\xi_{\text{F}}^2+\tau}{\tau}\right],\nonumber\\
		\varrho_2&=\bigg[\frac{\xi_{\text{F}}^2}{\tau},...,\frac{\xi_{\text{F}}^2+\tau-1}{\tau},\frac{\alpha_{\text{F}}}{\tau},...,\frac{\alpha_{\text{F}}+\tau-1}{\tau},\nonumber\\&
		~~~~\frac{\beta_{\text{F}}}{\tau},...,\frac{\beta_{\text{F}}+\tau-1}{\tau}\bigg],
		\nonumber\\
		\mathbb{D}_1&=\frac{\tau^{\left(\alpha_{\text{F}}+\beta_{\text{F}}-2\right)}\xi_{\text{F}}^2}{\left(2\pi\right)^{\tau-1}\Gamma\left(\alpha_{\text{F}}\right)\Gamma\left(\beta_{\text{F}}\right)}, 
		\mathbb{D}_2=\frac{\left(\alpha_{\text{F}} \beta_{\text{F}}\right)^\tau}{\tau^{2\tau}}. 
	\end{align}
	
	\subsection{THz Link}	
		To improve the robustness and the diversity performance of the THz link, we consider multiple antennas $N_{\text{r}}$ at the THz receiver with maximum ratio combiner (MRC).
	Here, the signal received from the source at the THz receiver of the AP is given by
	\begin{align}	
		\fontsize{10pt}{10pt}\selectfont	
		y_{{\text{T}}}(t)=\sqrt{P_{\text{T}}} h_{\text{l}}\sum_{j=1}^{N_{\text{r}}}h_{{\text{T}}_{{j}}}~ s(t)+n_{{\text{T}}}(t),
	\end{align}
	where $P_\text{T}$ is the total source power, $h_{\text{l}}$ is the path-loss, and $h_{\text{T}_{j}}=h_{\text{p}_j}h_{\text{f}_j}$ is the channel fading coefficient of the THz link at the $j^{\text{th}}$ antenna which consists of the combined impact of antenna misalignment $h_{\text{p}_j}$ and channel fading $h_{\text{f}_j}$. Further, $n_{\text{T}}$ represents the AWGN with  mean 0 and variance $\sigma_{\text{T}}^2$.
	After MRC, the instantaneous received SNR at the THz receiver is given as
	\begin{align}
		\gamma_{\text{T}}=\sum_{j=1}^{N_\text{r}}\gamma_{\text{T}_j}=\frac{P_{\text{T}}h_{\text{l}}^2||h_\text{T}||^2}{\sigma_\text{T}^2},
	\end{align}
	where $||h_\text{T}||^2=h_{\text{p}}^2\sum_{k=1}^{N_\text{r}}|h_{\text{f}_j}|^2$ is the combined THz channel gain after MRC. Since the receiving antennas are close to each other and the THz transmitter and receiver are not very far, the pointing error for each antenna link can be considered the same. The gain of the THz link's channel coefficient  $h_{\text{f}_j}$ is statistically characterized by generalized and easily tractable $\alpha-\mu$ fading which has been experimentally validated in  \cite{papasotiriou2021experimentally} with excellent fitting accuracy. At the THz receiver, the sum of $N_\text{r}$ random variables (RVs) with $\alpha-\mu$ distribution can also be well approximated with a single $\alpha-\mu$ RV. Further, the PDF of the misalignment error is Rayleigh distributed as shown in (\ref{Ray_PDF}). Hence, $||h_{\text{T}}||$ which represents the channel coefficient having the impact of both the misalignment error and channel fading follows the PDF 
	\begin{align}\label{PDF1}
		f_{||h_{\text{T}}||}(x)= \mathbb{C}_1x^{\xi_\text{T}^2-1}\Gamma\left(\mathbb{C}_2,\mathbb{C}_3x^{\alpha}\right),
	\end{align}
	where $\Gamma\left(a,x\right)=\int_{x}^{\infty} t^{a-1}\exp\left(-t\right)\text{d}t$ represents the upper incomplete Gamma function. Further, $\mathbb{C}_1=\frac{\xi_\text{T}^2}{\left(A_{0,\text{T}}\right)^{\xi_\text{T}^2}}\frac{(N_\text{r}\mu)^{\xi_\text{T}^2/\alpha}}{\Omega^{\xi_\text{T}^2}\Gamma(N_\text{r}\mu)}$, $\mathbb{C}_2=\frac{\alpha N_\text{r}\mu-\xi_{\text{T}}^2}{\alpha}$, $\mathbb{C}_3=\frac{N_{\text{r}}\mu}{\left(A_{0,\text{T}}\right)^{\alpha}\Omega^{\alpha}}$, and $\Omega$ is the $\alpha$-root mean value of the fading channel.	
 The instantaneous received SNR of the THz link is given as 
$\gamma_\text{T}=\bar{\gamma}_{\text{T}}||h_{\text{T}}||^2$,		
where $\bar{\gamma}_{\text{T}}=\frac{P_\text{s}h_{\text{l}}^2}{\sigma_{\text{T}}^2}$ is the average received SNR of the THz link.
Hence, the change of variables can be applied as $||h_{\text{T}}||=\sqrt{(\gamma_{\text{T}}/\bar{\gamma}_{\text{T}})}$ in (\ref{PDF1}), and the PDF of the THz link's SNR  is represented as
\begin{align}\label{PDF_THz}
	\fontsize{10pt}{10pt}\selectfont
	f_{\gamma_{\text{T}}}(\gamma_{\text{T}})=\frac{\mathbb{C}_1}{2(N_\text{r}\bar{\gamma}_{\text{T}})^{\frac{\xi_{\text{T}}^2}{2}}}\gamma_{\text{T}}^{\frac{\xi_{\text{T}}^2}{2}-1}\Gamma\left(\mathbb{C}_2,\mathbb{C}_3\left(\frac{\gamma_{\text{T}}}{N_\text{r}\bar{\gamma}_{\text{T}}}\right)^{\frac{\alpha}{2}}\right),
\end{align}	
Performing  $F_{\gamma_{\text{T}}}(\gamma_{\text{T}})=\int_{0}^{\gamma_{\text{T}}}f_{\gamma_{\text{T}}}(\gamma)\text{d}\gamma$ with some mathematical computations,
the CDF of the THz link's SNR is derived as
\begin{align}\label{CDF_THz}
	\fontsize{10pt}{10pt}\selectfont
	F_{\gamma_{\text{T}}}(\gamma_{\text{T}})&=\frac{\mathbb{C}_1}{\alpha(\bar{\gamma})^{\frac{\xi_{\text{T}}^2}{2}}}\gamma_{\text{T}}^{\frac{\xi_{\text{T}}^2}{2}}\text{G}^{2,1}_{2,3}\left[\mathbb{C}_3\left(\frac{\gamma_{\text{T}}}{\bar{\gamma}}\right)^{\frac{\alpha}{2}}\Big{|}^{1-\frac{\xi_{\text{T}}^2}{\alpha},1}_{0,\mathbb{C}_2,-\frac{\xi_{\text{T}}^2}{\alpha}}\right],
\end{align}
where $\bar{\gamma}=N_\text{r}\bar{\gamma}_{\text{T}}$.
	
	\subsubsection{Path-loss Modeling}
	Path-loss $h_{\text{l}}$ is a deterministic function and can be calculated as\cite{bhardwaj2021performance,kokkoniemi2021line}
	\begin{align}\label{THzPloss}
		h_{\text{l}}=\frac{c\sqrt{G^{\text{T}}_{\text{t}}G^{\text{T}}_{\text{r}}}}{4\pi f_{\text{T}}d_{\text{T}}}\text{exp}\left(-\frac{1}{2}d_{\text{T}}\left(\sum_{\hat{i}}\mathbb{F}_{\hat{i}}(f_{\text{T}},\nu)+\mathbb{G}(f_{\text{T}},\nu)\right)\right),
	\end{align}
	where $G^{\text{T}}_{\text{t}}$ and $G^{\text{T}}_{\text{r}}$ are the respective transmit and receive antenna gains, $f_{\text{T}}$ is the operating frequency, and $d_{\text{T}}$ is the THz link's length. Further, polynomial $\mathbb{G}_{\hat{i}}(f_{\text{T}},\nu)$ is required to fit (\ref{THzPloss}) with the actual theoretical response.
	Also, $\mathbb{F}_{\hat{i}}(f_{\text{T}},\nu)$ represents six polynomials for six major absorption lines centered at 119, 183, 325, 380, 439, and 448 GHz frequencies, can be defined as
	\begin{align}
		\mathbb{F}_1(f_{\text{T}},\nu)&=\frac{\mathbb{H}_1(\nu)}{\mathbb{H}_2(\nu)+\left(\frac{f_{\text{T}}}{100c}-\hat{j}_1\right)^2},\nonumber\\
		\mathbb{F}_2(f_{\text{T}},\nu)&=\frac{\mathbb{H}_3(\nu)}{\mathbb{H}_4(\nu)+\left(\frac{f_{\text{T}}}{100c}-\hat{j}_2\right)^2},\nonumber\\
		\mathbb{F}_3(f_{\text{T}},\nu)&=\frac{\mathbb{H}_5(\nu)}{\mathbb{H}_6(\nu)+\left(\frac{f_{\text{T}}}{100c}-\hat{j}_3\right)^2},\nonumber\\
		\mathbb{F}_4(f_{\text{T}},\nu)&=\frac{\mathbb{H}_7(\nu)}{\mathbb{H}_8(\nu)+\left(\frac{f_{\text{T}}}{100c}-\hat{j}_4\right)^2},\nonumber\\
		\mathbb{F}_5(f_{\text{T}},\nu)&=\frac{\mathbb{H}_9(\nu)}{\mathbb{H}_{10}(\nu)+\left(\frac{f_{\text{T}}}{100c}-\hat{j}_5\right)^2},\nonumber\\
		\mathbb{F}_6(f_{\text{T}},\nu)&=\frac{\mathbb{H}_{11}(\nu)}{\mathbb{H}_{12}(\nu)+\left(\frac{f_{\text{T}}}{100c}-\hat{j}_6\right)^2},\nonumber\\
		\mathbb{G}(f_{\text{T}},\nu)&=\frac{\nu}{0.0157}\left(2\times10^{-4}+D_0f_{\text{T}}^{D_1}\right),
	\end{align}
	where
	\begin{align}
		\mathbb{H}_1(\nu)&=5.159\times10^{-5}(1-\nu)\nonumber\\&
		\times\left(-6.65\times10^{-5}(1-\nu)+0.0159\right),\nonumber\\
		\mathbb{H}_2(\nu)&=\left(-2.09\times10^{-4}(1-\nu)+0.05\right)^2,\nonumber\\
		\mathbb{H}_3(\nu)&=0.1925\nu\left(0.1350\nu+0.0318\right), \nonumber\\
		\mathbb{H}_4(\nu)&=\left(0.4241\nu+0.0998\right)^2, \nonumber\\
		\mathbb{H}_5(\nu)&= 0.2251\nu\left(0.1314\nu+0.0297\right), \nonumber\\
		\mathbb{H}_6(\nu)&=\left(0.4127\nu+0.0932\right)^2,\nonumber\\
		\mathbb{H}_7(\nu)&=2.053\nu\left(0.1717\nu+0.0306\right),\nonumber\\
		\mathbb{H}_8(\nu)&=\left(0.5394\nu+0.0961\right)^2,\nonumber\\
		\mathbb{H}_9(\nu)&=0.177\nu\left(0.0832\nu+0.0213\right),\nonumber\\
		\mathbb{H}_{10}(\nu)&=\left(0.2615\nu+0.0668\right)^2,\nonumber\\
		\mathbb{H}_{11}(\nu)&=2.146\nu\left(0.1206\nu+0.0277\right),\nonumber\\
		\mathbb{H}_{12}(\nu)&=\left(0.3789\nu+0.0871\right)^2,\nonumber
	\end{align}
	$\hat{j}_1=$3.96 cm$^{-1}$, $\hat{j}_2=$6.11 cm$^{-1}$, $\hat{j}_3=$10.84 cm$^{-1}$, $\hat{j}_4=$12.68 cm$^{-1}$, $\hat{j}_5=$14.65 cm$^{-1}$, $\hat{j}_6=$14.94 cm$^{-1}$, $D_0=0.915\times10^{-112}$, $D_1=9.42$, and $\nu=\frac{\chi}{100p}\mathbb{I}(T,p)$ with $\mathbb{I}(T,p)=6.1121\left(1.0007+3.46\times10^{-6}p\right)\text{exp}\left(\frac{17.502T}{240.97+T}\right)$.  
	
	\subsection{Terrestrial mmWave Access Links}
	We consider a terrestrial region, where multiple terrestrial MUs are to be served from the AP via mmWave access links. With DF relaying, the AP first decodes the signal coming through the hybrid FSO/THz link, re-encodes it (if the message decoding is successful), and finally transmits it to the terrestrial MUs. 
	Signal received through the hybrid FSO/THz link can only be decoded successfully if the instantaneous received SNR at the AP is greater than or equal to the threshold SNR. 
	
	For improved and robust data-transmission (received through the backhaul hybrid FSO/THz link) to the terrestrial MUs, $N_{\text{t}}$ transmit antennas are considered at the AP, while single antenna is considered at the MUs due to, e.g., the size limit. Let, $\hat{s}(t)$ be the decoded signal at the AP, then after maximal ratio transmission (MRT), the signal received to each MU through the mmWave access link  is represented as
	\begin{align}
		y_{\text{RU}}=\sqrt{P_\text{R} p_\text{l}}\sum_{k=1}^{N_\text{t}}h_{\text{kU}}\hat{s}(t)+n_{\text{U}}(t),
	\end{align}
	where $P_{\text{R}}$ is the transmit power at the AP, $p_{\text{l}}$ is the path-loss, $h_{\text{kU}}$ is the channel gain of $k^{\text{th}}$ antenna to an MU, and $n_{\text{U}}(t)$ is the AWGN with mean 0 and variance $\sigma_\text{R}^2$ associated with the access link.
	The gain of each access link is characterized by the Gamma distribution i.e., $\mathcal{G}(m, \Omega_\text{R})$, with $m$ fading severity and $\Omega_\text{R}$ average power. At the MU, the signals coming from $N_\text{t}$ transmit antennas are combined. Hence, the sum of $N_\text{t}$ Gamma RVs can also be well approximated with a single Gamma RV as $\approx \mathcal{G}(mN_\text{t}, \Omega_\text{R})$. Thus, the instantaneous received SNR at each MU is given as $\gamma_\text{R}=\frac{P_\text{R}p_\text{l}||h_\text{R}||^2}{\sigma_\text{R}^2}$, where $||h_\text{R}||^2=\sum_{k=1}^{N_\text{t}}|h_{\text{kU}}|^2$.
	Therefore, the  PDF and the CDF expressions of the received SNR to each MU are given as 
	\begin{align}\label{CPDF_Nak}
		f_{\gamma_{\text{RU}}}(\gamma_{\text{R}})&=\frac{1}{\Gamma\left(mN_\text{t}\right)}\left(\frac{m}{\bar{\gamma}_{\text{R}}}\right)^{mN_\text{t}}\gamma_{\text{R}}^{mN_\text{t}-1}\exp{\left(-\frac{m}{\bar{\gamma}_{\text{R}}}\gamma_{\text{R}}\right)},\nonumber\\
		F_{\gamma_{\text{RU}}}(\gamma_{\text{R}})&=1-\frac{1}{\Gamma\left(mN_\text{t}\right)}\Gamma\left(mN_\text{t},\frac{m}{\bar{\gamma}_{\text{R}}}\gamma_{\text{R}}\right),
	\end{align}
	respectively, where $\bar{\gamma}_{\text{R}}=\Omega_\text{R}\frac{P_\text{R}p_\text{l}}{\sigma_\text{R}^2}$ is the average received SNR. 
	
	\subsubsection{Path-loss Modeling} The path-loss $p_\text{l}$ associated with the mmWave access link is given as \cite{he2009bit}
	\begin{align}
		p_l[\text{dB}] = G^\text{R}_{\text{t}} + G^\text{R}_{\text{r}}-20\log_{10}\left(\frac{4\pi L_\text{R}}{\lambda_\text{R}}\right)-\left(a_{\text{oxy}}+a_{\text{rain}}\right)L_\text{R},
	\end{align}
	where $G^\text{R}_\text{t}$ and $G^\text{R}_\text{r}$ represent respectively the transmit and the receiver antenna gains, $L_\text{R}$ and $\lambda_\text{R}$ are the link distance and wavelength of the mmWave access link, respectively, and $a_{\text{oxy}}$ and $a_{\text{rain}}$ represent the oxygen absorption and the rain attenuation, respectively.
	
	\section{Performance Analysis with Hard Switching}
	\subsection{Outage Probability}
	With hard switching, the FSO link is active, if the FSO link's SNR  is greater than or equal to a predefined threshold i.e.,  $\gamma_{\text{F}}\geq\gamma_{\text{th}}$. However, with $\gamma_{\text{F}}<\gamma_{\text{th}}$, receiver generates a feedback to activate the THz link if $\gamma_{\text{T}}\geq\gamma_{\text{th}}$. 
	Hence, the instantaneous received SNR ($\gamma_\text{c}$) of the hybrid FSO/THz link is given as 
	\begin{align}
		\gamma_\text{c}=
		\begin{cases}
			\gamma_{\text{F}}, & \text {if  } \gamma_\text{F}\geq \gamma_{\text{th}}\\
			\gamma_{\text{T}}, & \text {if  } \gamma_\text{F} < \gamma_{\text{th}},~ \gamma_\text{T}\geq \gamma_{\text{th}}\\
			0, & \text {if  } \gamma_\text{F} < \gamma_{\text{th}},~ \gamma_\text{T}< \gamma_{\text{th}}.
		\end{cases}
	\end{align}
	
	The hybrid FSO/THz link suffers from outage if both the FSO and THz links' instantaneous SNRs ($\gamma_\text{F}$ and $\gamma_\text{T}$) are below $\gamma_{\text{th}}$.
	Therefore, the outage probability of the hybrid FSO/THz link  with hard switching is calculated as
	\begin{align}\label{out11}
		\mathcal{P}^{\text{H}}_{\gamma_\text{c}}(\gamma_{\text{th}})&=F^{\text{H}}_{\gamma_\text{c}}(\gamma_{\text{th}})=\mathcal{P}\left[\gamma_{\text{F}}< \gamma_{\text{th}}, \gamma_{\text{T}}<\gamma_{\text{th}}\right].
	\end{align}
	The FSO and the THz links are statistically independent, hence, the outage probability is obtained as
	\begin{align}\label{out12}
		\mathcal{P}^{\text{H}}_{\gamma_\text{c}}\left(\gamma_{\text{th}}\right)&=F_{\gamma_{\text{F}}} \left(\gamma_{\text{th}}\right)\times F_{\gamma_{\text{T}}} \left(\gamma_{\text{th}}\right).
	\end{align}
	
	Substituting (\ref{CDF_THz}) and (\ref{Gamma_CDF}) in (\ref{out12}), outage probability of the hybrid FSO/THz link is derived as
	\begin{align}\label{outF1}
		\mathcal{P}^{\text{H}}_{\gamma_\text{c}}\left(\gamma_{\text{th}}\right)&=\frac{\mathbb{C}_1\mathbb{D}_1}{\alpha(\bar{\gamma})^{\frac{\xi_\text{T}^2}{2}}}\gamma_{\text{th}}^{\frac{\xi_\text{T}^2}{2}}\text{G}^{2,1}_{2,3}\left[\mathbb{C}_3\left(\frac{\gamma_{\text{th}}}{\bar{\gamma}}\right)^{\frac{\alpha}{2}}\Big{|}^{1-\frac{\xi_\text{T}^2}{\alpha},1}_{0,\mathbb{C}_2,-\frac{\xi_\text{T}^2}{\alpha}}\right]\nonumber\\&
		\times
		\text{G}^{3\tau,1}_{\tau+1,3\tau+1}\left[\frac{\mathbb{D}_2\gamma_{\text{th}}}{\left(A_{0,\text{F}}\right)^\tau\delta_{\tau}}\Big{|}^{1, \varrho_1}_{\varrho_2, 0}\right].
	\end{align}
	
	Moreover, considering DF relaying, the E2E outage probability is given as
\begin{align}\label{OUT_F_hard}
	\fontsize{10pt}{10pt}\selectfont
	\mathcal{P}^{\text{H}}_{\text{E2E}}\left(\gamma^\text{R}_{\text{th}}\right)=1-	\underset{\overset{\text{Backhaul hybrid link}}{}}{\underbrace{\left(1-F^{\text{H}}_{\gamma_\text{c}}\left(\gamma_{\text{th}}\right)\right)}}~\underset{\overset{\text{mmWave access link}}{}}{\underbrace{\left(1-F_{\gamma_{\text{RU}}}\left(\gamma^\text{R}_{\text{th}}\right)\right)}},
\end{align}
	where $\gamma^\text{R}_{\text{th}}$ is the threshold SNR at the terrestrial MU.
	Substituting (\ref{outF1}) and (\ref{CPDF_Nak}) in (\ref{OUT_F_hard}), the E2E outage probability is derived as
	\begin{align}\label{OUT_final}
		&\mathcal{P}^{\text{H}}_{\text{E2E}}\left(\gamma^\text{R}_{\text{th}}\right)=1-\frac{1}{\Gamma\left(mN_\text{t}\right)}\Gamma\left(mN_\text{t},\frac{m}{\bar{\gamma}_{\text{R}}}\gamma^\text{R}_{\text{th}}\right)\nonumber\\&
		+\frac{\mathbb{C}_1\mathbb{D}_1\gamma_{\text{th}}^{\frac{\xi_\text{T}^2}{2}}}{\alpha\left(\bar{\gamma}\right)^{\frac{\xi_\text{T}^2}{2}}\Gamma(mN_\text{t})}\text{G}^{2,1}_{2,3}\left[\mathbb{C}_3\left(\frac{\gamma_{\text{th}}}{\bar{\gamma}}\right)^{\frac{\alpha}{2}}\Big{|}^{1-\frac{\xi_\text{T}^2}{\alpha},1}_{0,\mathbb{C}_2,-\frac{\xi_\text{T}^2}{\alpha}}\right]
		\nonumber\\&
		\times
		\text{G}^{3\tau,1}_{\tau+1,3\tau+1}\left[\frac{\mathbb{D}_2\gamma_{\text{th}}}{\left(A_{0,\text{F}}\right)^\tau\delta_{\tau}}\Big{|}^{1, \varrho_1}_{\varrho_2, 0}\right]\Gamma\left(mN_\text{t},\frac{m\gamma^\text{R}_{\text{th}}}{\bar{\gamma}_{\text{R}}}\right).
	\end{align}
	
	\subsection{Asymptotic Outage Probability}
	To derive the diversity order, we take high SNR approximation by assuming  the transmit SNR tends to infinity which results in  $\delta_{\tau},\bar{\gamma},\bar{\gamma}_\text{R}\rightarrow \infty$. In this case, (\ref{OUT_F_hard}) can be approximated as
	\begin{align}
		\mathcal{P}^{\text{H,A}}_{\text{E2E}}\left(\gamma^\text{R}_{\text{th}}\right)\approx F^\text{A}_{\gamma_\text{c}}\left(\gamma_{\text{th}}\right)+F^\text{A}_{\gamma_{\text{RU}}}\left(\gamma^\text{R}_{\text{th}}\right).
	\end{align}
	
	{\textit{Corollary 1:} By using the identities (\ref{FSOapp}), (\ref{THz_Asym}), and (\ref{RF_asym}) as obtained in Appendix A, asymptotic outage probability  of the considered system is approximated as
		\begin{align}\label{Out_Asym}
			&\mathcal{P}_{\text{E2E}}^{\text{H,A}}\left(\gamma^\text{R}_{\text{th}}\right)\approx
			\Bigg[\mathbb{D}_1\sum_{p=1}^{3\tau}\left(\frac{\mathbb{D}_2\gamma_{\text{th}}}{\left(A_{0,\text{F}}\right)^{\tau}\delta_{\tau}}\right)^{\varrho_8,p}\nonumber\\&
			\times	\frac{\prod_{\underset{q\neq p}{q=1}}^{3\tau} \Gamma\left(\varrho_{8,q}-\varrho_{8,p}\right)\prod_{\underset{q=1}{}}^{1}\Gamma\left(1-\varrho_{7,q}+\varrho_{7,p}\right)}{\prod_{\underset{q=2}{}}^{r+1}\Gamma\left(\varrho_{7,q}-\varrho_{8,p}\right)\prod_{\underset{3\tau+1}{}}^{3\tau+1}\Gamma\left(1-\varrho_{8,q}+\varrho_{8,p}\right)}\Bigg]\nonumber\\&
			\times\left[\frac{\mathbb{C}_1\Gamma\left(\mathbb{C}_2\right)}{\xi_{\text{T}}^2}\left(\frac{\gamma_{\text{th}}}{\bar{\gamma}}\right)^{\frac{\xi_\text{T}^2}{2}}-
			\frac{\mathbb{C}_1\mathbb{C}_3^{\mathbb{C}_2}}{\mathbb{C}_2\left(\alpha N_{\text{r}}\mu\right)}\left(\frac{\gamma_{\text{th}}}{\bar{\gamma}}\right)^{\frac{\alpha N_\text{r}\mu}{2}}\right]\nonumber\\&
			+\frac{1}{\Gamma\left(mN_\text{t}+1\right)}\left(\frac{m \gamma^\text{R}_{\text{th}}}{\bar{\gamma}_\text{R}}\right)^{mN_\text{t}}.
		\end{align} 
		\textbf{Proof: See Appendix A.}
		
		Here,  $\varrho_7=\left[1,\varrho_1\right]$ and $\varrho_8=\left[\varrho_2,0\right]$. From (\ref{FSOapp}), it is clear that at high SNRs, the FSO link's performance depends upon the atmospheric turbulence parameters, pointing error parameter, and on the FSO detector's type. Hence, the diversity order is obtained by $\min\left(\frac{\xi_{\text{F}}^2}{\tau},\frac{\alpha_{\text{F}}}{\tau},\frac{\beta_{\text{F}}}{\tau}\right)$.  From (\ref{THz_Asym}), at high SNRs, the THz link's performance depends upon the misalignment error parameter, $\alpha$, $\mu$ fading parameters, and on the number of RF antennas $N_\text{r}$ at the THz receiver.
		Hence, the diversity order of the THz link is decided by  $\min\left(\frac{\xi_{\text{T}}^2}{2},\frac{\alpha N_{\text{r}}\mu}{2}\right)$. Since the FSO and THz links are working in parallel, the diversity order of the hybrid FSO/THz link will be
		\begin{align}
			&G^{\text{Hyb}}_{\text{d}}=\min\bigg(\left(\frac{\xi_{\text{F}}^2}{\tau}+\frac{\xi_{\text{T}}^2}{2}\right),\left(\frac{\alpha_{\text{F}}}{\tau}+\frac{\xi_{\text{T}}^2}{2}\right),\left(\frac{\beta_{\text{F}}}{\tau}+\frac{\xi_{\text{T}}^2}{2}\right),\nonumber\\&
			\left(\frac{\xi_{\text{F}}^2}{\tau}+\frac{\alpha N_{\text{r}}\mu}{2}\right),\left(\frac{\alpha_{\text{F}}}{\tau}+\frac{\alpha N_{\text{r}}\mu}{2}\right),\left(\frac{\beta_{\text{F}}}{\tau}+\frac{\alpha N_{\text{r}}\mu}{2}\right)\bigg).
		\end{align} 
		Further, the mmWave access link's performance is limited by the fading severity parameter $m$ and by the number of RF antennas at the AP which is clear from   (\ref{RF_asym}). Hence, the diversity order of the access link is $mN_{\text{t}}$. Therefore, the E2E diversity order of the considered dual-hop communication system is limited by
		\begin{align}
			&G^{\text{E2E}}_{\text{d}}=\min\bigg(\left(\frac{\xi_{\text{F}}^2}{\tau}+\frac{\xi_{\text{T}}^2}{2}\right),\left(\frac{\alpha_{\text{F}}}{\tau}+\frac{\xi_{\text{T}}^2}{2}\right),\left(\frac{\beta_{\text{F}}}{\tau}+\frac{\xi_{\text{T}}^2}{2}\right),\nonumber\\&
			\left(\frac{\xi_{\text{F}}^2}{\tau}+\frac{\alpha N_{\text{r}}\mu}{2}\right),
			\left(\frac{\alpha_{\text{F}}}{\tau}+\frac{\alpha N_{\text{r}}\mu}{2}\right),\left(\frac{\beta_{\text{F}}}{\tau}+\frac{\alpha N_{\text{r}}\mu}{2}\right),mN_{\text{t}}\bigg).
		\end{align} 
		\subsection{Ergodic Capacity}
		Ergodic capacity (bits/s/Hz) of a communication link is obtained by taking the statistical expectation of the mutual information which is expressed as
		\begin{align}\label{capacity}
			\mathcal{C}\left(\gamma_{\text{th}}\right)=\int_{\gamma_{\text{th}}}^{\infty}\text{log}_2\left(1+\gamma\right)f_{\gamma}\left(\gamma\right) \text{d}\gamma,
		\end{align}
		where $f_{\gamma}\left(\gamma\right)$ represents the PDF of the SNR of the considered link.
		For the hybrid FSO/THz link with hard switching, the ergodic capacity is obtained as
		\begin{align}\label{Cap_Hard}
			\mathcal{C}^{\text{H}}_{\text{Hyb}}\left(\gamma_{\text{th}}\right)=	\mathcal{C}^{\text{F}}\left(\gamma_{\text{th}}\right)+\mathcal{P}_{\gamma_{\text{F}}}\left(\gamma_{\text{th}}\right)\mathcal{C}^{\text{T}}\left(\gamma_{\text{th}}\right),
		\end{align}
		where $\mathcal{P}_{\gamma_{\text{F}}}\left(\gamma_{\text{th}}\right)=F_{\gamma_{\text{F}}}\left(\gamma_{\text{th}}\right)$ is the CDF of the FSO link (\ref{Gamma_CDF}). Further, $\mathcal{C}^{\text{F}}\left(\gamma_{\text{th}}\right)$ and $\mathcal{C}^{\text{T}}\left(\gamma_{\text{th}}\right)$ are the ergodic capacity of the FSO and the THz links, respectively, as determined in the following.

		\subsubsection{Capacity of the FSO Link}
		$\mathcal{C}^{\text{F}}\left(\gamma_{\text{th}}\right)$ can be derived as
		\begin{align}\label{cap_F}
			\mathcal{C}^{\text{F}}\left(\gamma_{\text{th}}\right)=\int_{\gamma_{\text{th}}}^{\infty}\text{log}_2\left(1+\Xi\gamma\right)f_{\gamma_{\text{F}}}\left(\gamma\right)\text{d}\gamma,
		\end{align}
		where  $\Xi=\frac{e}{2\pi}$ for IM/DD ($\tau=2$) and $\Xi=1$ for the heterodyne detection ($\tau=1$). Note that   (\ref{cap_F}) is the lower-bound for IM/DD, however (\ref{cap_F}) is exact for the heterodyne detection.
		Finally, $\mathcal{C}^{\text{F}}\left(\gamma_{\text{th}}\right)$ is solved as
		\begin{align}\label{Cap_FSO}
			\mathcal{C}^{\text{F}}\left(\gamma_{\text{th}}\right)=\Theta_1 - \Theta_2,
		\end{align} 
		where identities $\Theta_1$ and $\Theta_2$ are respectively derived as
		\begin{align}\label{I1F}
			\Theta_1=\frac{\mathbb{D}_1}{\text{ln}(2)}\text{G}^{3\tau+2,1}_{\tau+2,3\tau+2}\left[\frac{\mathbb{D}_2}{\Xi\left(A_{0,\text{F}}\right)^\tau\delta_{\tau}}\Big{|}^{0,1,\varrho_1}_{\varrho_2,0,0}\right],
		\end{align}
		and 
		\begin{align}\label{I2F}
			\Theta_2&=\frac{\mathbb{D}_1\varpi}{\text{ln}(2)}\Bigg\{\left(\Xi\gamma_{\text{th}}\right)^{\frac{1}{\varpi}}\text{G}^{3\tau,1}_{\tau+1,3\tau+1}\left[\frac{\mathbb{D}_2\gamma_{\text{th}}}{\left(A_{0,\text{F}}\right)^\tau\delta_{{\tau}}}\Big{|}^{1-\frac{1}{\varpi},\varrho_1}_{\varrho_2,-\frac{1}{\varpi}}\right]\nonumber\\&
			-\text{G}^{3\tau,1}_{\tau+1,3\tau+1}\left[\frac{\mathbb{D}_2\gamma_{\text{th}}}{\left(A_{0,\text{F}}\right)^\tau\delta_{\tau}}\Big{|}^{1,\varrho_1}_{\varrho_2,0}\right]
			\Bigg\}.
		\end{align}
		Here, $\varpi$ is an arbitrary large number.
		
		\textbf{Proof: See Appendix B.}
		
		\subsubsection{Capacity of the THz Link}
		Capacity of the THz link $\mathcal{C}^{\text{T}}\left(\gamma_{\text{th}}\right)$ can be derived as
		\begin{align}\label{cap_T}
			\mathcal{C}^{\text{T}}\left(\gamma_{\text{th}}\right)&=\int_{\gamma_{\text{th}}}^{\infty}\text{log}_2\left(1+\gamma\right)f_{\gamma_{\text{T}}}\left(\gamma\right)\text{d}\gamma,\nonumber\\&=
			\Theta_3-\Theta_{4},
		\end{align}
		where
		\begin{align}\label{I3F}
			\Theta_3&=\frac{\mathbb{C}_12^{\left(\mathbb{C}_2-\frac{3}{2}\right)}}{\text{ln}\left(2\right)(\bar{\gamma})^{\frac{\xi_\text{T}^2}{2}}\alpha\left(2\pi\right)^{\alpha-\frac{1}{2}}}\text{G}^{4+2\alpha,\alpha}_{2+2\alpha,4+2\alpha}\left[\frac{\mathbb{C}_3^2}{4\bar{\gamma}^{\alpha}}\Big{|}^{\varrho_3}_{\varrho_4}\right],
		\end{align}
		and
		\begin{align}\label{I4F}
			\Theta_{4}&=\frac{\mathbb{C}_1 \varpi\gamma_{\text{th}}^{\left(\frac{\xi_\text{T}^2}{2}+\frac{1}{\varpi}\right)}}{\text{ln}\left(2\right)\alpha\left(\bar{\gamma}\right)^{\frac{\xi_\text{T}^2}{2}}} \text{G}^{2,1}_{2,3}\left[\mathbb{C}_3\left(\frac{\gamma_{\text{th}}}{\bar{\gamma}}\right)^{\frac{\alpha}{2}}\Big{|}^{1-\left(\frac{\xi_\text{T}^2}{\alpha}+\frac{2}{\alpha\varpi}\right),1}_{0,\mathbb{C}_2,-\left(\frac{\xi_\text{T}^2}{\alpha}+\frac{2}{\alpha\varpi}\right)}\right]\nonumber\\&
			-\frac{\mathbb{C}_1 \varpi\gamma_{\text{th}}^{\frac{\xi_\text{T}^2}{2}}}{\text{ln}(2)\alpha\left(\bar{\gamma}\right)^{\frac{\xi_\text{T}^2}{2}}} \text{G}^{2,1}_{2,3}\left[\mathbb{C}_3\left(\frac{\gamma_{\text{th}}}{\bar{\gamma}}\right)^{\frac{\alpha}{2}}\Big{|}^{1-\frac{\xi_\text{T}^2}{\alpha},1}_{0,\mathbb{C}_2,-\frac{\xi_\text{T}^2}{\alpha}}\right].
		\end{align}
		Here, $\varrho_3=\left[\frac{1-\frac{\xi_{\text{T}}^2}{2}-1}{\alpha},...,\frac{\alpha-\frac{\xi_{\text{T}}^2}{2}-1}{\alpha},\frac{1-\frac{\xi_{\text{T}}^2}{2}}{\alpha},...,\frac{\alpha-\frac{\xi_{\text{T}}^2}{2}}{\alpha},\frac{1}{2},1\right]$, 
		$\varrho_4=\left[0,\frac{1}{2},\frac{\mathbb{C}_2}{2},\frac{\mathbb{C}_2+1}{2},\frac{-\frac{\xi_{\text{T}}^2}{2}}{\alpha},...,\frac{\alpha-\frac{\xi_{\text{T}}^2}{2}-1}{\alpha},\frac{-\frac{\xi_{\text{T}}^2}{2}}{\alpha},...,\frac{\alpha-\frac{\xi_{\text{T}}^2}{2}-1}{\alpha}\right]$.

		\textbf{Proof: See Appendix B.}
		
		\subsubsection{Capacity of the mmWave Access Link}
		Capacity of the mmWave access link $\mathcal{C}^{\text{R}}\left(\gamma_{\text{th}}\right)$ is derived as
		\begin{align}\label{cap_R}
			\mathcal{C}^{\text{R}}\left(\gamma^{\text{R}}_{\text{th}}\right)&=\int_{\gamma^\text{R}_{\text{th}}}^{\infty}\text{log}_2\left(1+\gamma\right)f_{\gamma_{\text{RU}}}(\gamma)\text{d}\gamma,\nonumber\\&=
			\Theta_{5}-\Theta_{6},
		\end{align}
		where
		\begin{align}
			\Theta_{5}&=\frac{\left({m}/{\bar{\gamma}_{\text{R}}}\right)^{mN_\text{t}}}{\text{ln}\left(2\right)\Gamma\left(mN_\text{t}\right)}\int_{0}^{\infty}\text{ln}\left(1+\gamma\right)\gamma^{mN_\text{t}-1}\exp{\left(-\frac{m}{\bar{\gamma}_{\text{R}}}\gamma\right)}\text{d}\gamma,\nonumber\\
			\Theta_{6}&=\frac{\left({m}/{\bar{\gamma}_{\text{R}}}\right)^{mN_\text{t}}}{\text{ln}\left(2\right)\Gamma\left(mN_\text{t}\right)}\int_{0}^{\gamma^\text{R}_{\text{th}}}\text{ln}\left(1+\gamma\right)\gamma^{mN_\text{t}-1}\exp{\left(-\frac{m}{\bar{\gamma}_{\text{R}}}\gamma\right)}\text{d}\gamma.
		\end{align}
			Substituting $\text{ln}\left(1+\gamma\right)=\text{G}^{1,2}_{2,2}\left[\gamma\big{|}^{1,1}_{1,0}\right]$  in $\Theta_{5}$ and applying \cite[(7.813)]{gradshteyn2000table} with some mathematical computations, $\Theta_{5}$ is solved as
		\begin{align}
			\Theta_{5}&=\frac{1}{\text{ln}\left(2\right)\Gamma\left(mN_\text{t}\right)}\text{G}^{1,3}_{3,2}\left[\frac{1}{m/\bar{\gamma}_\text{R}}\Big{|}^{\left(-mN_\text{t}+1\right),1,1}_{1,0}\right].
		\end{align}
		Replacing $\text{ln}\left(1+\gamma\right)$ with its meijer-G equivalent in $\Theta_{6}$, using the exponential series expansion, and applying  \cite[(07.34.21.0084.01)]{wolframe}, $\Theta_{6}$ is solved as
		\begin{align}
			\Theta_{6}&=\frac{1}{\text{ln}(2)\Gamma\left(mN_\text{t}\right)}\sum_{j_3=0}^{\infty}\frac{\left(-1\right)^{j_3}}{j_3!}\left(\frac{m\gamma^\text{R}_{\text{th}}}{\bar{\gamma}_\text{R}}\right)^{mN_\text{t}+j_3}\nonumber\\&
			\times\text{G}^{1,3}_{3,3}\left[\gamma^\text{R}_{\text{th}}\Big{|}^{1-\left(mN_\text{t}+j_3\right),1,1}_{1,0,-\left(mN_\text{t}+j_3\right)}\right].
		\end{align}
		
		Finally, the E2E capacity at the terrestrial MU is given as
		\begin{align}
			\mathcal{C}^{\text{H}}_{\text{E2E}}=\text{min}\left(\mathcal{C}^{\text{H}}_{\text{Hyb}}\left(\gamma_{\text{th}}\right),\mathcal{C}^{\text{R}}\left(\gamma^\text{R}_{\text{th}}\right)\right),
		\end{align} 
		where $\mathcal{C}_{\text{Hyb}}^{\text{H}}\left(\gamma_{\text{th}}\right)$ and $\mathcal{C}^{\text{R}}\left(\gamma^\text{R}_{\text{th}}\right)$ are derived in (\ref{Cap_Hard}) and (\ref{cap_R}), respectively.
		
		\subsection{ABER Analysis}
		A generalized ABER expression for variety of modulation schemes is given by
		\begin{align}\label{BER_main}
			\mathcal{B}_e(\gamma_{\text{th}})= \sum_{\hat{p}=1}^{n_0}\int_{\gamma_{\text{th}}}^{\infty}\mathcal{P}\left(e|\gamma\right)
			f_{\gamma}\left(\gamma\right)\text{d}\gamma,
		\end{align}
		where $f_{\gamma}\left(\gamma\right)$ represents the PDF of the SNR of the fading link and $\mathcal{P}\left(e|\gamma\right)= A\times\text{erfc}\left(\sqrt{B_{\hat{p}}\gamma}\right)$ is the conditional error probability of the AWGN channel. Here,  $\text{erfc}\left(x\right)=\frac{2}{\sqrt{\pi}}\int_{x}^{\infty}\exp\left(-t^2\right)\text{d}t$ represents the complementary error function. For different modulation schemes, the values of $n_0$, $A$, and $B_{\hat{p}}$  are given in Table \ref{tab_BER}. Note that IM/DD is used for the on-off keying (OOK) scheme, while for the binary phase shift keying (BPSK), M-ary phase shift keying (M-PSK), and M-ary quadrature amplitude modulation (M-QAM) schemes, heterodyne detection is used at the FSO receiver.
		{\renewcommand{\arraystretch}{1.3}
			\begin{table}[h]
				\caption {{Selection parameters for various modulation schemes \cite{zedini2020performance}.}}
				\label{tab_BER}
				\begin{tabular}{|l|lll|}
					\hline
					Modulation & $n_0$                   & $A$                            & $B_{\hat{p}}$                                       \\ \hline
					OOK        & 1                   & 1/2                                   & 1/2                                         \\
					BPSK       & 1                  & 1/2                                   & 1                                           \\
					M-PSK      & $\text{Max}\left(\frac{M}{4},1\right)$ & $\frac{1}{\text{Max}\left(2,\log_2\left(M\right)\right)}$ & $\text{sin}^2\left(\frac{\left(2\hat{p}-1\right)\pi}{M}\right)$ \\
					M-QAM      & $\frac{\sqrt{M}}{2}$        & $\frac{2}{\log_2\left(M\right)}\left(1-\frac{1}{\sqrt{M}}\right)$ & $\frac{3\left(2\hat{p}-1\right)^2}{2\left(M-1\right)}$                  \\ \hline
				\end{tabular}
			\end{table}
		}
		
	The ABER of the hybrid FSO/THz link is derived as
	\begin{align}\label{BER_Hybrid}
		\fontsize{10pt}{9pt}\selectfont
		\mathcal{B}^{\text{H}}_{\text{e,Hyb}}\left(\gamma_{\text{th}}\right)=\frac{1}{1-\mathcal{P}^{\text{H}}_{\gamma_\text{c}}\left(\gamma_{\text{th}}\right)}\left[\mathcal{B}_{\text{e,F}}\left(\gamma_{\text{th}}\right)+\mathcal{P}_{\gamma_\text{F}}\left(\gamma_{\text{th}}\right)\mathcal{B}_{\text{\text{e,T}}}\left(\gamma_{\text{th}}\right)\right].
	\end{align}
	Here, $\mathcal{P}^{\text{H}}_{\gamma_\text{c}}\left(\gamma_{\text{th}}\right)$ and $\mathcal{P}_{\gamma_{\text{F}}}\left(\gamma_{\text{th}}\right)$ represent the outage probability of the hybrid FSO/THz link (\ref{outF1}) and the FSO link (\ref{Gamma_CDF}), respectively. Further, $\mathcal{B}_{\text{e,F}}\left(\gamma_{\text{th}}\right)$ and $\mathcal{B}_{\text{e,T}}\left(\gamma_{\text{th}}\right)$ are the ABER of the FSO link (\ref{BER_FSO}) and the THz link ((\ref{BER_THz})), respectively, as determined in the following.
		\subsubsection{ABER of the FSO Link}
		For the FSO link, by substituting the PDF of the FSO link's SNR (\ref{Gamma_PDF}) in (\ref{BER_main}), we have
		\begin{align}\label{BER_F1}
			\mathcal{B}_{\text{e,F}}\left(\gamma_{\text{th}}\right)&= \sum_{\hat{p}=1}^{n_0}\frac{A\xi^2_\text{F}}{\tau\Gamma\left(\alpha_{\text{F}}\right)\Gamma\left(\beta_{\text{F}}\right)}\int_{\gamma_{\text{th}}}^{\infty}\frac{1}{\gamma}\text{erfc}\left(\sqrt{B_{\hat{p}}\gamma}\right)\nonumber\\&
			\times\text{G}^{3,0}_{1,3}\left[\frac{\alpha_{\text{F}} \beta_{\text{F}}}{A_{0,\text{F}}}\left(\frac{\gamma}{\delta_{\tau}}\right)^{\frac{1}{\tau}}\Big{|}^{\xi^2_\text{F}+1}_{\xi^2_\text{F},\alpha_{\text{F}},\beta_{\text{F}}}\right]\text{d}\gamma,
		\end{align}
		where $n_0$ is the summation limit for various modulation schemes, which is defined in Table \ref{tab_BER}.
		Here, (\ref{BER_F1}) can be expressed as
		\begin{align}\label{BER_FSO}
			\mathcal{B}_{\text{e,F}}\left(\gamma_{\text{th}}\right)=\sum_{\hat{p}=1}^{n_0}\left[\Theta_7\left(\hat{p}\right)+\Theta_8\left(\hat{p}\right)\right],
		\end{align} 
		where identities $\Theta_7\left(\hat{p}\right)$ and $\Theta_8\left(\hat{p}\right)$ are respectively derived as
		\begin{align}\label{BER_F3}
			\Theta_7\left(\hat{p}\right)&=\frac{A\mathbb{D}_1}{\sqrt{\pi}}
			\text{G}^{3\tau,2}_{\tau+2,3\tau+1}\left[\frac{\mathbb{D}_2}{B_{\hat{p}}\left(A_{0,\text{F}}\right)^\tau\delta_{\tau}}\Big{|}^{1,\frac{1}{2},\varrho_1}_{\varrho_2,0}\right],
		\end{align}
		and
		\begin{align}\label{BER_F12}
			\Theta_8\left(\hat{p}\right)&=A\mathbb{D}_1\Bigg\{\text{G}^{3\tau,1}_{\tau+1,3\tau+1}\left[\frac{\mathbb{D}_2\gamma_{\text{th}}}{(A_{0,\text{F}})^\tau\delta_{{\tau}}}\Big{|}^{1, \varrho_1}_{\varrho_2, 0}\right]\nonumber\\&
			-\frac{2}{\sqrt{\pi}}\sum_{j_1=0}^{\infty}\frac{(-1)^{j_1}B_{\hat{p}}^{\frac{2j_1+1}{2}}}{j_1!(2j_1+1)}\gamma_{\text{th}}^{\frac{2j_1+1}{2}}\nonumber\\&
			\times\text{G}^{3\tau,1}_{\tau+1,3\tau+1}\left[\frac{\mathbb{D}_2\gamma_{\text{th}}}{\left(A_{0,\text{F}}\right)^\tau\delta_{{\tau}}}\Big{|}^{1-\frac{2j_1+1}{2}, \varrho_1}_{\varrho_2, -\frac{2j_1+1}{2}}\right]\Bigg\}.
		\end{align}
		
		\textbf{Proof: See Appendix C.}

		\subsubsection{ABER of the THz Link}
		Substituting the PDF of the THz link's SNR  (\ref{PDF_THz}) in (\ref{BER_main}), we have
		\begin{align}\label{BER_T1}
			\mathcal{B}_{\text{e,T}}(\gamma_{\text{th}})&= \sum_{\hat{p}=1}^{n_0}
			\frac{A\mathbb{C}_1}{2(\bar{\gamma})^{\frac{\xi_{\text{T}}^2}{2}}}\int_{\gamma_{\text{th}}}^{\infty}\text{erfc}\left(\sqrt{B_{\hat{p}}\gamma}\right)\nonumber\\&
			\times\gamma^{\frac{\xi_{\text{T}}^2}{2}-1}\text{G}^{2,0}_{1,2}\left[\mathbb{C}_3\left(\frac{\gamma}{\bar{\gamma}}\right)^{\frac{\alpha}{2}}\Big{|}^{1}_{0,\mathbb{C}_2}\right]\text{d}\gamma.
		\end{align}
		Here, (\ref{BER_T1}) can be expressed as 
		\begin{align}\label{BER_THz}
			\mathcal{B}_{\text{e,T}}\left(\gamma_{\text{th}}\right)=\sum_{\hat{p}=1}^{n_0}\left[\Theta_9\left(\hat{p}\right)+\Theta_{10}\left(\hat{p}\right)\right],
		\end{align}
		where identities $\Theta_9\left(\hat{p}\right)$ and $\Theta_{10}\left(\hat{p}\right)$ are respectively derived as
		\begin{align}\label{BER_T1F}
			\Theta_9\left(\hat{p}\right)&=\frac{A\mathbb{C}_1}{2\sqrt{\pi}\left(\bar{\gamma}\right)^{\frac{\xi_{\text{T}}^2}{2}}}\frac{2^{\mathbb{C}_2-\frac{1}{2}}\alpha^{\frac{\xi_{\text{T}}^2}{2}-1}}{(2\pi)^{\frac{\alpha}{2}}B_{\hat{p}}^{\frac{\xi_{\text{T}}^2}{2}}}\nonumber\\&
			\times\text{G}^{4,2\alpha}_{2+2\alpha,4+\alpha}\left[\frac{\mathbb{C}_3^2}{\bar{\gamma}^{\alpha}}\frac{\alpha^{\alpha}}{4B_{\hat{p}}^{\alpha}}\Big{|}^{\varrho_5}_{\varrho_6}\right],
		\end{align}
		and
		\begin{align}\label{I10_Fin}
			\Theta_{10}\left(\hat{p}\right)&=\frac{A\mathbb{C}_1}{\left(\bar{\gamma}\right)^{\frac{\xi_{\text{T}}^2}{2}}\alpha}
			\Bigg\{\gamma_{\text{th}}^{\frac{\xi_{\text{T}}^2}{2}}  \text{G}^{2,1}_{2,3}\left[\mathbb{C}_3\left(\frac{\gamma_{\text{th}}}{\bar{\gamma}}\right)^{\frac{\alpha}{2}}\Big{|}^{1-\frac{\xi^2_\text{T}}{\alpha},1}_{0,\mathbb{C}_2,-\frac{\xi^2_\text{T}}{\alpha}}\right]\nonumber\\&
			-\frac{2}{\sqrt{\pi}}\sum_{j_1=0}^{\infty}\frac{\left(-1\right)^{j_1}B_{\hat{p}}^{\frac{2j_1+1}{2}}}{j_1!(2j_1+1)}\gamma_{\text{th}}^{\frac{\xi^2_\text{T}+2j_1+1}{2}}\nonumber\\&
			\times\text{G}^{2,1}_{2,3}\left[\mathbb{C}_3\left(\frac{\gamma_{\text{th}}}{\bar{\gamma}}\right)^{\frac{\alpha}{2}}\Big{|}^{1-\frac{\xi^2_\text{T}+2j_1+1}{\alpha},1}_{0,\mathbb{C}_2,-\frac{\xi^2_\text{T}+2j_1+1}{\alpha}}\right]\Bigg\}.
		\end{align}
		Here, $\varrho_5=\left[\frac{1-\frac{\xi_{\text{T}}^2}{2}}{\alpha},...,\frac{\alpha-\frac{\xi_{\text{T}}^2}{2}}{\alpha},\frac{1-\frac{\xi_{\text{T}}^2}{2}-\frac{1}{2}}{\alpha},...,\frac{\alpha-\frac{\xi_{\text{T}}^2}{2}-\frac{1}{2}}{\alpha},\frac{1}{2},1\right]$ and  $\varrho_6=\left[0,\frac{1}{2},\frac{\mathbb{C}_2}{2},\frac{\mathbb{C}_2+1}{2},\frac{1-\frac{\xi_{\text{T}}^2}{2}-1}{\alpha},...,\frac{\alpha-\frac{\xi_{\text{T}}^2}{2}-1}{\alpha}\right]$.
		
		\textbf{Proof: See Appendix C.}
		
		\subsubsection{ABER of the Access Link}
		
		Substituting the PDF of the access link SNR (\ref{CPDF_Nak}) in (\ref{BER_main}), we obtain
		\begin{align}\label{BER_R1}
			\mathcal{B}_{\text{e,R}}\left(\gamma^\text{R}_{\text{th}}\right)&= \sum_{\hat{p}=1}^{n_0}\frac{1}{\Gamma\left(mN_{\text{t}}\right)}\left(\frac{m}{\bar{\gamma}_{\text{R}}}\right)^{mN_{\text{t}}}\int_{\gamma^\text{R}_{\text{th}}}^{\infty}\text{erfc}\left(\sqrt{B_{\hat{p}}\gamma}\right)\nonumber\\&
			\times \gamma^{mN_{\text{t}}-1}\exp{\left(-\frac{m}{\bar{\gamma}_{\text{R}}}\gamma\right)}\text{d}\gamma,
		\end{align}
		and (\ref{BER_R1}) can be expressed as
		\begin{align}\label{BER_RF}
			\mathcal{B}_{\text{e,F}}\left(\gamma_{\text{th}}\right)=\sum_{\hat{p}=1}^{n_0}\left[\Theta_{11}\left(\hat{p}\right)+\Theta_{12}\left(\hat{p}\right)\right],
		\end{align}  
		with
		\begin{align}\label{BER_R1F}
			\Theta_{11}\left(\hat{p}\right)&=\frac{A}{\sqrt{\pi}\Gamma\left(mN_{\text{t}}\right)}\left(\frac{m}{\bar{\gamma}_{\text{R}}}\right)^{mN_{\text{t}}}
			\frac{B_{\hat{p}}^{\frac{1}{2}}\Gamma\left(mN_{\text{t}}+\frac{1}{2}\right)}{mN_{\text{t}}\left(B_{\hat{p}}+\frac{m}{\bar{\gamma}_{\text{R}}}\right)^{mN_{\text{t}}+\frac{1}{2}}}\nonumber\\&
			\times~_2F_1\left[1,mN_{\text{t}}+\frac{1}{2},mN_{\text{t}}+1,\frac{\frac{m}{\bar{\gamma}_{\text{R}}}}{B_{\hat{p}}+\frac{m}{\bar{\gamma}_{\text{R}}}}\right],
		\end{align}
		and
		\begin{align}\label{BER_R2F}
			\Theta_{12}\left(\hat{p}\right)&=\frac{A}{\sqrt{\pi}\Gamma\left(mN_{\text{t}}\right)}\sum_{j_1=0}^{\infty}\frac{\left(-1\right)^{j_1}}{j_1!}\left(\frac{m\gamma^\text{R}_{\text{th}}}{\bar{\gamma}_{\text{R}}}\right)^{mN_{\text{t}}+j_1}\nonumber\\&
			\times\text{G}^{2,1}_{2,3}\left[B_{\hat{p}}\gamma^\text{R}_{\text{th}}\Big{|}^{1-(mN_{\text{t}}+j_1),1}_{0,\frac{1}{2},-(mN_{\text{t}}+j_1)}\right],
		\end{align}
		where $_pF_q\left[a_1,...,a_p;b_1,...,b_q;x\right]=\sum_{j=0}^{\infty}\frac{x^j}{j!}\frac{\left(a_1\right)_j\left(a_2\right)_j...\left(a_p\right)_j}{\left(b_1\right)_j\left(b_2\right)_j...\left(b_q\right)_j}$ represents the hypergeometric function with $\left(a\right)_j=\frac{\Gamma\left(a+j\right)}{\Gamma\left(a\right)}$ being the Pochhammer symbol.
		
		\textbf{Proof: See Appendix C.}
		
			By substituting $\mathcal{B}_{\text{e,F}}\left(\gamma_{\text{th}}\right)$ and $\mathcal{B}_{\text{e,T}}\left(\gamma_{\text{th}}\right)$ from (\ref{BER_FSO}) and (\ref{BER_THz}), respectively in (\ref{BER_Hybrid}), the ABER of the hybrid link is obtained.
		\subsubsection{E2E ABER}
		For the hard switching case, the E2E ABER at the terrestrial MU is given as
		\begin{align}\label{BER_e2e}
			\mathcal{B}^{\text{H}}_{\text{e,E2E}}=\mathcal{B}^{\text{H}}_{\text{e,Hyb}}+\mathcal{B}_{\text{e,R}}-2\times\mathcal{B}^{\text{H}}_{\text{e,Hyb}}\times\mathcal{B}_{\text{e,R}}.
		\end{align}
		Substituting the identities $\mathcal{B}^{\text{H}}_{\text{e,Hyb}}$ and $\mathcal{B}_{\text{e,R}}$, derived in (\ref{BER_Hybrid}) and (\ref{BER_RF}), respectively, the E2E ABER (\ref{BER_e2e}) is obtained.
		
		\section{Performance Analysis with Soft Switching}
		In hard switching, single threshold is considered to switch between the THz and the FSO links which results in frequent back-and-forth switching. To avoid such frequent switchings and to increase the FSO link's active time period, a more practical approach is to use soft switching between the links by using dual-FSO thresholds. In this section, we study the network performance with soft switching at the AP. Here, to simplify the discussions, we present the results for the cases with two thresholds for the FSO link and one switching threshold for the THz link. This is because, we consider higher priority for the FSO link which, with a proper deployment, is expected to provide higher data rate, compared to the THz link. However, the results can be easily extended to the cases with two thresholds in the THz link and one or two thresholds in the FSO link.
		\subsection{Outage Probability}
		Figure \ref{region} illustrates the switching concept. Let us define
		$\gamma^{\text{F}}_{\text{th,U}}$ and $\gamma^{\text{F}}_{\text{th,L}}$ as the upper and lower FSO thresholds, respectively, and $\gamma^\text{T}_{\text{th}}$ is the THz link threshold.
		
		\begin{figure}[h]
			\centering
			\includegraphics[width=3.55in,height=2.5in]{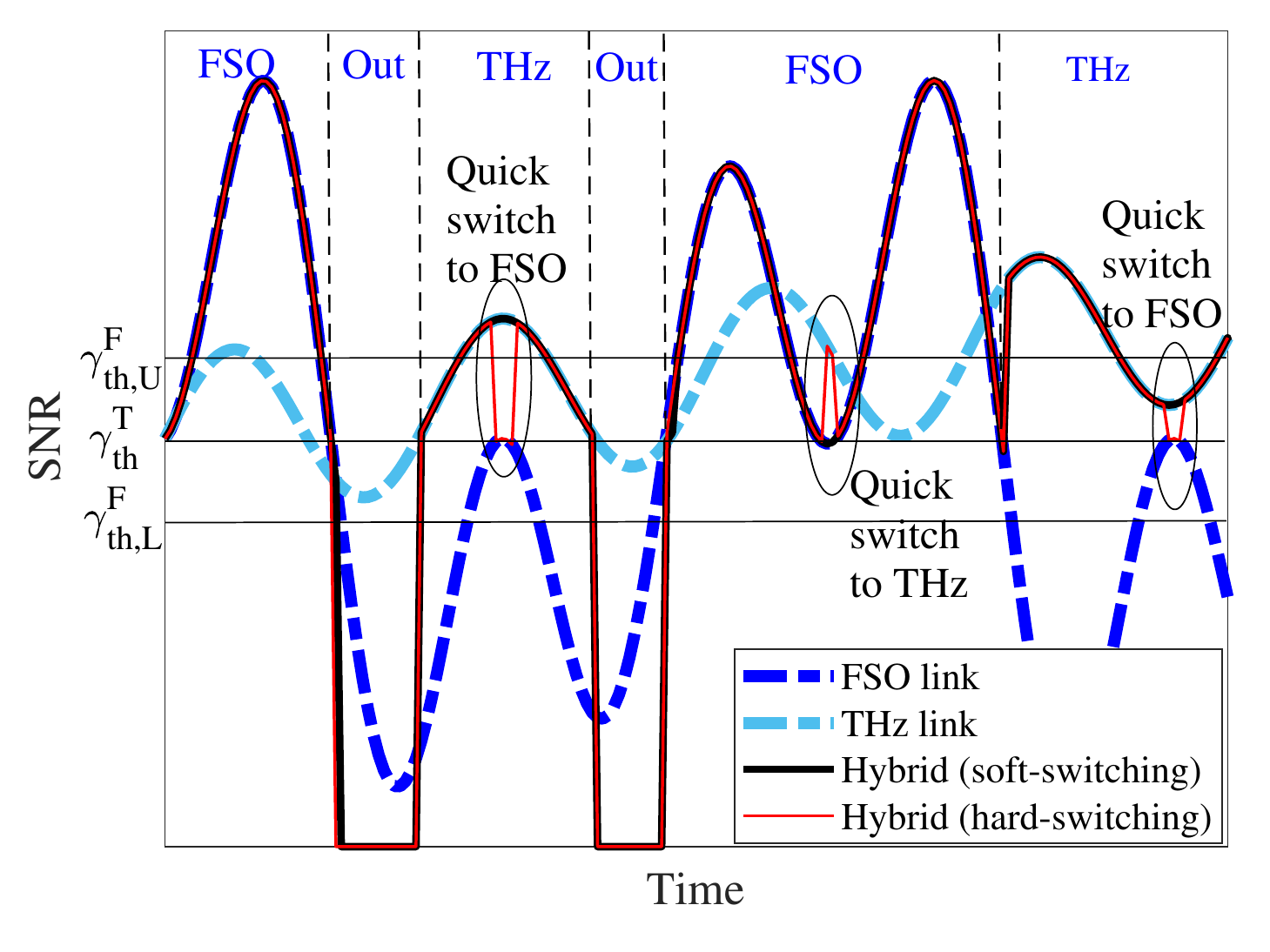}
			\caption{\small{An illustration of the switching-based hybrid FSO/THz system with both the soft and hard switching methods. Here, different regions represent the soft switching operation. Further, small ellipse in the figure shows a quick switch between the FSO and the THz links in hard switching. Note that for hard switching, there is only one threshold, i.e., $\gamma^{\text{F}}_{\text{th,U}}=\gamma^{\text{F}}_{\text{th,L}}=\gamma^{\text{T}}_{\text{th}}=\gamma_{\text{th}}$.}}
			\label{region}
		\end{figure}
	
		The soft switching operation is based on the following procedure:
		\begin{itemize}
			\item If $\gamma_{\text{F}} \geq \gamma^{\text{F}}_{\text{th,U}}$, FSO link is active.
			\item If $\gamma^{\text{F}}_{\text{th,L}} \leq \gamma_{\text{F}}\leq \gamma^{\text{F}}_{\text{th,U}}$, given $\gamma_{\text{F}} \geq \gamma^{\text{F}}_{\text{th,U}}$ previously, FSO link is active irrespective of the THz link. 
			\item If $\gamma^{\text{F}}_{\text{th,L}} \leq \gamma_{\text{F}}\leq \gamma^{\text{F}}_{\text{th,U}}$, given $\gamma_{\text{F}} < \gamma^{\text{F}}_{\text{th,L}}$ previously, FSO link is off and the THz link is active if $\gamma_{\text{T}} \geq \gamma^{\text{T}}_{\text{th}}$.
			\item If $\gamma_{\text{F}} < \gamma^{\text{F}}_{\text{th,L}}$ and  $\gamma_{\text{T}} \geq \gamma^{\text{T}}_{\text{th}}$, the THz link is active.
			\item If $\gamma^{\text{F}}_{\text{th,L}} \leq \gamma_{\text{F}}\leq \gamma^{\text{F}}_{\text{th,U}}$, given $\gamma_{\text{F}} < \gamma^{\text{F}}_{\text{th,L}}$ and  $\gamma_{\text{T}} < \gamma^{\text{T}}_{\text{th}}$, hybrid link is in outage.
			\item If $\gamma_{\text{F}} < \gamma^{\text{F}}_{\text{th,L}}$ and  $\gamma_{\text{T}} < \gamma^{\text{T}}_{\text{th}}$, hybrid link is in outage.
		\end{itemize}
		
		The state transition diagram of the hybrid FSO/RF link with dual FSO threshold is shown in Fig. \ref{State}.
		\begin{figure}[h]
			\centering
			\includegraphics[width=3.5in,height=2in]{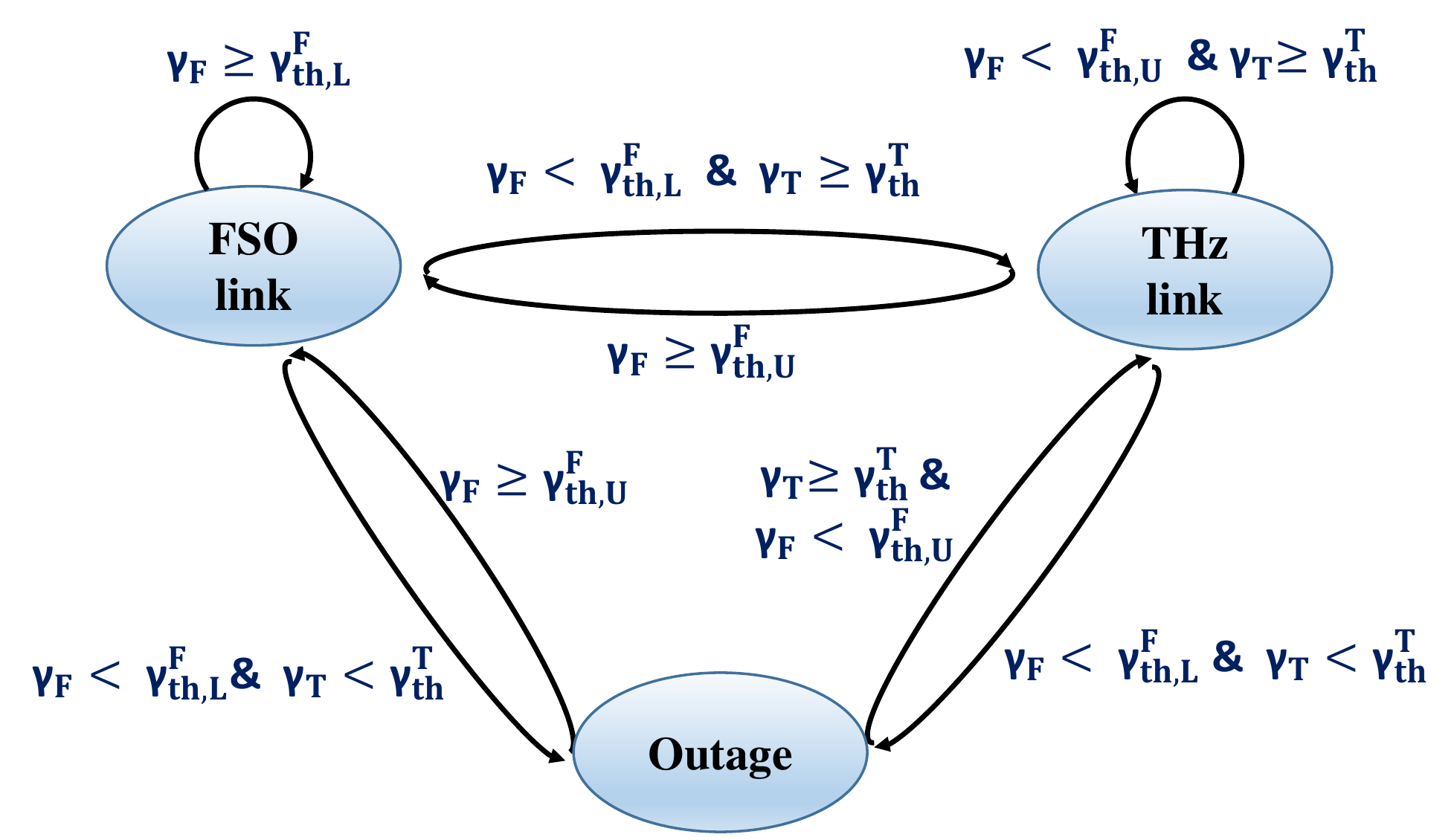}
			\caption{\small{State transition of the hybrid FSO/THz link in soft switching.}}
			\label{State}
		\end{figure}

		Considering the proposed setup, the hybrid FSO/THz link is in outage if both the FSO and THz links are off. The FSO link is off when $\gamma_{\text{F}}<\gamma^{\text{F}}_{\text{th,L}}$ or $\gamma^{\text{F}}_{\text{th,L}} \leq \gamma_{\text{F}} \leq \gamma^{\text{F}}_{\text{th,U}}$ given that $\gamma_{\text{F}}<\gamma^{\text{F}}_{\text{th,L}}$ previously.
		Hence, the probability that FSO link is off is given as 
		\begin{align}\label{out_FSO}
			\mathcal{P}^{\text{F}}_{\text{o}}\left(\gamma^{\text{F}}_{\text{th,U}},\gamma^{\text{F}}_{\text{th,L}}\right)=\mathcal{P}^{\text{F}}_{\text{low}}+\mathcal{P}^\text{F}_{\text{med}}\left(\frac{\mathcal{P}^{\text{F}}_{\text{low}}}{\mathcal{P}^{\text{F}}_{\text{low}}+\mathcal{P}^{\text{F}}_{\text{hig}}}\right),
		\end{align} 
		where $\mathcal{P}^{\text{F}}_{\text{low}}=\mathcal{P}^{\text{F}}\left(\gamma^{\text{F}}_{\text{th,L}}\right)$ denotes the probability that $\gamma_{\text{F}}<\gamma^{\text{F}}_{\text{th,L}}$. The term $\mathcal{P}^\text{F}_{\text{med}}=\left[\mathcal{P}^{\text{F}}\left(\gamma^{\text{F}}_{\text{th,U}}\right)-\mathcal{P}^{\text{F}}\left(\gamma^{\text{F}}_{\text{th,L}}\right)\right]$ denotes the probability that $\gamma^{\text{F}}_{\text{th,L}} \leq \gamma_{\text{F}}\leq \gamma^{\text{F}}_{\text{th,U}}$. Further, $\mathcal{P}^{\text{F}}_{\text{hig}}=\left[1-\mathcal{P}^{\text{F}}\left(\gamma^{\text{F}}_{\text{th,U}}\right)\right]$ denotes the probability that $\gamma_{\text{F}}\geq\gamma^{\text{F}}_{\text{th,U}}$.
		Here, $\left(\frac{\mathcal{P}^{\text{F}}_{\text{low}}}{\mathcal{P}^{\text{F}}_{\text{low}}+\mathcal{P}^{\text{F}}_{\text{hig}}}\right)$ is the probability that $\gamma^{\text{F}}_{\text{th,L}} \leq \gamma_{\text{F}} \leq \gamma^{\text{F}}_{\text{th,U}}$, given $\gamma_{\text{F}}<\gamma^{\text{F}}_{\text{th,L}}$ previously. Probabilities $\mathcal{P}^\text{F}_{\text{low}}$, $\mathcal{P}^\text{F}_{\text{med}}$, and $\mathcal{P}^\text{F}_{\text{hig}}$ can be obtained from (\ref{Gamma_CDF}) by substituting the appropriate thresholds.

		The probability that the THz link is in outage is given by
		\begin{align}\label{out_Hyb}
			\mathcal{P}^{\text{T}}_{\text{o}}\left(\gamma^{\text{T}}_{\text{th}}\right) = \mathcal{P}\left(\gamma_{\text{T}} < \gamma^{\text{T}}_{\text{th}}\right)=F_{\gamma_{\text{T}}}\left(\gamma^{\text{T}}_{\text{th}}\right).
		\end{align}
		Therefore, the outage probability of the hybrid FSO/THz link in soft switching case is given as
		\begin{align}\label{Out_soft}
			\mathcal{P}^{\text{S}}_{\text{Hyb}}=\mathcal{P}^{\text{F}}_{\text{o}}\left(\gamma^{\text{F}}_{\text{th,U}},\gamma^{\text{F}}_{\text{th,L}}\right) \times \mathcal{P}^\text{T}_\text{o}\left(\gamma^{\text{T}}_{\text{th}}\right).
		\end{align}
		
		\subsection{E2E Outage Probability}
		With selective DF relaying, the E2E outage probability for the  soft switching method is obtained by
		\begin{align}
			\mathcal{P}^{\text{S}}_{\text{E2E}}\left(\gamma^\text{R}_{\text{th}}\right)=1-\left(1-\mathcal{P}^{\text{S}}_{\text{Hyb}}\right)\times\left(1-\mathcal{P}^{\text{R}}_\text{o}\left(\gamma^\text{R}_{\text{th}}\right)\right),
		\end{align}
		where $\mathcal{P}^{\text{R}}_\text{o}\left(\gamma^\text{R}_{\text{th}}\right)=	F_{\gamma_{\text{RU}}}\left(\gamma^\text{R}_{\text{th}}\right)$ is obtained in (\ref{CPDF_Nak}).

		\subsection{Asymptotic Outage Probability}
		For the soft switching case, the FSO link's outage probability (\ref{out_FSO}) can be further simplified as
		\begin{align}\label{out_FSO_New}
			\mathcal{P}^{\text{F}}_{\text{o}}\left(\gamma^{\text{F}}_{\text{th,U}},\gamma^{\text{F}}_{\text{th,L}}\right)=\left(\frac{\mathcal{P}^{\text{F}}_{\text{low}}}{\mathcal{P}^{\text{F}}_{\text{low}}+\mathcal{P}^{\text{F}}_{\text{hig}}}\right),
		\end{align} 
		Substituting the high SNR approximated CDF of FSO link (\ref{FSOapp}), the asymptotic outage probability of the FSO link for the soft switching is obtained as
		\begin{align}\label{out_FSO_App}
			\mathcal{P}^{\text{F,A}}_\text{o}\left(\gamma^{\text{F}}_{\text{th,U}},\gamma^{\text{F}}_{\text{th,L}}\right)\approx\left(\frac{\mathcal{P}^{\text{F,A}}_{\text{low}}}{\mathcal{P}^{\text{F,A}}_{\text{low}}+\mathcal{P}^{\text{F,A}}_{\text{hig}}}\right).
		\end{align} 
		At high SNR, the asymptotic outage probability of the soft switching based hybrid FSO/THz link is given as
		\begin{align}
			\mathcal{P}^{\text{S,A}}_{\text{Hyb}}\approx\mathcal{P}^{\text{F,A}}_\text{o}\left(\gamma^{\text{F}}_{\text{th,U}},\gamma^{\text{F}}_{\text{th,L}}\right) \times \mathcal{P}^{\text{T,A}}_\text{o}\left(\gamma^\text{T}_{\text{th}}\right),
		\end{align}
		where $\mathcal{P}^{\text{T,A}}_\text{o}\left(\gamma^\text{T}_{\text{th}}\right)$ is the asymptotic outage probability of the THz link as derived in (\ref{THz_Asym}).
		Therefore, the asymptotic E2E outage probability at the MU is obtained as
		\begin{align}
			\mathcal{P}^{\text{S,A}}_{\text{E2E}}\left(\gamma^\text{R}_{\text{th}}\right)\approx \mathcal{P}^{\text{S,A}}_{\text{Hyb}}+\mathcal{P}^{\text{R,A}}_\text{o}\left(\gamma^\text{R}_{\text{th}}\right),
		\end{align}
		where $\mathcal{P}^{\text{R,A}}_\text{o}(\gamma^\text{R}_{\text{th}})$ is the  asymptotic outage probability of the access link as obtained in (\ref{RF_asym}).
		
		It is observed that the diversity order of the considered system will remain the same for both the soft and hard switching cases. However, as explained in the following, a marginal SNR gain is achieved in soft switching as compared to the hard switching case. 
		\subsection{Ergodic Capacity}
		For the soft switching based hybrid FSO/THz system, the ergodic capacity is obtained as
		\begin{align}\label{Cap_hyb_s}
			\mathcal{C}^{\text{S}}_{\text{Hyb}}&=	\mathcal{C}^{\text{F}}\left(\gamma^{\text{F}}_{\text{th,U}}\right)+\mathcal{P}^{\text{F}}_{\text{o}}\left(\gamma^{\text{F}}_{\text{th,U}},\gamma^{\text{F}}_{\text{th,L}}\right)\mathcal{C}^{\text{T}}\left(\gamma_{\text{th,T}}\right)\nonumber\\&
			+\left[\mathcal{C}^{\text{F}}\left(\gamma^{\text{F}}_{\text{th,L}}\right)-\mathcal{C}^{\text{F}}\left(\gamma^{\text{F}}_{\text{th,U}}\right)\right]\frac{\mathcal{P}^{\text{F}}_{\text{hig}}}{\mathcal{P}^{\text{F}}_{\text{low}}+\mathcal{P}^{\text{F}}_{\text{hig}}}.
		\end{align}
		By substituting $\mathcal{C}^{\text{F}}\left(\gamma^{\text{F}}_{\text{th,U}}\right)$ and $\mathcal{C}^{\text{F}}\left(\gamma^{\text{F}}_{\text{th,L}}\right)$ from (\ref{Cap_FSO}), $\mathcal{P}^{\text{F}}_{\text{o}}\left(\gamma^{\text{F}}_{\text{th,U}},\gamma^{\text{F}}_{\text{th,L}}\right)$ from (\ref{out_FSO}), $\mathcal{P}^\text{F}_{\text{low}}$  and $\mathcal{P}^\text{F}_{\text{hig}}$ from (\ref{Gamma_CDF}), and $\mathcal{C}^{\text{T}}\left(\gamma_{\text{th,T}}\right)$ from (\ref{cap_T})
		in (\ref{Cap_hyb_s}), the ergodic capacity of the soft switching-based hybrid FSO/THz link is obtained.
		Further, the E2E capacity is given as
		\begin{align}
			\mathcal{C}^{\text{S}}_{\text{E2E}}=\text{min}\left(\mathcal{C}^{\text{S}}_{\text{Hyb}},\mathcal{C}^{\text{R}}(\gamma^\text{R}_{\text{th}})\right),
		\end{align} 
		where $\mathcal{C}^{\text{S}}_{\text{Hyb}}$ and $\mathcal{C}^{\text{R}}\left(\gamma^\text{R}_{\text{th}}\right)$ are derived in (\ref{Cap_hyb_s}) and (\ref{cap_R}), respectively.
		\vspace{-1em}
		\subsection{ABER Analysis}
		For the considered soft switching-based hybrid FSO/THz system, the ABER of the hybrid FSO/THz link for generalized digital modulation schemes is given as
		\begin{align}
				\fontsize{10pt}{8pt}\selectfont
				\mathcal{B}^{\text{S}}_{\text{e,Hyb}}&=\frac{1}{1-\mathcal{P}^{\text{S}}_{\text{Hyb}}}\Bigg[\mathcal{B}_{\text{e,F}}\left(\gamma^{\text{F}}_{\text{th,U}}\right) 
				+ 	\mathcal{P}^{\text{F}}_{\text{o}}\left(\gamma^{\text{F}}_{\text{th,U}},\gamma^{\text{F}}_{\text{th,L}}\right)\mathcal{B}_{\text{e,T}}\left(\gamma^\text{T}_{\text{th}}\right)\nonumber\\&
				+\left(\mathcal{B}_{\text{e,F}}\left(\gamma^{\text{F}}_{\text{th,L}}\right)-\mathcal{B}_{\text{e,F}}\left(\gamma^{\text{F}}_{\text{th,U}}\right)\right)\frac{\mathcal{P}^{\text{F}}_{\text{hig}}}{\mathcal{P}^{\text{F}}_{\text{low}}+\mathcal{P}^{\text{F}}_{\text{hig}}}\Bigg].
			\end{align}
		Here, $\mathcal{B}_{\text{e,F}}\left(\gamma\right)$ and $\mathcal{B}_{\text{e,T}}\left(\gamma\right)$ represent the ABER of the FSO link  (\ref{BER_FSO}) and  the THz link (\ref{BER_THz}), respectively. Also, substituting $\mathcal{P}^{\text{F}}_{\text{o}}\left(\gamma^{\text{F}}_{\text{th,U}},\gamma^{\text{F}}_{\text{th,L}}\right)$ from (\ref{out_FSO}), $\mathcal{P}^\text{F}_{\text{low}}$  and $\mathcal{P}^\text{F}_{\text{hig}}$ from (\ref{Gamma_CDF}), and $\mathcal{P}^{\text{S}}_{\text{Hyb}}$ from (\ref{Out_soft}), we obtain the ABER of the hybrid FSO/THz link.
		Further, the ABER of the terrestrial access link $\left(\mathcal{B}_{\text{e,R}}\left(\gamma^\text{T}_{\text{th}}\right)\right)$ is derived in (\ref{BER_RF}).
		Hence, the E2E ABER at each terrestrial MU is obtained by
		\begin{align}\label{BERSoft_e2e}
			\mathcal{B}^{\text{S}}_{\text{e,E2E}}=\mathcal{B}^{\text{S}}_{\text{e,Hyb}}+\mathcal{B}_{\text{e,R}}-2\times\mathcal{B}^{\text{S}}_{\text{e,Hyb}}\times\mathcal{B}_{\text{e,R}}.
		\end{align}
		
			\begin{table*}[]
			\centering 	
			\caption{Parameter settings for the FSO, the THz, and the mmWave access links.}
			\begin{tabular}{|ll|ll|ll|}
				\hline
				\multicolumn{2}{|c|}{FSO link}                                    &\multicolumn{2}{c|}{THz link}                                    & \multicolumn{2}{c|}{mmWave access link}                               \\ \hline
				\multicolumn{1}{|l|}{Parameters}                      & Value     &	\multicolumn{1}{l|}{Parameters}                      & Value     & \multicolumn{1}{l|}{Parameters}                      & Value      \\ \hline
				\multicolumn{1}{|l|}{Wavelength ($\lambda_\text{F}$)}       & 1550 nm &	\multicolumn{1}{l|}{Carrier frequency ($f_\text{T}$)}       & 119 GHz   & \multicolumn{1}{l|}{Carrier frequency ($f_\text{R}$)}       & 28 GHz     \\ 
				\multicolumn{1}{|l|}{Strong turbulence: $C_n^2$} & 1$\times 10^{-12}$m$^{{-2}/{3}}$    &	\multicolumn{1}{l|}{Transmit antenna Gain ($G_\text{t}^\text{T}$)} & 55 dBi    & \multicolumn{1}{l|}{Transmit antenna Gain ($G_\text{t}^\text{R}$)} & 44 dBi     \\ 
				\multicolumn{1}{|l|}{Moderate turbulence: $C_n^2$} & 5$\times 10^{-13}$m$^{{-2}/{3}}$    &	\multicolumn{1}{l|}{Receive antenna Gain ($G_\text{r}^\text{T}$)}  & 55 dBi    & \multicolumn{1}{l|}{Receiver antenna Gain ($G_\text{r}^\text{R}$)} & 44 dBi     \\ 
				\multicolumn{1}{|l|}{Atmospheric turb. (Str.): $\alpha_{\text{F}}$,$\beta_{\text{F}}$ }      & 4.343, 2.492 &\multicolumn{1}{l|}{Atmospheric pressure ($p$)}      & 101325 Pa & \multicolumn{1}{l|}{Oxygen absorption ($a_{\text{oxy}}$)}   & 15.1 dB/km \\ 
				\multicolumn{1}{|l|}{Atmospheric turb. (Mod.): $\alpha_{\text{F}}$,$\beta_{\text{F}}$ }               & 5.838, 4.249    &	\multicolumn{1}{l|}{Temperature ($T$)}               & 298 K     & \multicolumn{1}{l|}{Rain attenuation $(a_{\text{rain}})$}             & 0 dB/km      \\ 
				\multicolumn{1}{|l|}{Opt.-to-elect. conversion ($\eta)$}      & 1    &\multicolumn{1}{l|}{Relative humidity ($\chi$)}      & 50$\%$    & \multicolumn{1}{l|}{Link length ($L_{\text{R}}$)}                                &    100 m       \\ 
				\multicolumn{1}{|l|}{Link length ($L_{\text{F}}$)}             & 200 m    &\multicolumn{1}{l|}{Link length ($L_{\text{T}}$)}             & 200 m     & \multicolumn{1}{l|}{}                                &            \\ 
				\multicolumn{1}{|l|}{Receiver radius (${a_{\text{F}}}$)}             & 20 cm    &\multicolumn{1}{l|}{Receiver radius (${a_{\text{T}}}$)}             &  ${\lambda_{\text{T}}\sqrt{G^{\text{T}}_{\text{t}}}}/{\left(2\pi\right)}$    & \multicolumn{1}{l|}{}                                &            \\ 
				\multicolumn{1}{|l|}{Beamwidth (${\omega_{\text{L,F}}}$)}             & 40 cm    &\multicolumn{1}{l|}{Beamwidth (${\omega_{\text{L,T}}}$)}             & 50 cm     & \multicolumn{1}{l|}{}                                &            \\ 
				\multicolumn{1}{|l|}{Jitter standard deviation ($\sigma_{\text{j,F}}$)}             & 5 cm    &\multicolumn{1}{l|}{Jitter standard deviation ($\sigma_{\text{j,T}}$)}             &  6 cm    & \multicolumn{1}{l|}{}                                &            \\ \hline
			\end{tabular}
			\label{ParametersN}
			\vspace{-1em}
		\end{table*}
		\section{Simulation Results}
			In this section, numerical results from the derived expressions are obtained and are compared with the simulation results. Here, the results are presented for the cases with transmit SNR which, in the log-domain, is defined as $10\log_{10}\left({P_{\text{i}}/\sigma_{\text{i}}^2}\right)$ with ${P}_{\text{i}}$ and $\sigma_{\text{i}}^2$ being the transmit power and the AWGN variance, respectively of the $\text{i}$ link. Considering the links long-term channel characteristics, it is straightforward to represent the results in terms of received SNR.				
		For both the downlink and uplink scenario, $P_\text{F}=P_\text{T}=P_\text{R}=P$ are considered.
		Further, identical AWGN variances are considered for all the links, i.e., $\sigma^2_{\text{o}}=\sigma^2_{\text{T}}=\sigma^2_{\text{R}}=\sigma^2=1$. Also, we performed the E2E analysis for the $n^{\text{th}}$ MU. However, this can be easily extended to the independently distributed multiple MUs by changing the threshold SNR at the AP to accommodate the multiple MUs.  In all figures, except for Fig. \ref{detComp} and Fig. \ref{BER_COMP}  which consider both IM/DD and heterodyne detection, we consider heterodyne detection ($\tau=1$). 
		Parameters settings for the FSO, the THz, and the mmWave access links are shown in Table \ref{ParametersN}.
		\begin{figure}[h]
			\centering
			\includegraphics[width=3.52in,height=2.5in]{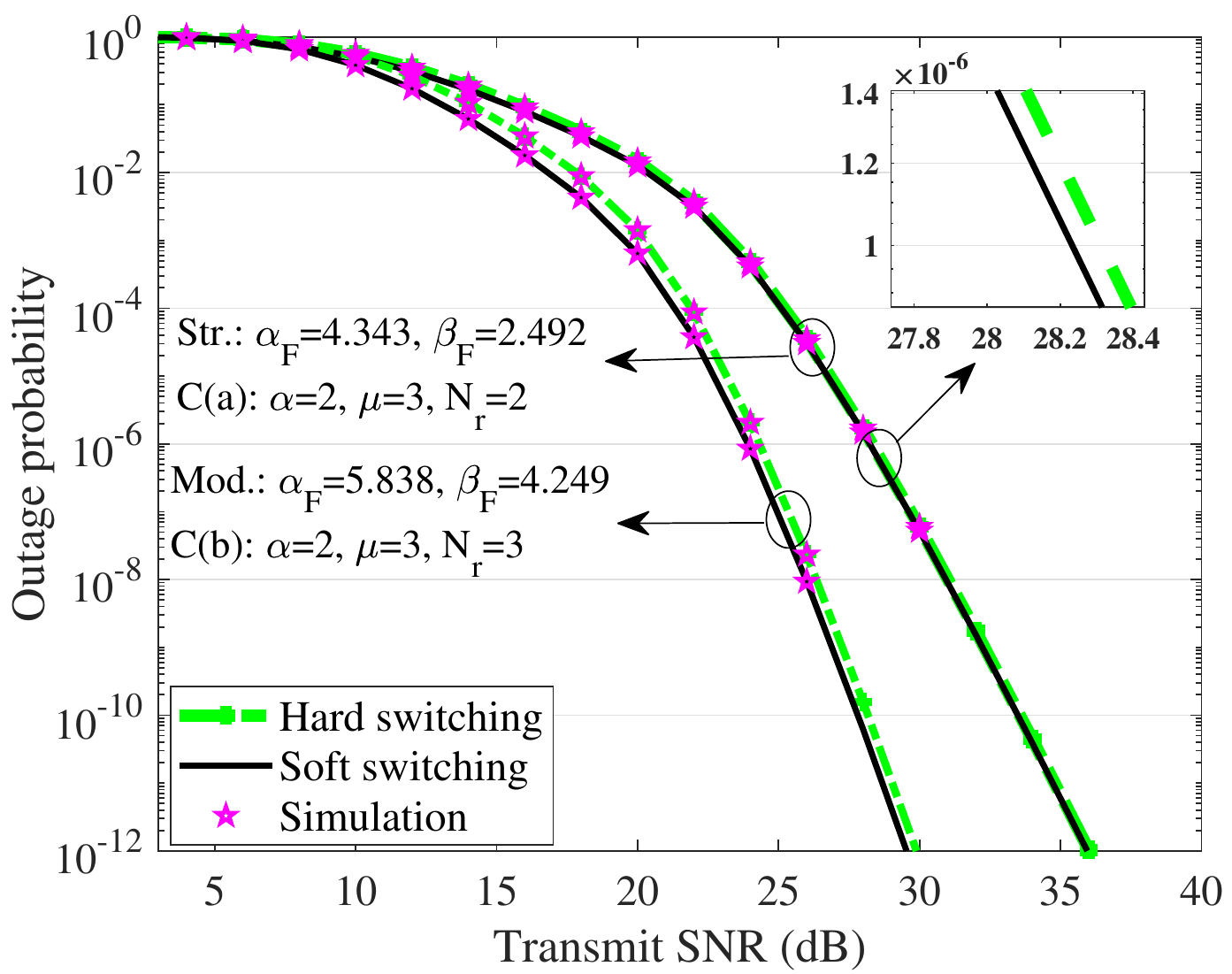}
			\caption{\small{Comparison of the outage probability of the hybrid FSO/THz link for the hard versus soft switching, considering $L_{\text{F}}=L_{\text{T}}=$200 m and $L_{\text{R}}=$100 m.}}
			\label{Comp1}
		\end{figure}
		\begin{figure*}[h]
			\hfill
			\subfigure[]{\includegraphics[width=8.8cm,height=6.5cm]{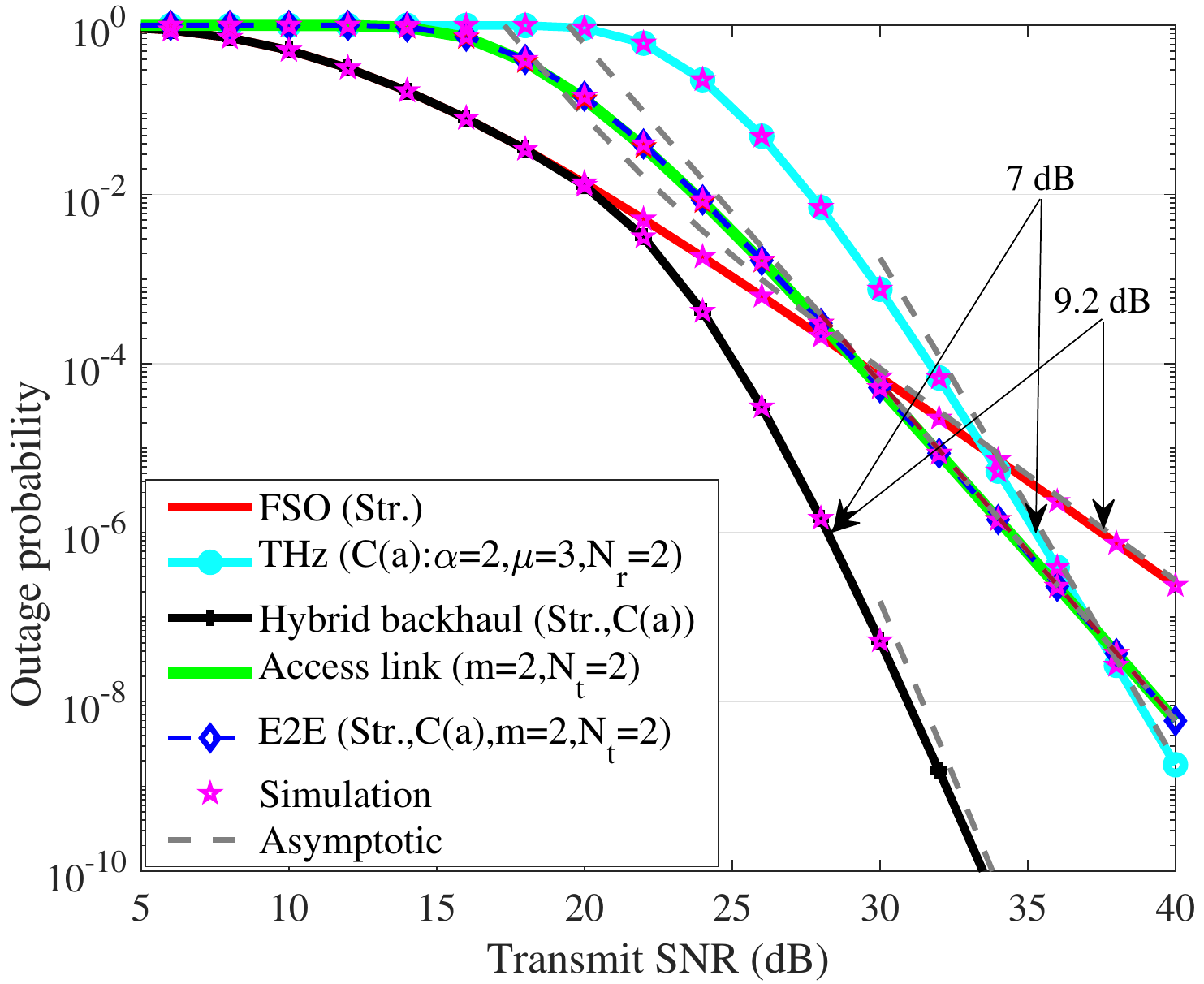}
				\label{O1}}
			\hfill
			\subfigure[]{\includegraphics[width=8.8cm,height=6.5cm]{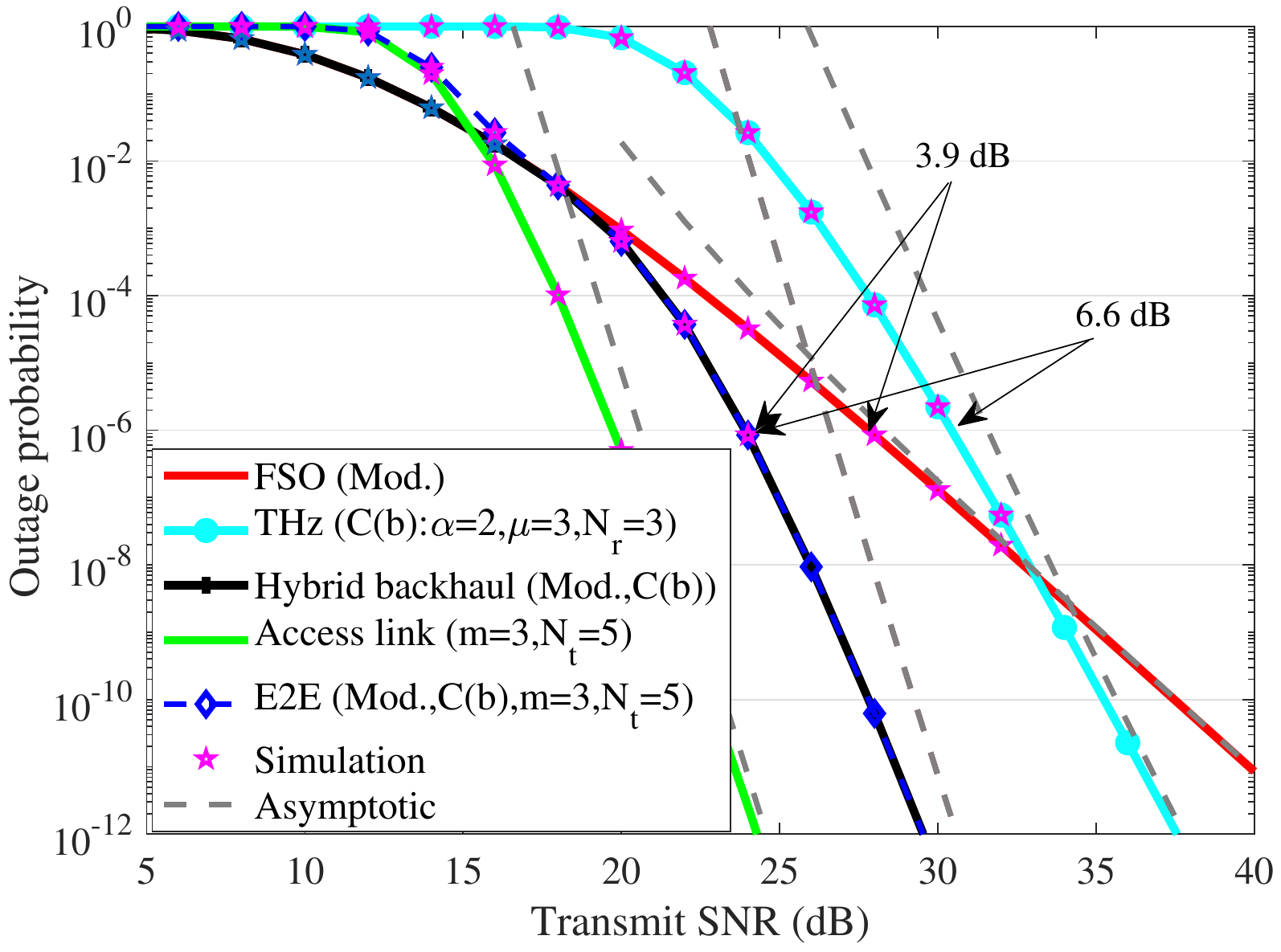}
				\label{O3}}
			\hfill
			\vspace{-1em}
			\caption{\small{Outage probability versus the transmit SNR, considering $L_{\text{F}}=L_{\text{T}}=200$ m and $L_{\text{R}}=100$ m.}}
			\label{Out_sep_N}	
			\hrulefill
		\end{figure*}
	
		\subsection{Outage Probability and Diversity Order Analysis}	
		
		In Fig. \ref{Comp1}, we compare the outage probability of the hybrid FSO/THz link for the soft and hard switching methods and verify the results through simulations. We consider strong (Str. in the figures) and moderate turbulence (Mod. in the figures) cases for the FSO link. To achieve approximately similar performance in the THz link as compared to the FSO link, $\text{case(a)}: \alpha=2, \mu=3, N_\text{r}=2$ and $\text{case(b)}: \alpha=2, \mu=3, N_\text{r}=3$ are used for strong and moderate turbulence cases, respectively. \par
		
		Considering the cases with strong turbulence for the FSO link and case(a) for the THz link,
		soft switching provides approximately 0.1 dB gain over hard switching case to achieve an outage probability of $10^{-6}$, which is shown in  Fig. \ref{Comp1} with a magnified figure. We will achieve a marginal gain but most importantly the problem of frequent switching is minimized. 
		Considering the cases with moderate turbulence for FSO link and case(b) for the THz link,
		soft switching provides approximately 0.4 dB gain over the hard switching case to achieve an outage probability of $10^{-6}$. 
		
		From the results, we observe that the soft switching provides marginal gain, in
		terms of outage probability, over the hard switching, and reduces the frequent back-and-forth switching.
		Thus, from a practical point of view, for a hybrid communication system with parallel links, soft switching
		between the links may be of interest.  Since the performance of hard switching is almost similar to the soft switching, we have shown the rest of the results for the soft switching case, to avoid the repetition, except for Fig. \ref{ABER_comp} which studies the ABER for different switching methods.\par

		In Fig. \ref{Out_sep_N}, theoretical, simulation, and asymptotic results of individual, hybrid FSO/THz, and E2E outage probability are studied for various combinations of selection parameters for the downlink communication. 
		With an outage probability of $10^{-6}$, the transmission power of the FSO link needs to be increased by around 10 dB to guarantee the same quality-of-service in the strong turbulence case as in the case with moderate turbulence. 
		For an outage probability of $10^{-6}$, THz link achieves nearly 4.9 dB gain when moves from case(a) to case(b) while deploying one extra antenna at the THz receiver. 
		In Fig. \ref{O1}, for the access link, $m=2, N_{\text{t}}=2$ case is considered while in the backhaul hybrid FSO/THz link, we consider strong turbulence for FSO link  and case(a) for the THz link.
		As seen in Fig. \ref{O1} (and, also Fig. \ref{O3}), for the FSO, THz, hybrid, access, and E2E links the high SNR approximations are tight which validate the diversity order derivations. With an outage probability of $10^{-6}$, the hybrid backhaul link provides around 7 dB and 9.2 dB gains over the THz and the FSO links, respectively. 
		Also, the E2E outage probability follows the access link's performance because the backhaul hybrid FSO/THz link outperforms the access link (with $m=2, N_{\text{t}}=2$) by 6.25 dB. Hence, the diversity order is limited by $mN_{\text{t}}$. By increasing the number of antennas at the AP, the E2E performance and diversity order at the terrestrial MUs can be improved.\par
		In Fig. \ref{O3}, we consider moderate turbulence for the FSO link and case(b) for the THz link. In this case, with an outage probability of $10^{-6}$, the hybrid backhaul link's performance improves by  4.3 dB  as compared to the case considered in Fig. \ref{O1}.  For $m=3,N_{\text{t}}=5$, the access link achieves around 6.6 dB and 13 dB gains over the $m=2,N_{\text{t}}=3$ and $m=2,N_{\text{t}}=2$ cases, respectively. 
		However, in Fig. \ref{O3}, the E2E outage performance follows the hybrid link's performance, since the access link's performance is improved significantly and experiences around 4 dB better SNR compared to the backhaul hybrid FSO/THz link. Hence, the E2E diversity order is  $\min\Big(\left(\frac{\xi_{\text{F}}^2}{\tau}+\frac{\xi_{\text{T}}^2}{2}\right),\left(\frac{\alpha_{\text{F}}}{\tau}+\frac{\xi_{\text{T}}^2}{2}\right),\left(\frac{\beta_{\text{F}}}{\tau}+\frac{\xi_{\text{T}}^2}{2}\right),\left(\frac{\xi_{\text{F}}^2}{\tau}+\frac{\alpha N_{\text{r}}\mu}{2}\right),\\
		\left(\frac{\alpha_{\text{F}}}{\tau}+\frac{\alpha N_{\text{r}}\mu}{2}\right),\left(\frac{\beta_{\text{F}}}{\tau}+\frac{\alpha N_{\text{r}}\mu}{2}\right)\Big)$ in this case.   
		The FSO and the THz links in parallel improve the performance of the backhaul hybrid FSO/THz link for all strong, moderate, and weak turbulence/fading conditions as compared to the individual FSO or THz links' performance. Hence, the switching based hybrid paradigm guarantees reliable and robust high data-rate backhaul communication even in severe weather conditions. 
		
				\begin{figure}[]
						\centering
						\includegraphics[width=3.5in,height=2.5in]{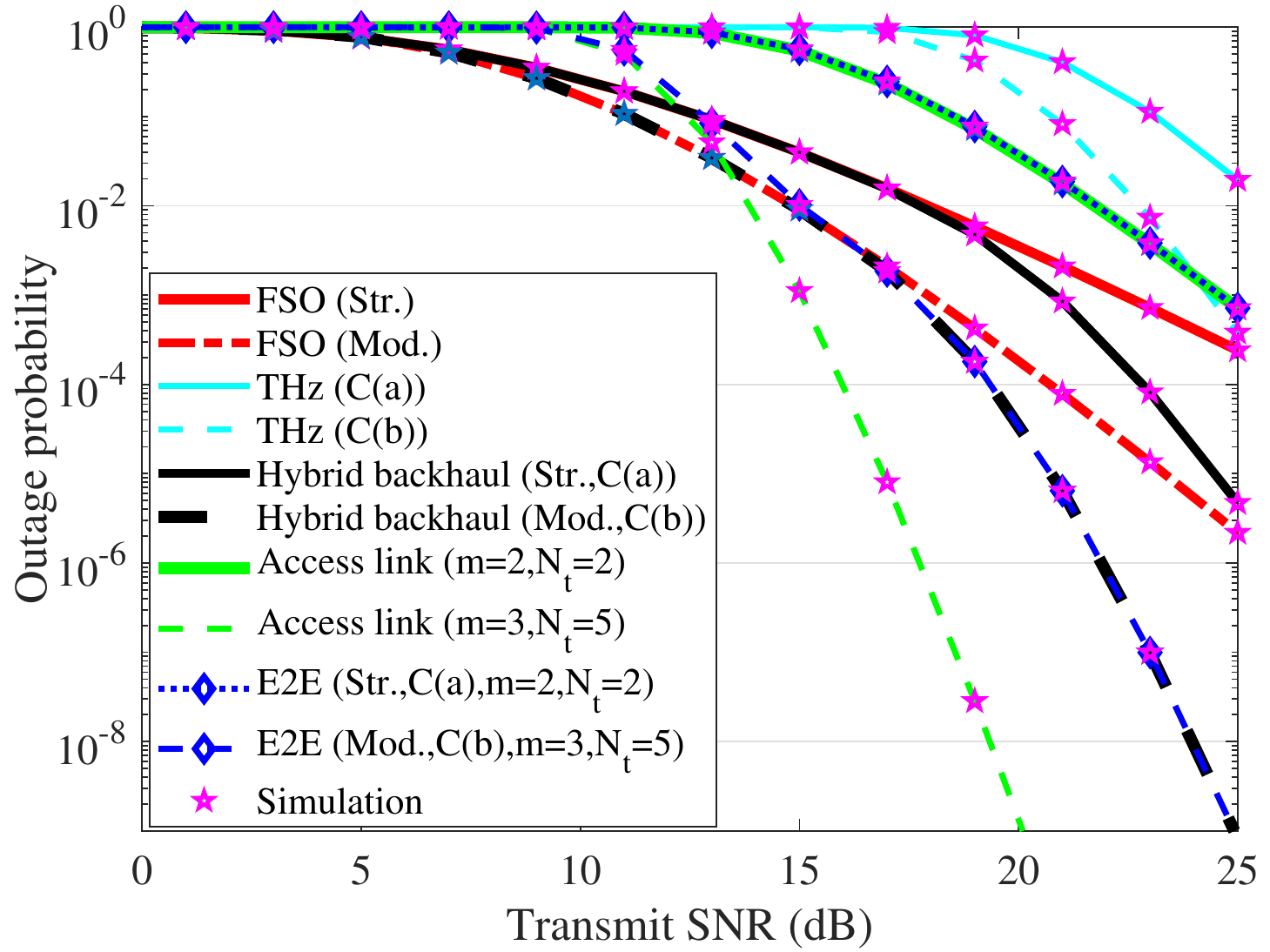}
						\vspace{-1em}
						\caption{\small{Outage probability versus the transmit SNR for uplink scenario.}}
						\label{Out_uplink}
					\end{figure}

		While we present the theoretical results for downlink transmission, the same procedure can be applied for uplink communication.	In Fig. \ref{Out_uplink}, theoretical results of individual and E2E outage probability are studied for uplink scenario for various combinations of selection parameters. As seen in Fig. \ref{Out_uplink}, the performance in uplink scenario is poor as compared to the downlink scenario (Fig. \ref{Out_sep_N}) for all the considered cases because less power is available at the MU as a source terminal. 
		For the  hybrid backhaul link, we consider strong turbulence for FSO link and case(a) for the THz link. For this, with an outage probability of $10^{-3}$, the hybrid backhaul link provides around 1.5 dB and 9.3 dB gains as compared to the cases where we only use the FSO or the THz links, respectively. For $m=2, N_\text{t}=2$ case, the access link lags by 2 dB as compared to the hybrid backhaul link for $10^{-3}$ outage probability, consequently the E2E link follows the access link's performance.
		Similarly, by considering moderate turbulence for FSO link and case(b) for the THz link, the hybrid backhaul link provides around 0.5 dB and 3.7 dB gains for an outage probability of $10^{-3}$, as compared to the FSO or the THz links, respectively. However, for $m=3, N_\text{t}=5$ case, the access link provides around  3 dB gain over the hybrid backhaul link for $10^{-3}$ outage probability. Consequently, the E2E link follows the hybrid backhaul link's performance.
		Here, we observe that the hybrid backhaul link's performance is mainly limited by the THz link, since the 200 m long THz link's performance is quite poor with limited source power for both cases.
		
		\begin{figure}[]
		\centering
		\includegraphics[width=3.5in,height=2.5in]{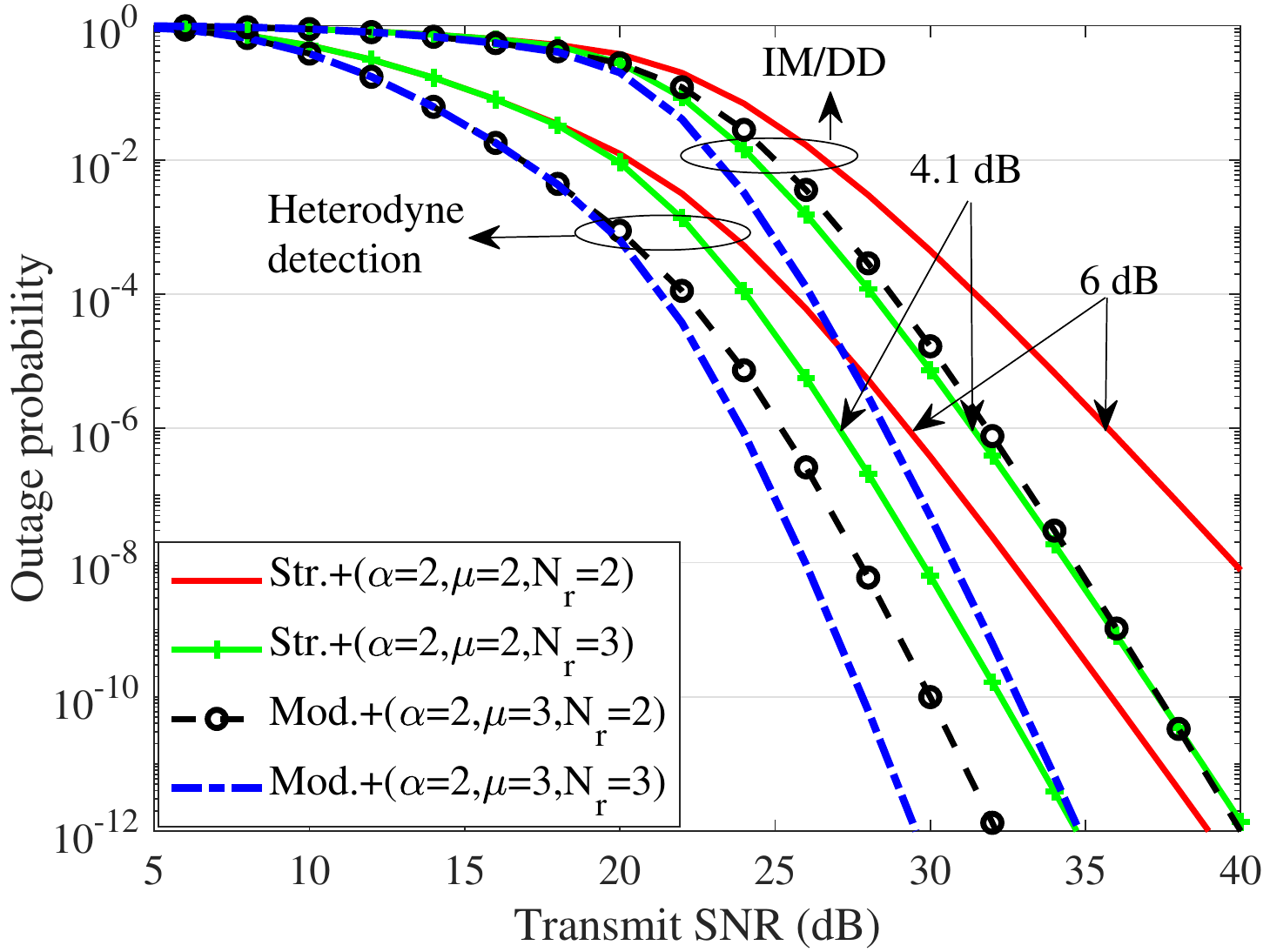}
		\vspace{-1em}
		\caption{\small{Performance of the hybrid FSO/THz link for both IM/DD and heterodyne detection at the FSO receiver.}}
		\label{detComp}
	\end{figure}	
		In Fig. \ref{detComp}, we compare the outage probability of the hybrid FSO/THz backhaul link for both IM/DD and heterodyne detection at the FSO receiver.		
		In Fig. \ref{detComp}, for the analysis, we consider strong turbulence for the FSO link with $\text{C}(1): \alpha=2,\mu=2,N_{\text{r}}=2$ and $\text{C}(2): \alpha=2,\mu=2,N_{\text{r}}=3$ cases for the THz link. Similarly, we consider moderate turbulence for the FSO link with $\text{C}(3): \alpha=2,\mu=3,N_{\text{r}}=2$ and $\text{C}(4): \alpha=2,\mu=3,N_{\text{r}}=3$ cases for the THz link. From Fig. \ref{detComp}, the outage
		performance is better in case of heterodyne detection as compared to  the IM/DD, because in heterodyne detection, the atmospheric turbulence is handled more effectively than the IM/DD \cite{zedini2016performance}.  
		Considering strong turbulence for the FSO link with $\text{C}(1)$ and $\text{C}(2)$ cases for the THz link, we receive around 6 dB and 4.1 dB gains in heterodyne detection over the IM/DD, respectively. 
		Note that the FSO link has more than 10 dB gain in heterodyne detection case as compared to the IM/DD. However, with the parallel THz link, the hybrid FSO/THz backhaul link's performance improves even with IM/DD at the FSO receiver, and for all the mentioned cases in Fig. \ref{detComp},  the gains limit to 4-7 dB  for the heterodyne detection over the IM/DD technique.
			
			\begin{figure}[h]
					\centering
					\includegraphics[width=3.52in,height=2.5in]{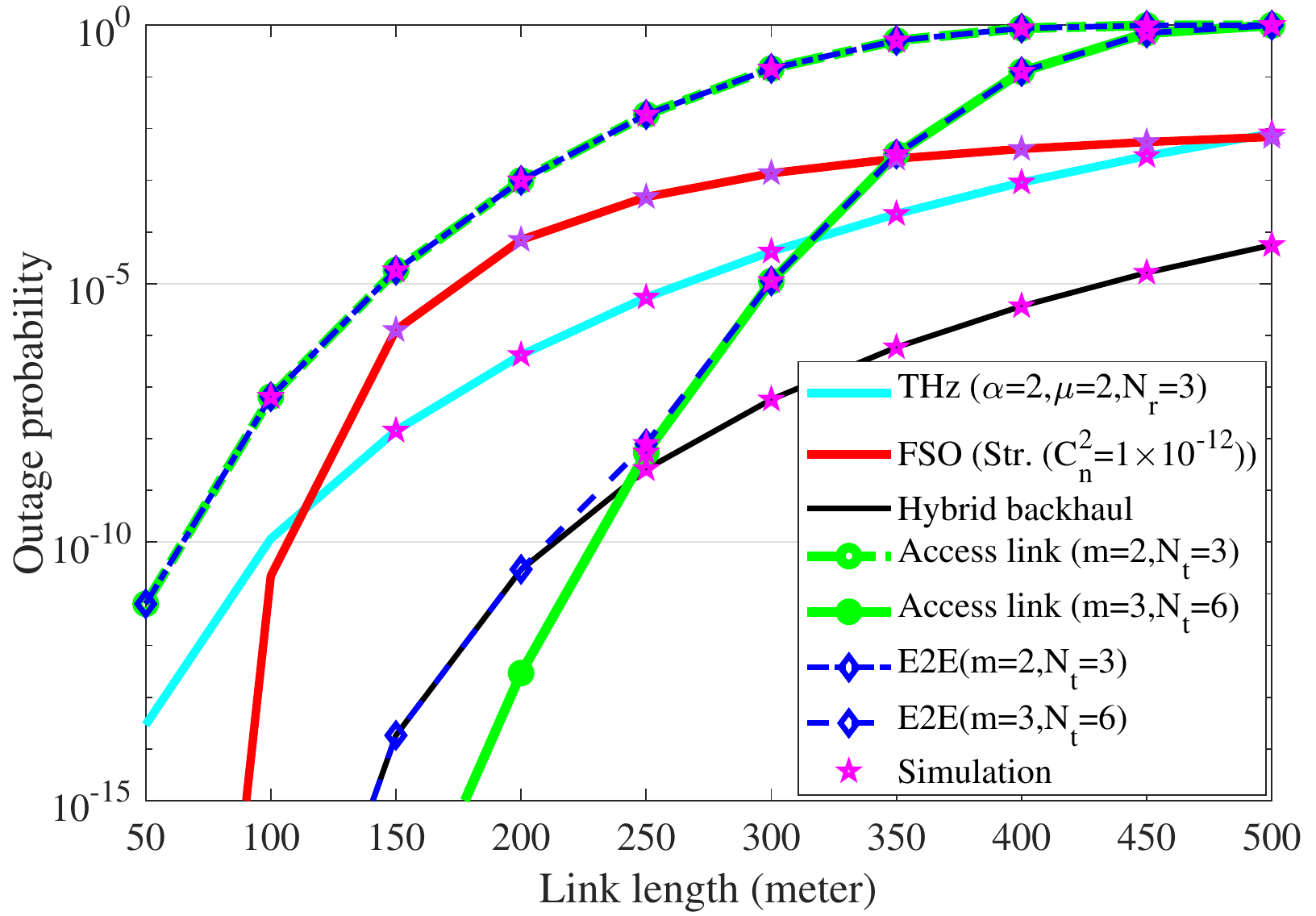}
					\caption{\small{Outage probability versus the link length by considering 30 dB transmit SNR.}}
					\label{distance}
				\end{figure}

		In Fig. \ref{distance}, a comparison of the outage probability in the individual, hybrid FSO/THz, and E2E links is shown versus the link length by considering 30 dB transmit SNR.
		In Fig. \ref{distance}, strong turbulence for the FSO link ($C_n^2=1\times10^{-12} \text{m}^{-2/3}$) and $\alpha=2,\mu=2, N_{\text{r}}=3$ for the THz link are considered. Further, $m=2,N_{\text{t}}=3$ and $m=3, N_{\text{t}}=6$ are considered for the mmWave access link. Initially, the FSO link performs far superior for very short distance. However, with increasing distance, the atmospheric turbulence and atmospheric attenuation limit the performance of the FSO link which degrades gradually with respect to 50 m increase in the distance. However, for the THz link, the path-loss is quite high at 119 GHz frequency, and even 50 m increase in link distance reduces the THz link's performance significantly.
		Even though the FSO and the THz links performances are poor after some distances, the hybrid FSO/THz backhaul link provides far superior performance throughout the considered range, and guarantees the reliability of the backhaul communication.\par
		
		From Fig. \ref{distance}, for $m=2,N_{\text{t}}=3$ case, it is  clear  that the access link's performance decreases significantly with respect to the link distance, and it is in outage after 300 m distance. For this case, the E2E performance follows the  access link's performance, hence, the MUs are in outage after 300 m distance.  By deploying more antennas at the AP and considering $m=3, N_{\text{t}}=6$ case, the access link's performance improves  significantly. Therefore, from Fig. \ref{distance},  for $m=3, N_{\text{t}}=6$ case, we observe that till 250 m link distance, the E2E performance follows the hybrid link's performance, thereafter follows the access link's performance. This is due to the fact that increased number of antennas provides significant performance  improvement to the access link.  With an increased distance, path-loss is also high which reduces the access link's performance as compared to the hybrid link at high distance, and consequently limits the E2E outage performance. Hence, deploying more antennas at the AP improves the access link's performance. Consequently, it enhances the coverage to MUs which are comparatively far from AP and also the E2E performance of the communication system. Also, this gives an idea about the number of antennas required at the AP to serve the MUs situated within a certain radius from the AP.
		
		\begin{figure}[h]
		\centering
		\includegraphics[width=3.52in,height=2.5in]{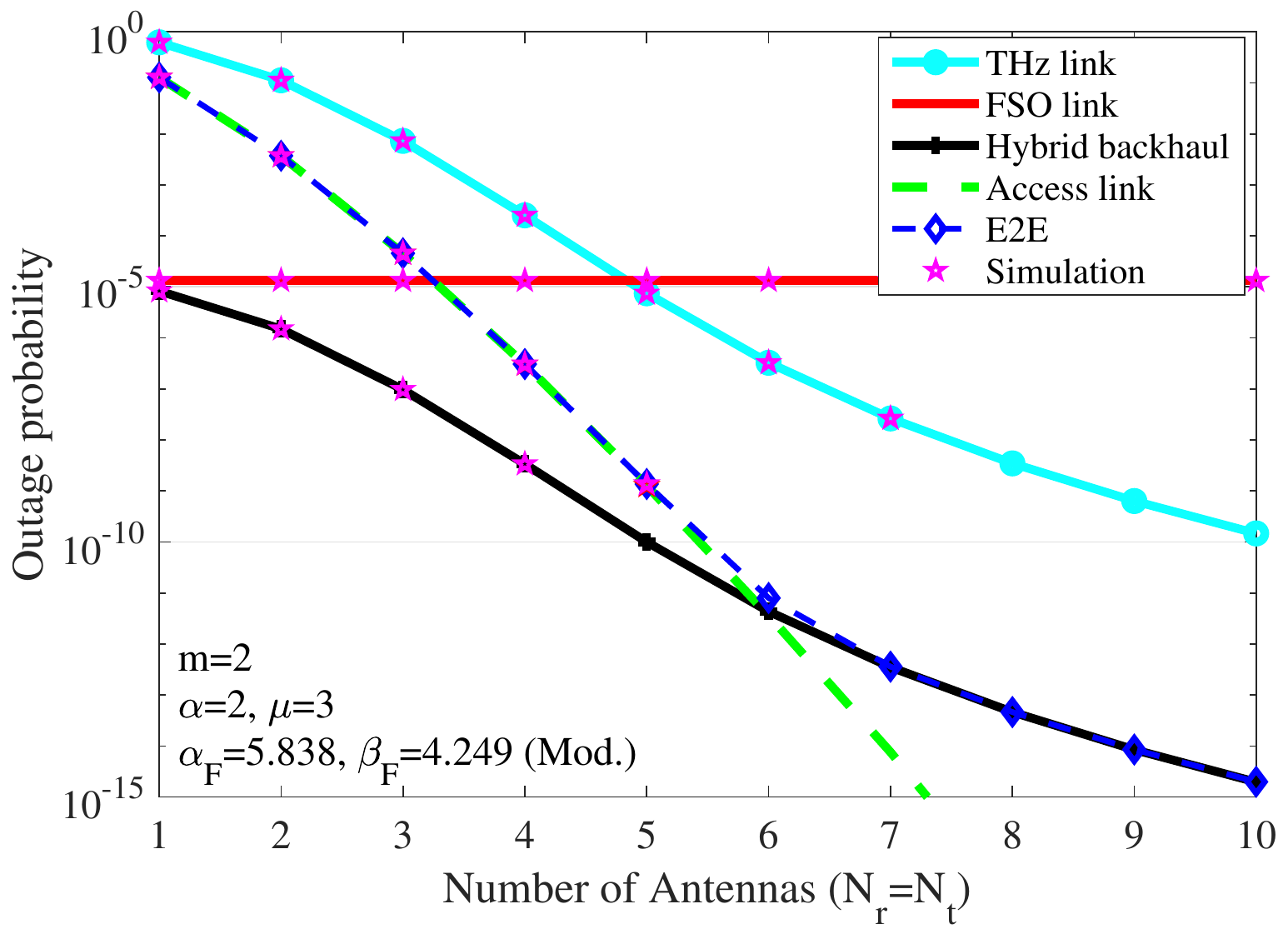}
		\caption{\small{Impact of the number of antennas on the outage probability while considering 25 dB transmit SNR.}}
		\label{Ant}
	\end{figure}

		In Fig. \ref{Ant}, the impact of the number of antennas is shown on the hybrid FSO/THz link and E2E outage performance. For the analysis, the same number of antennas at the THz receiver and AP transmitter ($N_{\text{r}}=N_{\text{t}}$) are considered. Moderate turbulence for the FSO link, $\alpha=2, \mu=3$ for the THz link, and $m=2$ for the access link are considered. Further, 25 dB transmit SNR and $L_\text{F}=L_\text{T}=$200 m, $L_{\text{R}}=$100 m are considered. From Fig. \ref{Ant}, it is observed that the THz link's performance improves with the increase in $N_{\text{r}}$ by an order of $10^{-2}$  approximately for each antenna. This corresponds to the improved hybrid FSO/THz link's outage performance. With the increase in number of antennas at the AP $N_{\text{t}}$, the access link's performance improves significantly by an order of $10^{-3}$  for each antenna. Hence, from  Fig. \ref{Ant}, it is clear that till 5 antennas, the E2E outage performance follows the access link. However, with $\geq6$ antennas, the access link's performance surpasses the hybrid FSO/THz  link's performance, and the E2E outage performance follows the hybrid link. It is observed that for $m=2$, considering  3 antennas at the AP can provide the E2E performance up-to the received backhaul hybrid FSO/THz link with strong turbulence and case(a) for the FSO and the THz links, respectively. Similarly, for $m=2$, considering  5 antennas at the AP can provide the E2E performance up-to the received backhaul hybrid FSO/THz link with moderate turbulence and case(b) for the FSO and the THz links, respectively. Hence, by increasing the number of antennas at the THz receiver, one can obtain a performance comparable to the FSO link. At the same time, the RF antennas at the AP improves the mmWave access link's performance to reach the almost same performance as in the backhaul link.
		
		\begin{figure}[h]
			\centering
			\includegraphics[width=3.52in,height=2.5in]{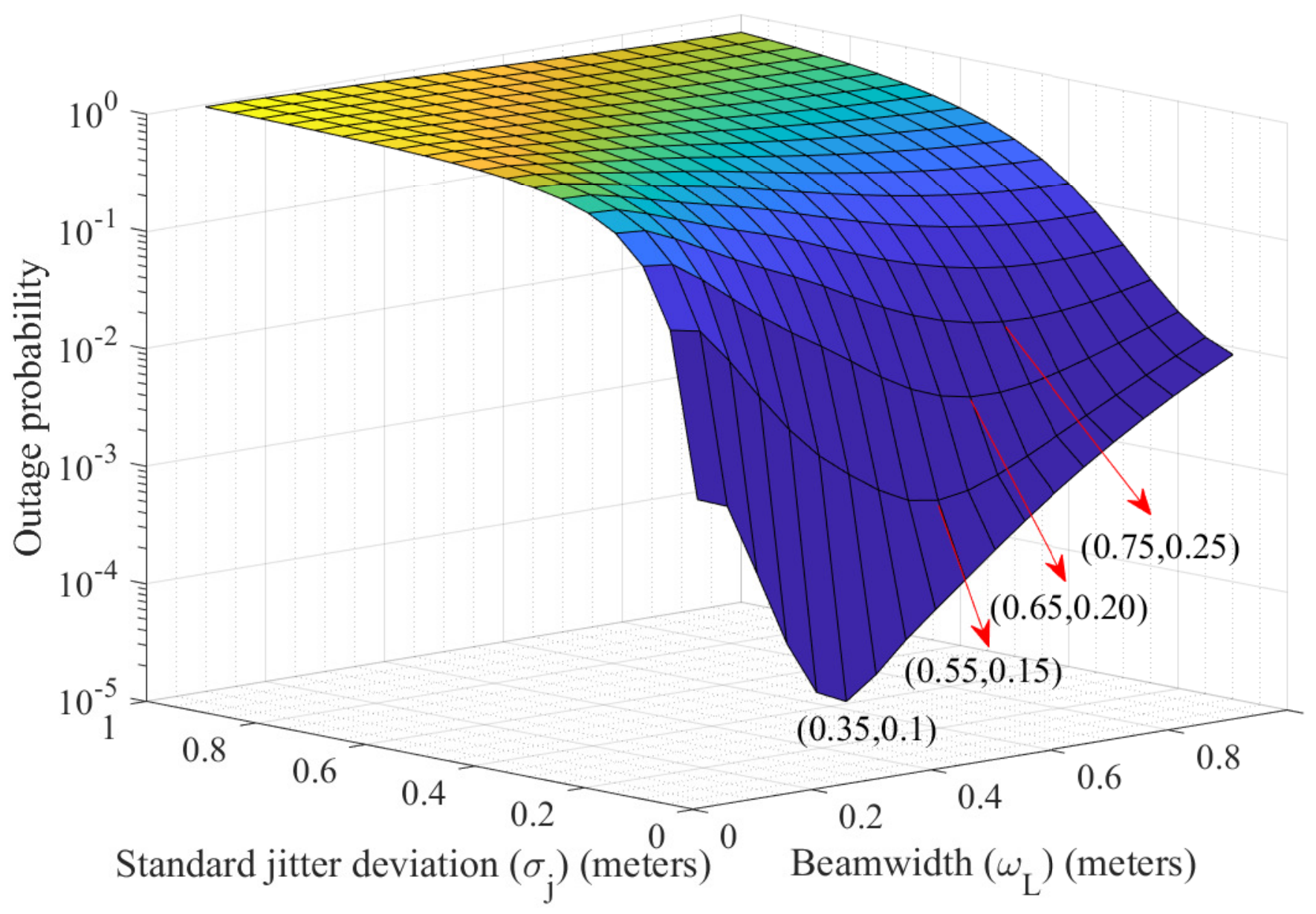}
			\vspace{-1em}
			\caption{\small{The impact of beamwidth and jitter standard deviation on the outage probability of the hybrid FSO/THz backhaul link while considering common receiver radius=10 cm and transmit SNR=30 dB.}}
			\label{Imp_Point1}
		\end{figure}
		In Fig. \ref{Imp_Point1}, the combined impact of beamwidth and jitter standard deviation is shown on the outage performance of the backhaul hybrid FSO/THz link. For this, equal beamwidth and jitter standard deviation are considered for both the FSO and THz links, i.e., $\omega_{\text{L,F}}=\omega_{\text{L,T}}=\omega_\text{L}$ and $\sigma_{\text{j,F}}=\sigma_{\text{j,T}}=\sigma_\text{j}$, and 10 cm receiver aperture radius is considered for both links. Further, moderate turbulence for the FSO link and $\alpha=2, \mu=3, N_{\text{r}}=3$ for the THz link are considered. \par
		From  Fig. \ref{Imp_Point1}, it is observed that for a given transmit power with an increase in jitter standard deviation $\sigma_\text{j}$, the pointing/misalignment error increases. Hence, the coverage performance decreases significantly and the FSO and THz links are in outage for high values of $\sigma_\text{j}$. Further, with the increase in beamwidth $\omega_{\text{L,i}}$, the equivalent beamwidth $\omega_{\text{Leq},\text{i}}$ increases which gives rise to the parameter $\xi_\text{i}={\omega_{\text{L}_{\text{eq}},\text{i}}}/{\left(2\sigma_{\text{j,i}}\right)}$. With the increase in $\omega_{\text{L,i}}$, the outage performance of the FSO and the THz links (and consequently the hybrid FSO/THz link) should increase. However, with an increase in the beamwidth $\omega_{\text{L,i}}$, the parameter $v_{0,\text{i}}$ decreases and hence, the received power $A_{0,\text{i}}=[\text{erf}(v_{0,\text{i}})]^2$ decreases. 
		Thus, for a given transmit power, there is a trade-off between the beamwidth and the jitter standard deviation for optimum performance, and the minimum outage probability is obtained with a specific value of the beamwidth for each jitter standard deviation.
		The beamwidth should be increased with the increasing jitter standard deviation to achieve an optimum outage performance. From Fig. \ref{Imp_Point1}, for 0.1, 0.15, 0.2, 0.25 meter jitter standard deviations, the optimum outage performance can be achieved with 0.35, 0.55, 0.65, 0.75 meter beamwidth, respectively. However, further increase in beamwidth will reduce the outage performance of the FSO, the THz, and consequently the hybrid FSO/THz links, since less power is collected at the receiver.
		
			\begin{figure}[h]
			\centering
			\includegraphics[width=3.52in,height=2.5in]{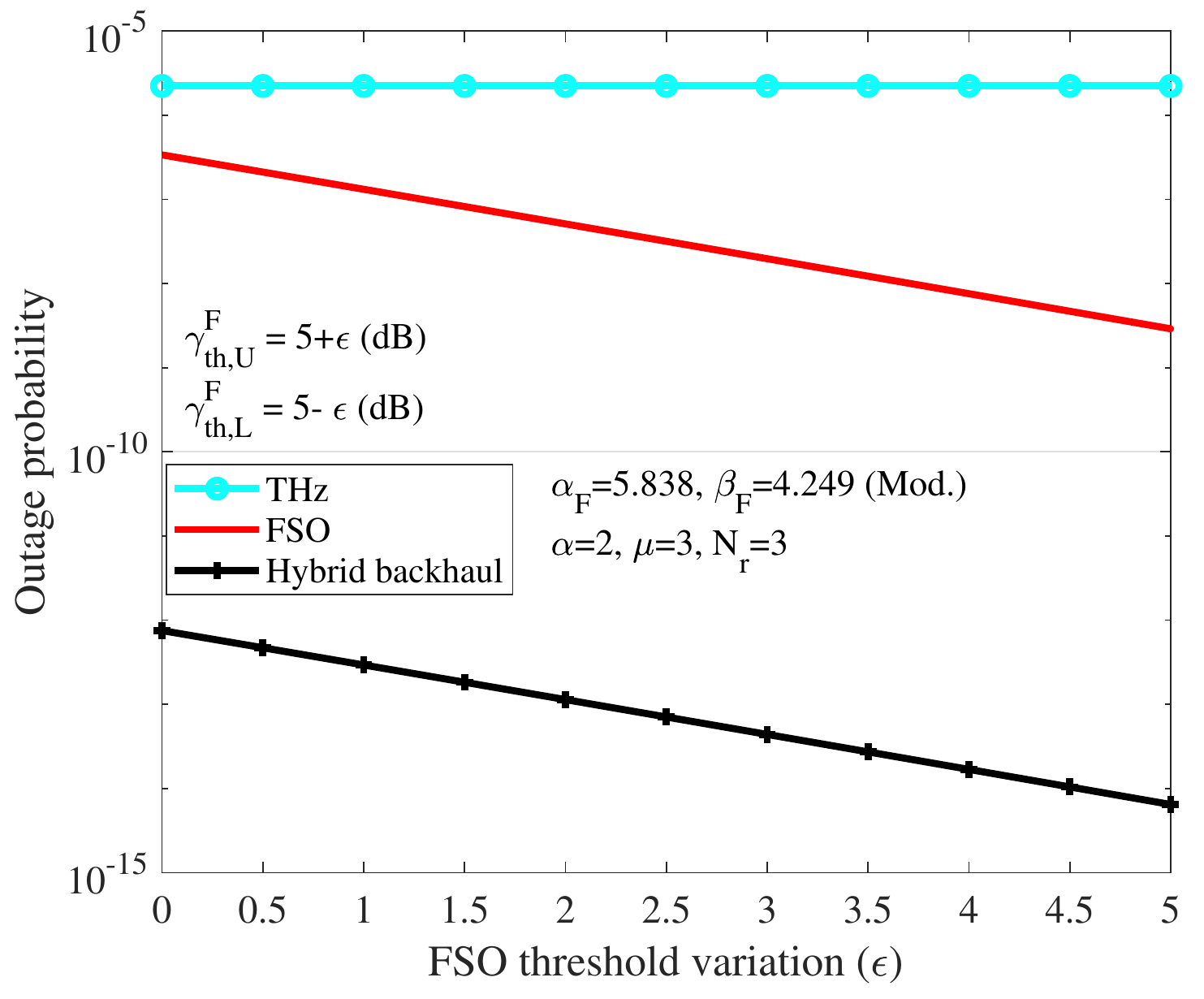}
			\caption{Impact of threshold SNR on the outage probability of hybrid FSO/THz link, considering 30 dB transmit SNR.}
			\label{FSOThr}
		\end{figure}
		In Fig. \ref{FSOThr}, the impact of change in upper and lower FSO thresholds is shown on the outage performance of the backhaul hybrid FSO/THz link. For  analysis, moderate turbulence for the FSO link and $\alpha=2,\mu=3,N_{\text{r}}=3$ are considered. Further, the upper and lower FSO thresholds vary as $\gamma^{\text{F}}_{\text{th,U}}=(5+\epsilon)$ dB and $\gamma^{\text{F}}_{\text{th,L}}=(5-\epsilon)$ dB, respectively, where $\epsilon$ represents the variation parameter for the FSO threshold. From Fig. \ref{FSOThr}, we observe that the THz link's performance remain unchanged. However, with the increase in $\epsilon$, the gap between the upper and lower FSO thresholds increases which improves the outage performance of the FSO link. As a result, the hybrid FSO/THz link's performance improves accordingly which further reduces the probability of back-and-forth switching in the hybrid FSO/THz link.
				
			\subsection{ABER Analysis}	
		
		\begin{figure}[h]
			\centering
			\includegraphics[width=3.52in,height=2.5in]{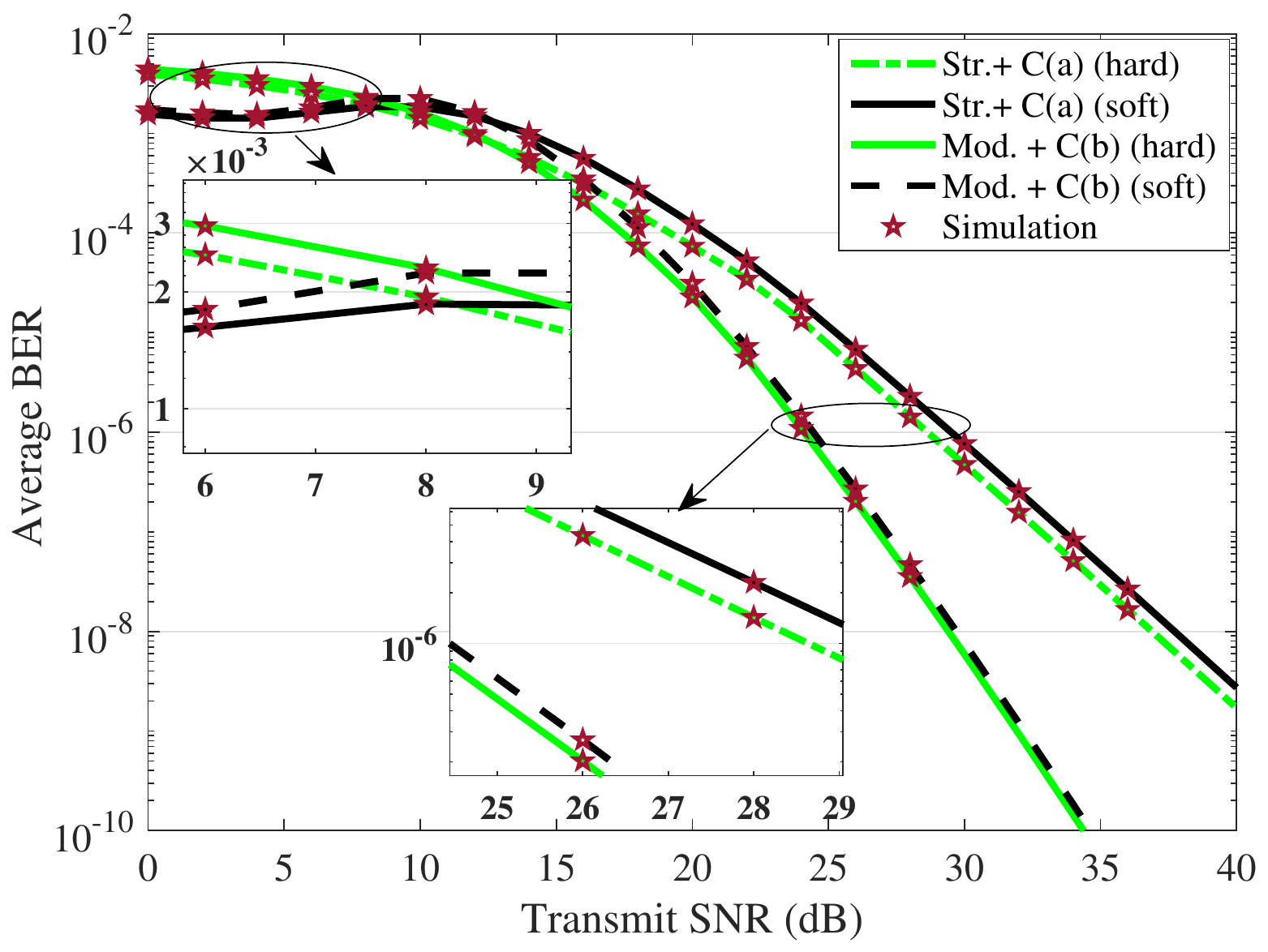}
			\caption{\small{ABER comparison of the soft and hard switching for the hybrid FSO/THz backhaul link, considering strong and moderate turbulence cases for the FSO link and $\text{case(a)}: \alpha=2, \mu=3, N_{\text{r}}=2$ and $\text{case(b)}: \alpha=2, \mu=3, N_{\text{r}}=3$ for the THz link.}}
			\label{ABER_comp}
		\end{figure}
		In Fig. \ref{ABER_comp}, the ABER results of the hybrid FSO/THz link for the soft and hard switching are compared and verified through the simulation results. We have shown two magnified figures at low and medium SNRs to compare the ABER results of the soft and hard switching methods.
		Considering the strong  and moderate cases for the hybrid FSO/THz link,
		the ABER performance for the soft switching degrades respectively by 0.91 dB and 0.35 dB as compared to the hard switching case to achieve an ABER probability of $10^{-6}$, which is shown in  Fig. \ref{ABER_comp} with a magnified figure in the medium SNR regime. This is due to the fact that in soft switching case, the FSO link is mostly active at high SNRs even if THz link's SNR may be larger than the FSO link when FSO link's SNR is in between the upper and lower thresholds. However, with the soft switching, the frequent switching between the FSO and THz links is reduced significantly at the cost of marginal lag in the ABER performance.
		Further, at low SNRs, the soft switching provides better ABER performance as compared to the hard switching which is shown in  Fig. \ref{ABER_comp} with a magnified figure at low SNRs. This is due to the fact that at low SNRs, the FSO link is inactive most of the times. This forces the hybrid link to switch to the better quality THz link (or to the outage if THz link's SNR is also poor). Conclusively, at the low (resp. high) SNRs, the soft (resp.
		hard) switching provides better ABER performance, compared to the hard (resp. soft) switching.
		
				\begin{figure}[t]
						\centering
						\includegraphics[width=3.52in,height=2.5in]{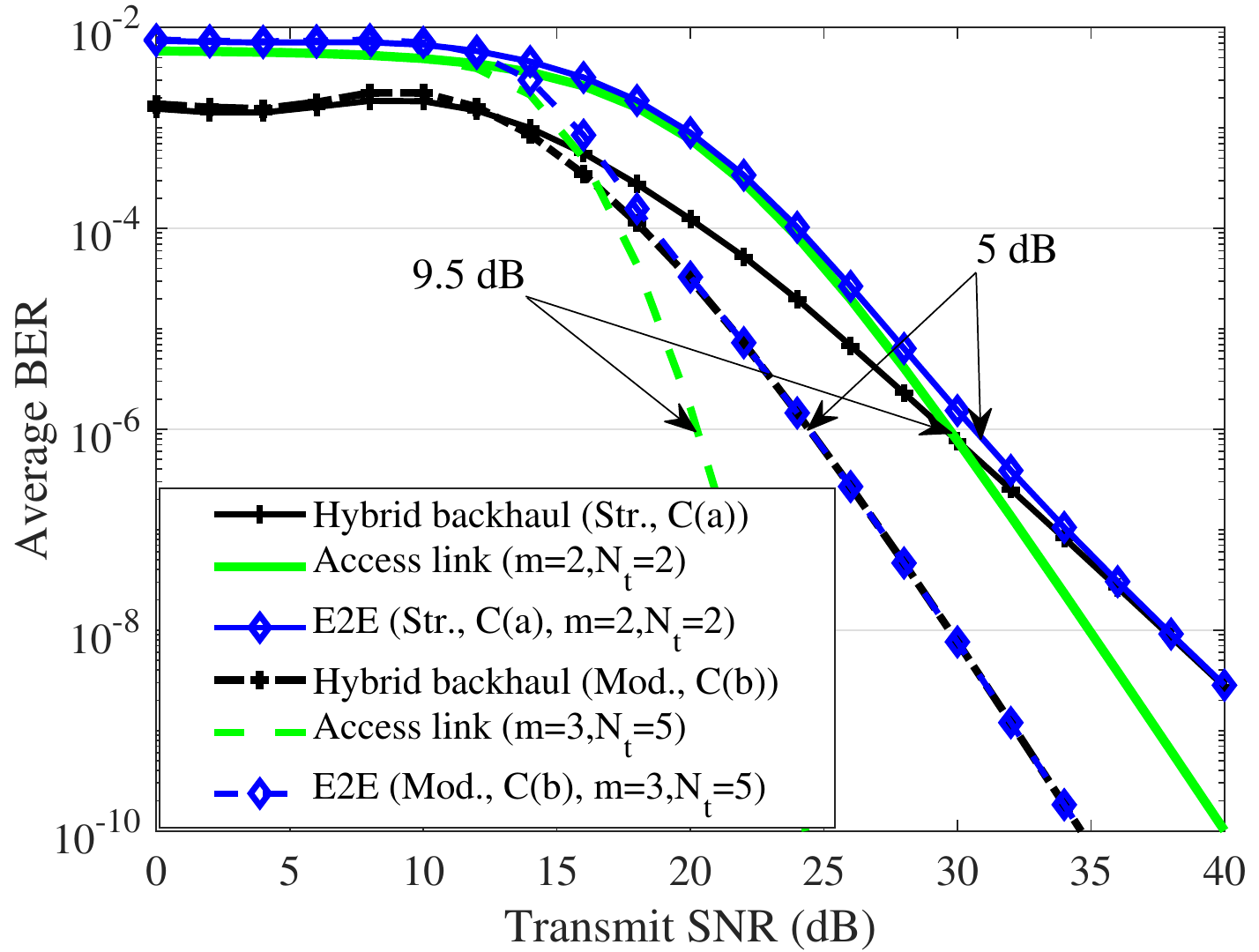}
						\caption{\small{ABER versus the transmit SNR for various cases.}}
						\label{BER_comp_F}
					\end{figure}

		In Fig. \ref{BER_comp_F}, the analytical ABER results of the hybrid FSO/THz and the access links are shown for various cases and their impact is highlighted on the E2E ABER performance of the considered system. 
		To achieve an ABER probability of $10^{-6}$, one needs to increase the transmit SNR by 5 dB to achieve the same quality-of-service in the strong case as that obtained in the moderate turbulence case. On the other hand, the access link provides nearly 9.5 dB gain when moves from $m=2,N_{\text{t}}=2$ to $m=3,N_{\text{t}}=5$ to achieve $10^{-6}$ ABER probability. Considering the strong case for the hybrid link and $m=2, N_{\text{t}}=2$ for the access link, the hybrid link outperforms the access link below 35 dB SNR, and the E2E ABER performance follows the access link's performance. At high SNRs, however, the access link's performance surpasses the hybrid link's performance, and the E2E performance is dominated by the hybrid link. 
		Considering the moderate case for the hybrid link and $m=3, N_{\text{t}}=5$ for the access link, the ABER performance of access link improves significantly due to the increased number of antennas at the AP. Hence, the E2E ABER performance follows the hybrid link for the cases with $>20$ dB SNR.  
		
		\begin{figure}[h]
		\centering
		\includegraphics[width=3.52in,height=2.5in]{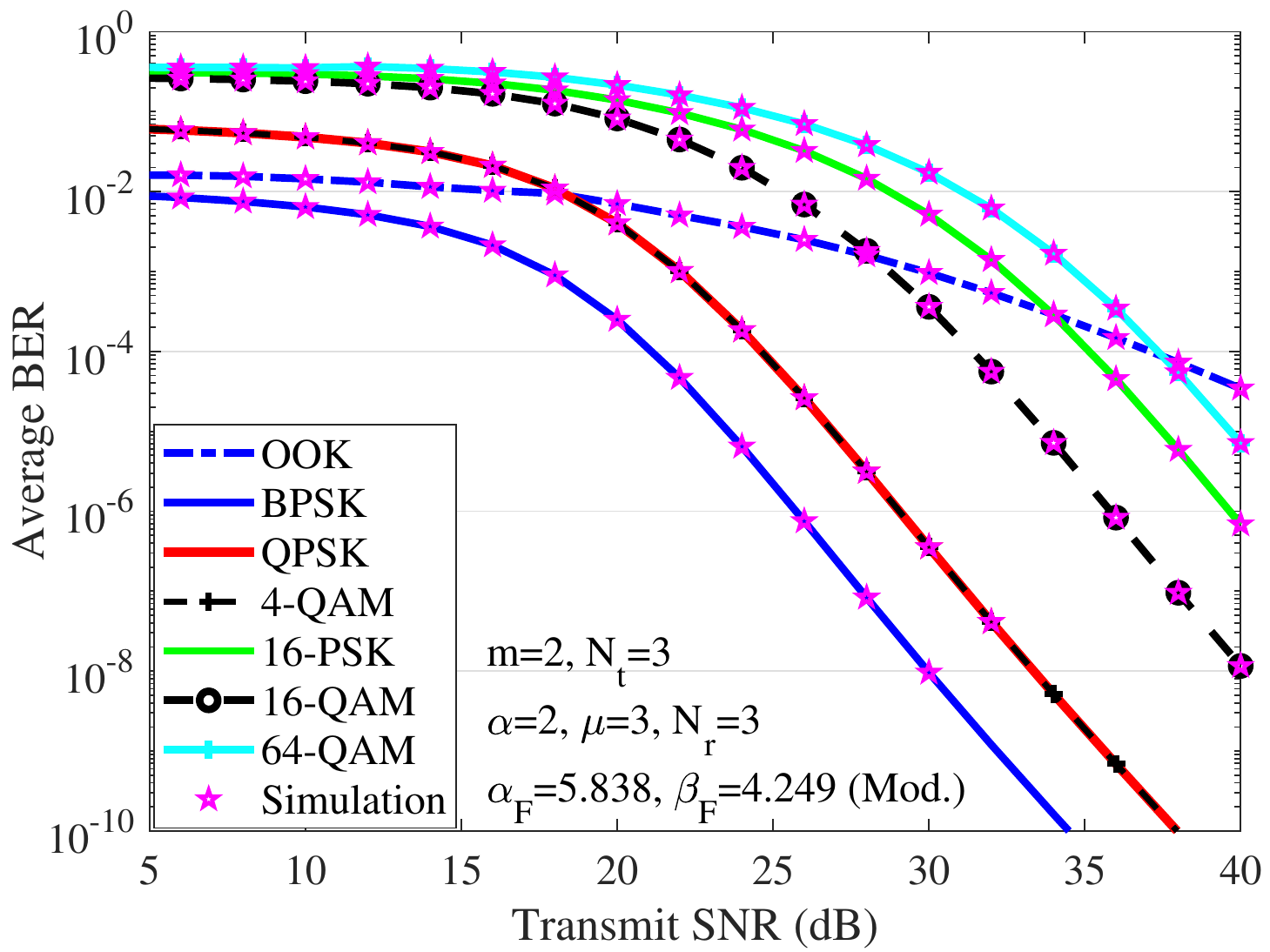}
		\caption{\small{The E2E ABER for various modulation schemes.}}
		\label{BER_COMP}
	\end{figure}
		In Fig. \ref{BER_COMP}, the E2E ABER for various modulation schemes are compared  and verified with the simulation results. For the analysis, moderate turbulence for the FSO link, $\alpha=2,\mu=3,N_{\text{r}}=3$ for the THz link, and $m=2,N_{\text{t}}=3$ for the access link are considered.  For the IM/DD at the FSO receiver, we consider OOK for the FSO link while BPSK is considered for the THz and mmWave access links. For the heterodyne detection at the FSO receiver, rest of the modulation schemes are considered for the FSO as well as the THz and mmWave access links. From Fig. \ref{BER_COMP}, we observe
		that to achieve an ABER of $10^{-4}$,  BPSK provides around 16 dB gain as compared to OOK. This is due to the fact that the heterodyne detection at the FSO receiver improves the ABER performance significantly as compared to the IM/DD, because the heterodyne detection can handle the atmospheric turbulence effects more efficiently \cite{zedini2016performance}. Further, with the increase in constellation order, we can achieve higher data-rates at the cost of increased BER. There is always a trade-off between the required data-rates and acceptable BER performance which depends upon the requirements of the communication system. The increase in BER is due to the fact, that for the high constellation order, the Euclidean distance between the constellation points increases to maintain the constant transmit power which increases the BER. The ABER performance of QPSK and 4-QAM is the same. However, with the increase in constellation order, the M-QAM provides better performance as compared to the M-PSK. This is due to the fact that in M-PSK, only the phase of the constellation points is modulated. Hence, with increasing constellation order, the constellation points are quite close to each other which increases the BER. On the other hand, in M-QAM, both the amplitude and phase of the constellation points are modulated. Hence, for the same constellation order, the constellation points are better placed in M-QAM than in M-PSK which results in better BER performance in M-QAM than in M-PSK. To achieve an ABER of $10^{-4}$, 16-QAM provides around 3.7 dB gain over 16-PSK. Further, increasing the constellation order from 16 to 64, the ABER performance of QAM is reduced by 5.9 dB to achieve an ABER of $10^{-4}$.
		\begin{figure*}
			\hfill
			\subfigure[Comparison of the FSO and THz links' capacity, obtained through the closed-form and integral-form of identities $\Theta_2$ and $\Theta_{4}$.]{\includegraphics[width=8.8cm,height=6.8cm]{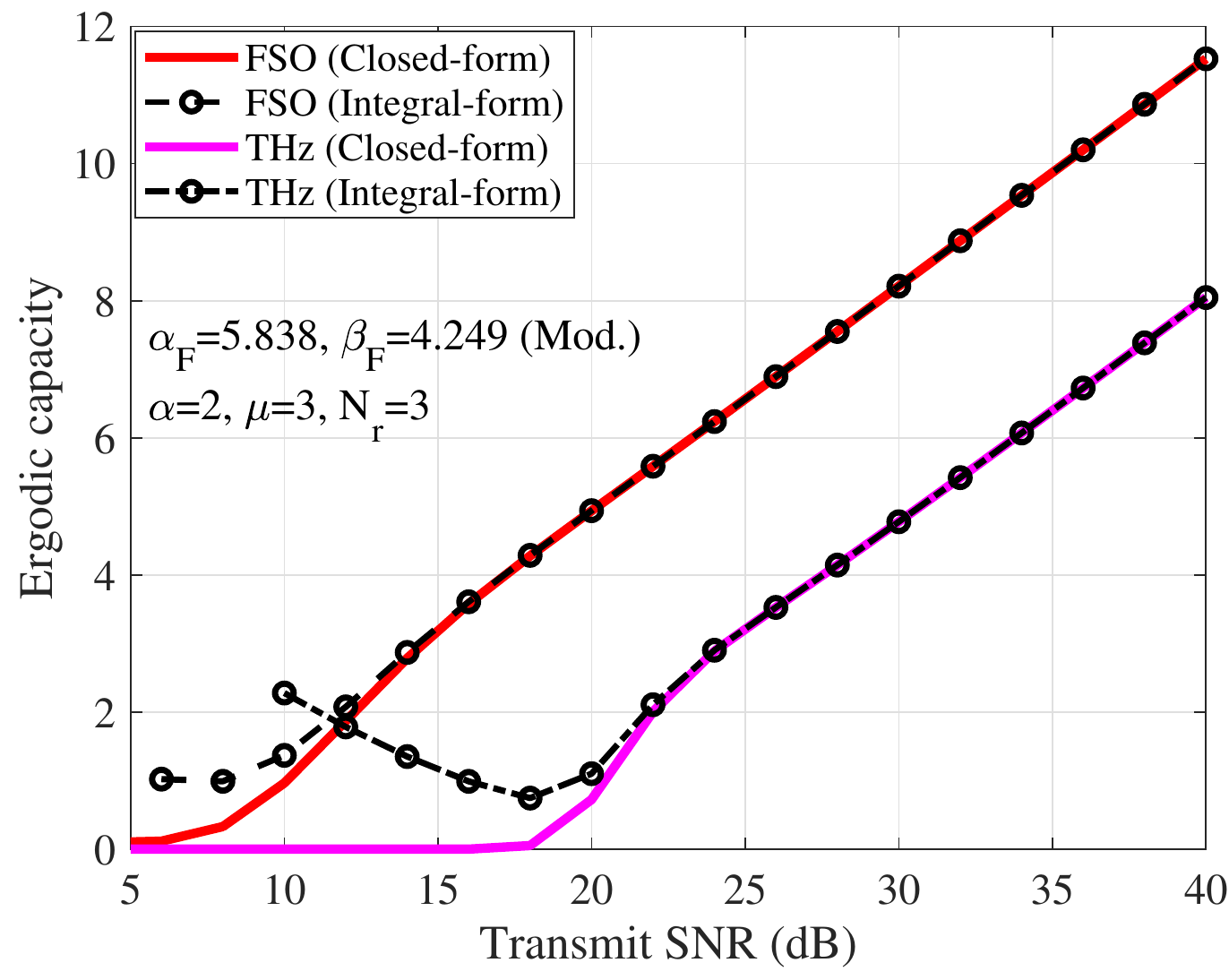}
				\label{Cap1}}	
			\hfill
			\subfigure[Hybrid FSO/THz backhaul link's capacity along with the E2E ergodic capacity.]{\includegraphics[width=8.8cm,height=6.8cm]{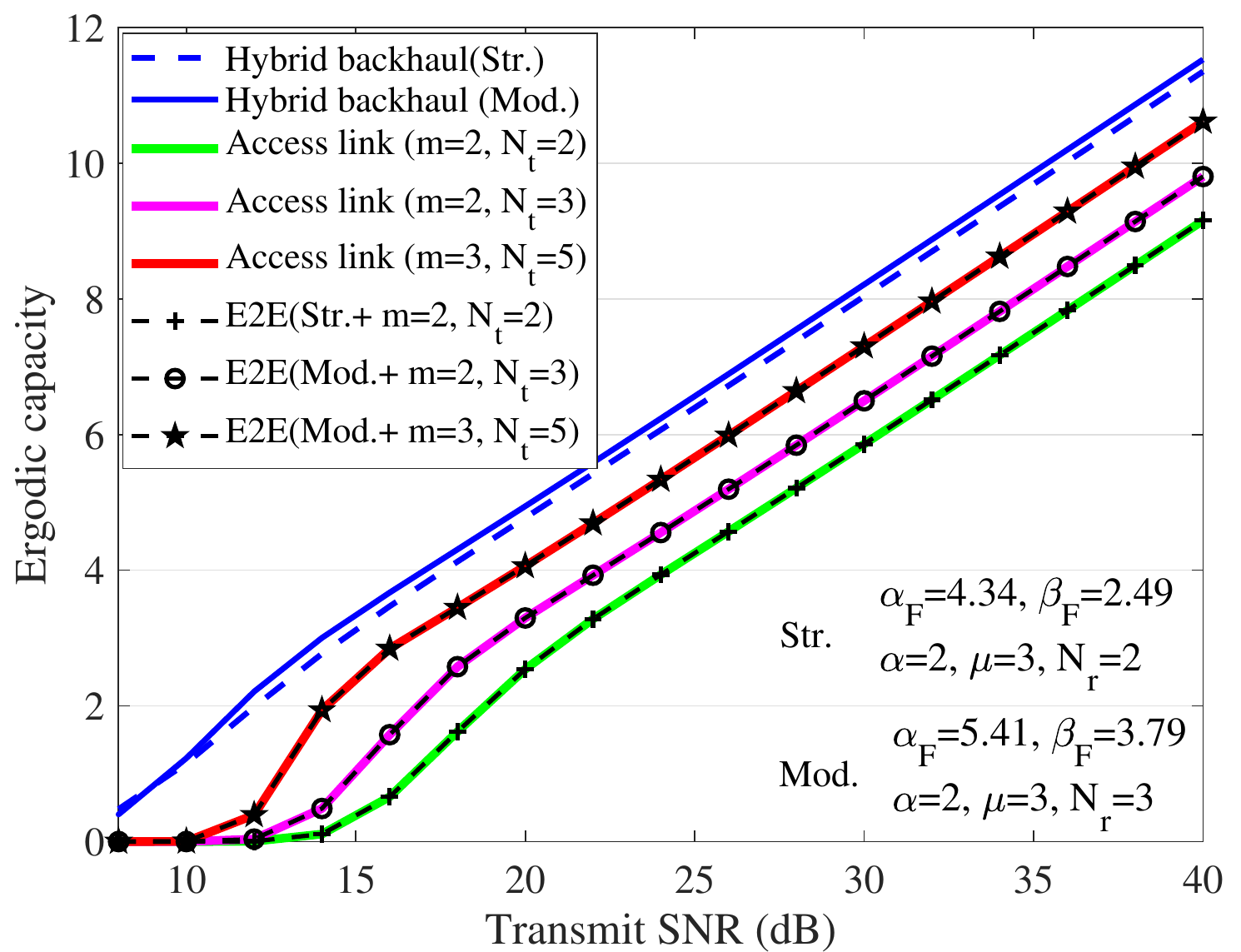}
				\label{Cap2}}
			\hfill
			\caption{\small{Ergodic capacity versus the transmit SNR.}}
			\label{Cap_main}	
			\hrulefill
			\vspace{-1em}
		\end{figure*}
				
		\subsection{Ergodic Capacity Analysis}			
		In Fig. \ref{Cap_main}, we study the ergodic capacity versus the transmit SNR. In Fig. \ref{Cap1}, the ergodic capacity results of the FSO and  the THz links are shown by considering moderate turbulence for the FSO link and $\alpha=2, \mu=3, N_{\text{r}}=3$ for the THz link. A comparison of capacity results is performed by considering the closed-form and integral-form of identities $\Theta_2$ and $\Theta_4$, i.e., (\ref{I2F}) and (\ref{I4F}), for the FSO and the THz links, respectively. It is observed that the closed-form results match well with the integral-form results at medium and high SNRs but differ at low SNRs for both the FSO and THz links. This is due to the fact that while calculating the capacity for the FSO and THz links, the identities $\Theta_2$ and $\Theta_4$ (integrals from 0 to a finite threshold) are solved by taking the high SNR approximation of  $\text{ln}(1+\gamma)\approx \text{ln}(\gamma)$. 
		\par
		
		In Fig. \ref{Cap2}, the ergodic capacity results of the hybrid FSO/THz link and the access link are obtained for various cases and their impacts on the E2E capacity are shown. We consider strong  and moderate cases for the hybrid FSO/THz link. It is observed that at moderate/high SNRs the ergodic capacity decreases by 0.18 bits/s/Hz when moving from the moderate case to the strong turbulence case. Also, it is observed that 
		0.65 and 0.8 bits/s/Hz gains are achieved while moving from $m=2, N_{\text{t}}=2$ to $m=2, N_{\text{t}}=3$ and $m=2, N_{\text{t}}=3$ to $m=3, N_{\text{t}}=5$, respectively.
		However, the hybrid FSO/THz link has higher capacity than the access link for all the considered cases, hence, the E2E capacity at MUs is restricted to the access link's capacity. Interestingly, the results indicate that, with different parameter settings, at moderate/high SNRs the ergodic capacity of the individual, hybrid, and E2E links scale with the transmit SNR, in the log-domain, linearly.
		
		\section{Conclusions}
		In this work, we consider a hybrid FSO/THz based backhaul network to provide high-data rates to the terrestrial MUs via mmWave access links. The atmospheric turbulence and pointing errors are considered in the FSO link, while the THz link may suffer from short-term fading and misalignment errors. We investigate the performance of the proposed scheme for different soft and hard switching methods. 			
		The analytical and the simulation results show that the soft switching provides marginal gain, in terms of outage probability, over the hard switching,	however reduces the frequent back-and-forth switching. Also, the atmospheric turbulence parameters and
		pointing error directly affect the diversity order of the FSO link, while the fading parameters, misalignment and the number of antennas affect the diversity order in the THz link. For a broad range of parameter settings, the E2E capacity is restricted by the rate achieved in the access links, as the bottleneck of the network performance. Finally, at
		the low (resp. high) SNRs, the soft (resp. hard) switching provides better ABER performance, compared to the hard (resp. soft) switching.  
		
		\appendices
		\numberwithin{equation}{section}
		\section{Proof of Asymptotic Outage Probability}
		Meijer-G function is the dominant term in the outage probability of the FSO link (\ref{Gamma_CDF}). To obtain the asymptotic outage probability at high SNRs, it can be simplified as 
		\cite[(07.34.06.0001.01)]{wolframe}
		\begin{align} \label{FSOapp}
			&F^\text{A}_{\gamma_{\text{F}}}(\gamma_{\text{F}})\approx \mathbb{D}_1\sum_{p=1}^{3\tau}\Big(\frac{\mathbb{D}_2\gamma_{\text{F}}}{(A_{0,\text{F}})^\tau\delta_{{\tau}}}\Big)^{\varrho_8,p}\nonumber\\&
			\times	\frac{\prod_{\underset{q\neq p}{q=1}}^{3\tau} \Gamma(\varrho_{8,q}-\varrho_{8,p})\prod_{\underset{q=1}{}}^{1}\Gamma(1-\varrho_{7,q}+\varrho_{8,p})}{\prod_{\underset{q=2}{}}^{\tau+1}\Gamma(\varrho_{7,q}-\varrho_{8,p})\prod_{\underset{q=3\tau+1}{}}^{3\tau+1}\Gamma(1-\varrho_{8,q}+\varrho_{8,p})},
		\end{align} 
		where  $\varrho_7=[1,\varrho_1]$ and $\varrho_8=[\varrho_2,0]$.
		From, (\ref{FSOapp}), it is observed that the asymptotic outage probability of the FSO link can be seen as $\approx (G_\text{c}\times \delta_{{\tau}})^{-G_\text{d}}$, where $G_\text{c}$ and $G_\text{d}$ are the coding and diversity gains of the FSO link. Therefore, from (\ref{FSOapp}), the diversity gain is obtained as $\min\left(\frac{\xi^2_\text{F}}{\tau},\frac{\alpha_{\text{F}}}{\tau},\frac{\beta_{\text{F}}}{\tau}\right)$.
		
		Considering the high SNR approximation of upper incomplete gamma function as $\Gamma(m,x)\underset{\overset{x\rightarrow 0}{}}{\approx}\left(\Gamma(m)-\frac{x^m}{m}\right)$, at high SNR, the PDF of the THz link (\ref{PDF_THz}) can be approximated as
		\begin{align}\label{PDF_THz_Asym}
			f^\text{A}_{\gamma_{\text{T}}}(\gamma_{\text{T}})\approx\frac{\mathbb{C}_1}{2(\bar{\gamma})^{\frac{\xi_{\text{T}}^2}{2}}}\gamma_{\text{T}}^{\frac{\xi_{\text{T}}^2}{2}-1}\left(\Gamma(\mathbb{C}_2)-\frac{\mathbb{C}_3^{\mathbb{C}_2}}{\mathbb{C}_2}\left(\frac{\gamma_{\text{T}}}{\bar{\gamma}}\right)^{\frac{\mathbb{C}_2\alpha}{2}}\right).
		\end{align}	
		Performing  $F^\text{A}_{\gamma_{\text{T}}}(\gamma_{\text{T}})=\int_{0}^{\gamma_{\text{T}}}f^\text{A}_{\gamma_{\text{T}}}(\gamma)\text{d}\gamma$ with some mathematical computations,
		the CDF of the THz link at high SNRs is derived as
		\begin{align}\label{THz_Asym}
			F^\text{A}_{\gamma_{\text{T}}}(\gamma_{\text{T}})\approx\frac{\mathbb{C}_1\Gamma(\mathbb{C}_2)}{\xi_{\text{T}}^2}\left(\frac{\gamma_{\text{T}}}{\bar{\gamma}}\right)^{\frac{\xi_{\text{T}}^2}{2}}-
			\frac{\mathbb{C}_1\mathbb{C}_3^{\mathbb{C}_2}}{\mathbb{C}_2\left(\alpha N_{\text{r}}\mu\right)}\left(\frac{\gamma_{\text{T}}}{\bar{\gamma}}\right)^{\frac{\alpha N_{\text{r}}\mu}{2}}.
		\end{align}	
		From (\ref{THz_Asym}), the asymptotic outage probability of the THz link can be seen as $\approx(G_\text{c}\times\bar{\gamma})^{-G_\text{d}}$, and its diversity gain can be obtained as $\min\left(\frac{\xi_{\text{T}}^2}{2},\frac{\alpha N_\text{r}\mu}{2}\right)$.\par
		
		Considering the high SNR approximation of $\Gamma(m,x)\underset{\overset{x\rightarrow 0}{}}{\approx}\left(\Gamma(m)-\frac{x^m}{m}\right)$, the CDF and the PDF of the Nakagami-m distributed mmWave access link at the $n^{th}$ MU is approximated as
		\begin{align}\label{RF_asym}
			f^A_{\gamma_{\text{MU}}}(\gamma_{\text{R}})&\approx\frac{1}{\Gamma\left(mN_{\text{t}}\right)}\left(\frac{m}{\bar{\gamma}_{\text{R}}}\right)^{mN_{\text{t}}}\gamma_{\text{R}}^{mN_{\text{t}}-1},\nonumber\\  F^A_{\gamma_{\text{MU}}}(\gamma_{\text{R}})&\approx\frac{1}{\Gamma\left(mN_{\text{t}}\right)}\left(\frac{m}{\bar{\gamma}_{\text{R}}}\gamma_{\text{R}}\right)^{mN_{\text{t}}},
		\end{align}
		respectively.
		From (\ref{RF_asym}), the diversity gain of the mmWave access link is $ m N_{\text{t}}$ which is directly dependent on the number of antennas $N_{\text{t}}$ at the AP and mmWave access link's fading severity $m$.
		
		\section{Proof of Identities $\Theta_1$, $\Theta_2$, $\Theta_3$, and $\Theta_4$}
		$\Theta_1$ and $\Theta_2$ in (\ref{Cap_FSO}) can be respectively given as
		\begin{align}\label{I1I2}
			\Theta_1&= \frac{\xi^2_{\text{F}}}{\text{ln}(2)\tau\Gamma(\alpha_{\text{F}})\Gamma(\beta_{\text{F}})}\int_{0}^{\infty}\frac{1}{\gamma}\text{ln}(1+\Xi\gamma)\nonumber\\&
			\times\text{G}^{3,0}_{1,3}\left[\frac{\alpha_{\text{F}} \beta_{\text{F}}}{A_{0,\text{F}}}\left(\frac{\gamma}{\delta_{{\tau}}}\right)^{\frac{1}{\tau}}\Big{|}^{\xi^2_{\text{F}}+1}_{\xi^2_{\text{F}},\alpha_{\text{F}},\beta_{\text{F}}}\right]\text{d}\gamma,\nonumber\\
			\Theta_2&= \frac{\xi^2_{\text{F}}}{\text{ln}(2)\tau\Gamma(\alpha_{\text{F}})\Gamma(\beta_{\text{F}})}\int_{0}^{\gamma_{\text{th}}}\frac{1}{\gamma}\text{ln}(1+\Xi\gamma)\nonumber\\&
			\times\text{G}^{3,0}_{1,3}\left[\frac{\alpha_{\text{F}} \beta_{\text{F}}}{A_{0,\text{F}}}\left(\frac{\gamma}{\delta_{{\tau}}}\right)^{\frac{1}{\tau}}\Big{|}^{\xi^2_{\text{F}}+1}_{\xi^2_{\text{F}},\alpha_{\text{F}},\beta_{\text{F}}}\right]\text{d}\gamma.
		\end{align}
		
		$\Theta_1$ can be written as
		\begin{align}
			\Theta_1&= \frac{\xi^2_{\text{F}}}{\text{ln}(2)\tau\Gamma(\alpha_{\text{F}})\Gamma(\beta_{\text{F}})}\int_{0}^{\infty}\frac{1}{\gamma}\text{G}^{2,1}_{2,2}\left[\Xi\gamma\Big{|}^{1,1}_{1,0}\right]\nonumber\\&
			\times\text{G}^{3,0}_{1,3}\left[\frac{\alpha_{\text{F}} \beta_{\text{F}}}{A_{0,\text{F}}}\left(\frac{\gamma}{\delta_{{\tau}}}\right)^{\frac{1}{\tau}}\Big{|}^{\xi^2_{\text{F}}+1}_{\xi^2_{\text{F}},\alpha_{\text{F}},\beta_{\text{F}}}\right]\text{d}\gamma.
		\end{align}
		
		Applying \cite[(07.34.21.0013.01)]{wolframe} and after some mathematical computations, $\Theta_1$ is derived as (\ref{I1F}).
				
		Using high SNR approximation of $\text{ln}(1+\Xi\gamma)\approx\text{ln}(\Xi\gamma)$, $\Theta_2$ can be written as
		\begin{align}
			\Theta_2&= \frac{\xi^2_{\text{F}}}{\text{ln}(2)\tau\Gamma(\alpha_{\text{F}})\Gamma(\beta_{\text{F}})}\int_{0}^{\gamma_{\text{th}}}\frac{1}{\gamma}\text{ln}(\Xi\gamma)\nonumber\\&
			\times\text{G}^{3,0}_{1,3}\left[\frac{\alpha_{\text{F}} \beta_{\text{F}}}{A_{0,\text{F}}}\left(\frac{\gamma}{\delta_{{\tau}}}\right)^{\frac{1}{\tau}}\Big{|}^{\xi^2_{\text{F}}+1}_{\xi^2_{\text{F}},\alpha_{\text{F}},\beta_{\text{F}}}\right]\text{d}\gamma.
		\end{align}
		Approximating $\text{ln}(x)\approx\varpi\left(x^{\frac{1}{\varpi}}-1\right)$ for higher arguments, where $\varpi$ is an arbitrary large number, $\Theta_2$ can be solved as (\ref{I2F}).
		
		
		Identities $\Theta_3$ and $\Theta_{4}$ in (\ref{cap_T}) can be respectively given as
		\begin{align}\label{I3I4}
			\Theta_3&=\frac{\mathbb{C}_1}{2\text{ln}(2)(\bar{\gamma})^{\frac{\xi_{\text{T}}^2}{2}}}\int_{0}^{\infty}\text{ln}\left(1+\gamma\right)\gamma^{\frac{\xi_{\text{T}}^2}{2}-1}\nonumber\\&
			\times
			\Gamma\left(\mathbb{C}_2,\mathbb{C}_3\left(\frac{\gamma}{\bar{\gamma}}\right)^{\frac{\alpha}{2}}\right)\text{d}\gamma,\nonumber\\
			\Theta_{4}&=\frac{\mathbb{C}_1}{2\text{ln}(2)(\bar{\gamma})^{\frac{\xi_{\text{T}}^2}{2}}}\int_{0}^{\gamma_{\text{th}}}\text{ln}\left(1+\gamma\right)\gamma^{\frac{\xi_{\text{T}}^2}{2}-1}
			\nonumber\\&
			\times
			\Gamma\left(\mathbb{C}_2,\mathbb{C}_3\left(\frac{\gamma}{\bar{\gamma}}\right)^{\frac{\alpha}{2}}\right)\text{d}\gamma.
		\end{align}
		
		$\Theta_3$ can be further modified as
		\begin{align}\label{I3F1}
			\Theta_3&=\frac{\mathbb{C}_1}{2\text{ln}(2)(\bar{\gamma})^{\frac{\xi_{\text{T}}^2}{2}}}\int_{0}^{\infty}\text{G}^{2,1}_{2,2}\left[\gamma\Big{|}^{1,1}_{1,0}\right]\gamma^{\frac{\xi_{\text{T}}^2}{2}-1}\nonumber\\&
			\times\Gamma\left(\mathbb{C}_2,\mathbb{C}_3\left(\frac{\gamma}{\bar{\gamma}}\right)^{\frac{\alpha}{2}}\right)\text{d}\gamma.
		\end{align}
		
		Applying \cite[(07.34.21.0013.01)]{wolframe} and after some mathematical computations, $\Theta_3$ can be derived as (\ref{I3F}).
		Using high SNR approximation of $\text{ln}(1+\gamma)\approx\text{ln}(\gamma)$, $\Theta_{4}$ can be written as
		\begin{align}
			\Theta_{4}&=\frac{\mathbb{C}_1}{2\text{ln}(2)(\bar{\gamma})^{\frac{\xi_{\text{T}}^2}{2}}}\int_{0}^{\gamma_{\text{th}}}\text{ln}(\gamma)\gamma^{\frac{\xi_{\text{T}}^2}{2}-1}\Gamma\left(\mathbb{C}_2,\mathbb{C}_3\left(\frac{\gamma}{\bar{\gamma}}\right)^{\frac{\alpha}{2}}\right)\text{d}\gamma.\nonumber\\
		\end{align}
		Approximating $\text{ln}(x)\approx\varpi\left(x^{\frac{1}{\varpi}}-1\right)$ for higher arguments and using the change of variables, $\Theta_{4}$ can be solved as (\ref{I4F}).
		
		\section{Proof of Identities $\Theta_7\left(\hat{p}\right)$, $\Theta_8\left(\hat{p}\right)$, $\Theta_9\left(\hat{p}\right)$, $\Theta_{10}\left(\hat{p}\right)$, $\Theta_{11}\left(\hat{p}\right)$, and $\Theta_{12}\left(\hat{p}\right)$}
		$\Theta_7$ and $\Theta_8$ in (\ref{BER_FSO}) can be respectively given as
		\begin{align}\label{I7I8}
			\Theta_7\left(\hat{p}\right)=\frac{A\xi^2_{\text{F}}}{\tau\Gamma(\alpha_{\text{F}})\Gamma(\beta_{\text{F}})}\int_{0}^{\infty}\frac{1}{\gamma}\text{erfc}\left(\sqrt{B_{\hat{p}}\gamma}\right)\nonumber\\
			\times\text{G}^{3,0}_{1,3}\left[\frac{\alpha_{\text{F}} \beta_{\text{F}}}{A_{0,\text{F}}}\left(\frac{\gamma}{\delta_{{\tau}}}\right)^{\frac{1}{\tau}}\Big{|}^{\xi^2_{\text{F}}+1}_{\xi^2_{\text{F}},\alpha_{\text{F}},\beta_{\text{F}}}\right]\text{d}\gamma,\nonumber\\
			\Theta_8\left(\hat{p}\right)=\frac{A\xi^2_{\text{F}}}{\tau\Gamma(\alpha_{\text{F}})\Gamma(\beta_{\text{F}})}\int_{0}^{\gamma_{\text{th}}}\frac{1}{\gamma}\text{erfc}\left(\sqrt{B_{\hat{p}}\gamma}\right)\nonumber\\
			\times\text{G}^{3,0}_{1,3}\left[\frac{\alpha_{\text{F}} \beta_{\text{F}}}{A_{0,\text{F}}}\left(\frac{\gamma}{\delta_{{\tau}}}\right)^{\frac{1}{\tau}}\Big{|}^{\xi^2_{\text{F}}+1}_{\xi^2_{\text{F}},\alpha_{\text{F}},\beta_{\text{F}}}\right]\text{d}\gamma.
		\end{align}
		 The meijer-G representation of the complementary error function is given as
		 \begin{align}\label{ErfcMeijerG}
		 \text{erfc}(\sqrt{x})=\frac{1}{\sqrt{\pi}}\text{G}^{2,0}_{1,2}\left[x\big{|}^{1}_{0,\frac{1}{2}}\right].
		 \end{align}
	  Using (\ref{ErfcMeijerG})$, \Theta_7$ can be written as
		\begin{align}\label{BER_F2}
			\Theta_7\left(\hat{p}\right)&=\frac{A\xi^2_{\text{F}}}{\tau\sqrt{\pi}\Gamma(\alpha_{\text{F}})\Gamma(\beta_{\text{F}})}\int_{0}^{\infty}\frac{1}{\gamma}\text{G}^{2,0}_{1,2}\left[B_{\hat{p}}\gamma\Big{|}^{1}_{0,\frac{1}{2}}\right]\nonumber\\&
			\times\text{G}^{3,0}_{1,3}\left[\frac{\alpha_{\text{F}} \beta_{\text{F}}}{A_{0,\text{F}}}\left(\frac{\gamma}{\delta_{{\tau}}}\right)^{\frac{1}{\tau}}\Big{|}^{\xi^2_{\text{F}}+1}_{\xi^2_{\text{F}},\alpha_{\text{F}},\beta_{\text{F}}}\right]\text{d}\gamma.
		\end{align}
		Applying  \cite[(07.34.21.0013.01)]{wolframe}, (\ref{BER_F2}) can be solved as (\ref{BER_F3}).
		
		To solve $\Theta_8\left(\hat{p}\right)$, first we take the Maclaurin series expansion of the complementary error function as
		\begin{align}\label{erfcapp}
			\text{erfc}(\sqrt{x})=1-\frac{2}{\sqrt{\pi}}\sum_{j_1=0}^{\infty}\frac{(-1)^{j_1}}{j_1!(2j_1+1)}x^{\frac{2j_1+1}{2}}.
		\end{align}
		Substituting (\ref{erfcapp}) in (\ref{I7I8}), $\Theta_8$ can be written as
		\begin{align}\label{BER_F11}
			\Theta_8\left(\hat{p}\right)&=A\frac{\xi^2_{\text{F}}}{\tau\Gamma(\alpha_{\text{F}})\Gamma(\beta_{\text{F}})}\int_{0}^{\gamma_{\text{th}}}\frac{1}{\gamma}\nonumber\\&
			\times\left[1-\frac{2}{\sqrt{\pi}}\sum_{j_1=0}^{\infty}\frac{(-1)^{j_1}}{j_1!(2j_1+1)}(B_{\hat{p}}\gamma)^{\frac{2j_1+1}{2}}\right]\nonumber\\&
			\times\text{G}^{3,0}_{1,3}\left[\frac{\alpha_{\text{F}} \beta_{\text{F}}}{A_{0,\text{F}}}\left(\frac{\gamma}{\delta_{{\tau}}}\right)^{\frac{1}{\tau}}\Big{|}^{\xi^2_{\text{F}}+1}_{\xi^2_{\text{F}},\alpha_{\text{F}},\beta_{\text{F}}}\right]\text{d}\gamma.
		\end{align}
		
		After using  \cite[(07.34.21.0084.01)]{wolframe} with some mathematical computations, $\Theta_8\left(\hat{p}\right)$ can be derived as (\ref{BER_F12}).
		
		$\Theta_9\left(\hat{p}\right)$ and $\Theta_{10}\left(\hat{p}\right)$ in (\ref{BER_THz}) can be respectively given as
		\begin{align}\label{I9I10}
			\Theta_9\left(\hat{p}\right)&=\frac{A\mathbb{C}_1}{2(\bar{\gamma})^{\frac{\xi_{\text{T}}^2}{2}}}\int_{0}^{\infty}\text{erfc}\left(\sqrt{B_{\hat{p}}\gamma}\right)\gamma^{\frac{\xi_{\text{T}}^2}{2}-1}\nonumber\\&
			\times\text{G}^{2,0}_{1,2}\left[\mathbb{C}_3\left(\frac{\gamma}{\bar{\gamma}}\right)^{\frac{\alpha}{2}}\Big{|}^{1}_{0,\mathbb{C}_2}\right]\text{d}\gamma,\nonumber\\
			\Theta_{10}\left(\hat{p}\right)&=\frac{A\mathbb{C}_1}{2(\bar{\gamma})^{\frac{\xi_{\text{T}}^2}{2}}}\int_{0}^{\gamma_{\text{th}}}\text{erfc}\left(\sqrt{B_{\hat{p}}\gamma}\right)\gamma^{\frac{\xi_{\text{T}}^2}{2}-1}\nonumber\\&
			\times\text{G}^{2,0}_{1,2}\left[\mathbb{C}_3\left(\frac{\gamma}{\bar{\gamma}}\right)^{\frac{\alpha}{2}}\Big{|}^{1}_{0,\mathbb{C}_2}\right]\text{d}\gamma.
		\end{align}
		
		Using (\ref{ErfcMeijerG}), $\Theta_9\left(\hat{p}\right)$ is modified as
		\begin{align}
			\Theta_9\left(\hat{p}\right)&=\frac{A\mathbb{C}_1}{2\sqrt{\pi}(\bar{\gamma})^{\frac{\xi_{\text{T}}^2}{2}}}\int_{0}^{\infty}\gamma^{\frac{\xi_{\text{T}}^2}{2}-1}\text{G}^{2,0}_{1,2}\left[B_{\hat{p}}\gamma\Big{|}^{1}_{0,\frac{1}{2}}\right]\nonumber\\&
			\times\text{G}^{2,0}_{1,2}\left[\mathbb{C}_3\left(\frac{\gamma}{\bar{\gamma}}\right)^{\frac{\alpha}{2}}\Big{|}^{1}_{0,\mathbb{C}_2}\right]\text{d}\gamma.
		\end{align}
		Applying \cite[(07.34.21.0013.01)]{wolframe} and after some mathematical computations, $\Theta_9$ can be derived as (\ref{BER_T1F}).
		
		To solve $\Theta_{10}\left(\hat{p}\right)$,  we substitute (\ref{erfcapp}) in (\ref{I9I10}). Hence, $\Theta_{10}\left(\hat{p}\right)$ is written as
		\begin{align}\label{I10_Sol1}
			\Theta_{10}\left(\hat{p}\right)&=\frac{A\mathbb{C}_1}{2(\bar{\gamma})^{\frac{\xi_{\text{T}}^2}{2}}}\int_{0}^{\gamma_{\text{th}}}\gamma^{\frac{\xi_{\text{T}}^2}{2}-1}\nonumber\\&
			\times\left[1-\frac{2}{\sqrt{\pi}}\sum_{j_1=0}^{\infty}\frac{(-1)^{j_1}}{j_1!(2j_1+1)}(B_{\hat{p}}\gamma)^{\frac{2j_1+1}{2}}\right]\nonumber\\&
			\times\text{G}^{2,0}_{1,2}\left[\mathbb{C}_3\left(\frac{\gamma}{\bar{\gamma}}\right)^{\frac{\alpha}{2}}\Big{|}^{1}_{0,\mathbb{C}_2}\right]\text{d}\gamma.
		\end{align}
		Applying \cite[(07.34.21.0084.01)]{wolframe} and after some mathematical computations, (\ref{I10_Sol1}) can be solved as (\ref{I10_Fin}).
		
		$\Theta_{11}\left(\hat{p}\right)$ and $\Theta_{12}\left(\hat{p}\right)$ in (\ref{BER_RF}) can be respectively given as
		\begin{align}\label{I11I12}
			&\Theta_{11}\left(\hat{p}\right)= \frac{A}{\Gamma(mN_{\text{t}})}\left(\frac{m}{\bar{\gamma}_{\text{R}}}\right)^{mN_{\text{t}}}\nonumber\\&
			\times
			\int_{0}^{\infty}\text{erfc}\left(\sqrt{B_{\hat{p}}\gamma}\right) \gamma^{mN_{\text{t}}-1}\exp{\left(-\frac{m}{\bar{\gamma}_{\text{R}}}\gamma\right)}\text{d}\gamma.\nonumber\\
			&\Theta_{12}\left(\hat{p}\right)= \frac{A}{\Gamma(mN_{\text{t}})}\left(\frac{m}{\bar{\gamma}_{\text{R}}}\right)^{mN_{\text{t}}}\nonumber\\&
			\times
			\int_{0}^{\gamma^\text{R}_{\text{th}}}\text{erfc}\left(\sqrt{B_{\hat{p}}\gamma}\right) \gamma^{mN_{\text{t}}-1}\exp{\left(-\frac{m}{\bar{\gamma}_{\text{R}}}\gamma\right)}\text{d}\gamma.
		\end{align}
		
		To solve $\Theta_{11}\left(\hat{p}\right)$, first $\text{erfc}(\cdot)$ is transformed into its incomplete gamma equivalent as $\text{erfc}(\sqrt{x})=\frac{1}{\sqrt{\pi}}\Gamma\left(\frac{1}{2},x\right)$. Hence, $\Theta_{11}\left(\hat{p}\right)$ is written as
		\begin{align}
			\Theta_{11}\left(\hat{p}\right)&=\frac{A}{\sqrt{\pi}\Gamma(mN_{\text{t}})}\left(\frac{m}{\bar{\gamma}_{\text{R}}}\right)^{mN_{\text{t}}}\int_{0}^{\infty}\Gamma\left(\frac{1}{2},{B_{\hat{p}}\gamma}\right)\nonumber\\&
			\times \gamma^{mN_{\text{t}}-1}\exp{\left(-\frac{m}{\bar{\gamma}_{\text{R}}}\gamma\right)}\text{d}\gamma.
		\end{align}
		Applying  \cite[(6.455)]{gradshteyn2000table}, $\Theta_{11}\left(\hat{p}\right)$ can be derived as (\ref{BER_R1F}).
		
		For $\Theta_{12}\left(\hat{p}\right)$, we substitute the exponential series expansion  and (\ref{ErfcMeijerG}) in (\ref{I11I12}). Hence,  $\Theta_{12}\left(\hat{p}\right)$ is written as
		\begin{align}\label{I12_Sol1}
			\Theta_{12}\left(\hat{p}\right)&= \frac{A}{\sqrt{\pi}\Gamma(mN_{\text{t}})}\sum_{j_2=0}^{\infty}\frac{(-1)^{j_2}}{j_2!}\left(\frac{m}{\bar{\gamma}_{\text{R}}}\right)^{mN_{\text{t}}+j_2}\nonumber\\&\int_{0}^{\gamma^\text{R}_{\text{th}}} \gamma^{mN_{\text{t}}+j_2-1}\text{G}^{2,0}_{1,2}\left[B_{\hat{p}}\gamma\Big{|}^{1}_{0,\frac{1}{2}}\right]\text{d}\gamma.
		\end{align}
		Applying \cite[(07.34.21.0084.01)]{wolframe} and after some mathematical computations, (\ref{I12_Sol1}) can be solved as (\ref{BER_R2F}).

		\bibliographystyle{IEEEtran}
		\tiny
		\bibliography{Ref_THz}

\begin{thebibliography}{10}
\providecommand{\url}[1]{#1}
\csname url@samestyle\endcsname
\providecommand{\newblock}{\relax}
\providecommand{\bibinfo}[2]{#2}
\providecommand{\BIBentrySTDinterwordspacing}{\spaceskip=0pt\relax}
\providecommand{\BIBentryALTinterwordstretchfactor}{4}
\providecommand{\BIBentryALTinterwordspacing}{\spaceskip=\fontdimen2\font plus
\BIBentryALTinterwordstretchfactor\fontdimen3\font minus
  \fontdimen4\font\relax}
\providecommand{\BIBforeignlanguage}[2]{{%
\expandafter\ifx\csname l@#1\endcsname\relax
\typeout{** WARNING: IEEEtran.bst: No hyphenation pattern has been}%
\typeout{** loaded for the language `#1'. Using the pattern for}%
\typeout{** the default language instead.}%
\else
\language=\csname l@#1\endcsname
\fi
#2}}
\providecommand{\BIBdecl}{\relax}
\BIBdecl

\bibitem{elayan2019terahertz}
H.~Elayan, O.~Amin, B.~Shihada, R.~M. Shubair, and M.-S. Alouini, ``Terahertz
  band: The last piece of {RF} spectrum puzzle for communication systems,''
  \emph{IEEE Open J. Commun. Soc.}, vol.~1, pp. 1--32, Nov. 2019.

\bibitem{yi2021design}
C.~Yi, D.~Kim, S.~Solanki, J.-H. Kwon, M.~Kim, S.~Jeon, Y.-C. Ko, and I.~Lee,
  ``Design and performance analysis of {THz} wireless communication systems for
  chip-to-chip and personal area networks applications,'' \emph{IEEE J. Sel.
  Areas Commun.}, vol.~39, no.~6, pp. 1785--1796, Apr. 2021.

\bibitem{rajatheva2020scoring}
N.~Rajatheva, I.~Atzeni, S.~Bicais, E.~Bjornson, A.~Bourdoux, S.~Buzzi,
  C.~D'Andrea, J.-B. Dore, S.~Erkucuk, M.~Fuentes \emph{et~al.}, ``Scoring the
  terabit/s goal: {B}roadband connectivity in {6G},'' \emph{arXiv preprint
  arXiv:2008.07220}, Aug. 2020.

\bibitem{li2021performance}
S.~Li and L.~Yang, ``Performance analysis of dual-hop {THz} transmission
  systems over $\alpha$-$\mu$ fading channels with pointing errors,''
  \emph{IEEE Internet Things J.}, vol.~9, no.~14, pp. 11\,772 -- 11\,783, Dec.
  2021.

\bibitem{trichili2020roadmap}
A.~Trichili, M.~A. Cox, B.~S. Ooi, and M.-S. Alouini, ``Roadmap to free space
  optics,'' \emph{J. Optical Society America B}, vol.~37, no.~11, pp.
  A184--A201, Nov. 2020.

\bibitem{singya2021performance}
P.~K. Singya and M.-S. Alouini, ``Performance of {UAV} assisted multiuser
  terrestrial-satellite communication system over mixed {FSO/RF} channels,''
  \emph{IEEE Trans. Aero. Electron. Syst.}, vol.~58, no.~2, pp. 781--796, Sep.
  2021.

\bibitem{singya2020performance}
P.~K. Singya, N.~Kumar, V.~Bhatia, and M.-S. Alouini, ``On the performance
  analysis of higher order {QAM} schemes over mixed {RF/FSO} systems,''
  \emph{IEEE Trans. Veh. Technol.}, vol.~69, no.~7, pp. 7366--7378, Apr. 2020.

\bibitem{bhatnagar2013performance}
M.~R. Bhatnagar and M.~Arti, ``{Performance analysis of hybrid
  satellite-terrestrial FSO cooperative system},'' \emph{IEEE Photon. Technol.
  Lett.}, vol.~25, no.~22, pp. 2197--2200, Sep. 2013.

\bibitem{zedini2016performance}
E.~Zedini, H.~Soury, and M.-S. Alouini, ``On the performance analysis of
  dual-hop mixed {FSO/RF} systems,'' \emph{IEEE Trans. Wireless Commun.},
  vol.~15, no.~5, pp. 3679--3689, May 2016.

\bibitem{makki2017performance}
B.~Makki, T.~Svensson, M.~Brandt-Pearce, and M.-S. Alouini, ``On the
  performance of millimeter wave-based {RF-FSO} multi-hop and mesh networks,''
  \emph{IEEE Tran. Wireless Commun.}, vol.~16, no.~12, pp. 7746--7759, Sep.
  2017.

\bibitem{makki2017performance1}
------, ``Performance analysis of {RF-FSO} multi-hop networks,'' in \emph{Proc.
  IEEE WCNC'2017}, San Francisco, CA, USA, Mar. 2017, pp. 1--6.

\bibitem{zedini2020performance}
E.~Zedini, A.~Kammoun, and M.-S. Alouini, ``Performance of multibeam very high
  throughput satellite systems based on {FSO} feeder links with {HPA}
  nonlinearity,'' \emph{IEEE Trans. Wireless Commun.}, vol.~19, no.~9, pp.
  5908--5923, Jun. 2020.

\bibitem{lee2020throughput}
J.-H. Lee, K.-H. Park, Y.-C. Ko, and M.-S. Alouini, ``Throughput maximization
  of mixed {FSO/RF} {UAV}-aided mobile relaying with a buffer,'' \emph{IEEE
  Trans. Wireless Commun.}, vol.~20, no.~1, pp. 683--694, Oct. 2020.

\bibitem{xu2021performance}
G.~Xu and Z.~Song, ``Performance analysis of a {UAV}-assisted {RF/FSO} relaying
  systems for internet of vehicles,'' \emph{IEEE Internet Things J.}, vol.~9,
  no.~8, pp. 5730--5741, Jan. 2021.

\bibitem{xu2020performance}
------, ``Performance analysis for mixed $\kappa$-$\mu$ fading and
  {M}-distribution dual-hop radio frequency/free space optical communication
  systems,'' \emph{IEEE Trans. Wireless Commun.}, vol.~20, no.~3, pp.
  1517--1528, Nov. 2020.

\bibitem{balti2019tractable}
E.~Balti and B.~K. Johnson, ``Tractable approach to {mmWaves} cellular analysis
  with {FSO} backhauling under feedback delay and hardware limitations,''
  \emph{IEEE Trans. Wireless Commun.}, vol.~19, no.~1, pp. 410--422, Oct. 2019.

\bibitem{balti2021joint}
------, ``On the joint effects of {HPA} nonlinearities and {IQ} imbalance on
  mixed {RF/FSO} cooperative systems,'' \emph{IEEE Trans. Commun.}, vol.~69,
  no.~11, pp. 7879--7894, Aug. 2021.

\bibitem{yang2018performance}
L.~Yang, J.~Yuan, X.~Liu, and M.~O. Hasna, ``{On the performance of LAP-based
  multiple-hop RF/FSO systems},'' \emph{IEEE Trans. Aero. Electron. Syst.},
  vol.~55, no.~1, pp. 499--505, Jul. 2018.

\bibitem{michailidis2018outage}
E.~T. Michailidis, N.~Nomikos, P.~Bithas, D.~Vouyioukas, and A.~G. Kanatas,
  ``Outage probability of triple-hop mixed {RF/FSO/RF} stratospheric
  communication systems,'' in \emph{Proc. IEEE SPACOMM'2018}, Athens, Greece,
  Apr. 2018, pp. 1--6.

\bibitem{singya2022Haps}
P.~K. Singya and M.-S. Alouini, ``Mixed {FSO/RF} based multiple {HAPs} assisted
  multiuser multiantenna terrestrial communication,'' \emph{Frontiers Commun.
  Netw.}, Mar. 2022.

\bibitem{hassan2019hybrid}
M.~Z. Hassan, M.~J. Hossain, J.~Cheng, and V.~C. Leung, ``Hybrid {RF/FSO}
  backhaul networks with statistical-{QoS}-aware buffer-aided relaying,''
  \emph{IEEE Trans. Wireless Commun.}, vol.~19, no.~3, pp. 1464--1483, Oct.
  2019.

\bibitem{zhang20203d}
S.~Zhang and N.~Ansari, ``{3D} drone base station placement and resource
  allocation with {FSO}-based backhaul in hotspots,'' \emph{IEEE Trans. Veh.
  Technol.}, vol.~69, no.~3, pp. 3322--3329, Jan. 2020.

\bibitem{douik2016hybrid}
A.~Douik, H.~Dahrouj, T.~Y. Al-Naffouri, and M.-S. Alouini, ``Hybrid
  radio/free-space optical design for next generation backhaul systems,''
  \emph{IEEE Trans. Commun.}, vol.~64, no.~6, pp. 2563--2577, Apr. 2016.

\bibitem{Nock2016}
K.~Nock, C.~Font, and M.~Rupar, ``Adaptive transmission algorithms for a
  hard-switched {FSO/RF} link,'' in \emph{Proc. IEEE MILCOM'2016}, Baltimore,
  MD, USA, Dec. 2016, pp. 877--881.

\bibitem{sharma2019switching}
S.~Sharma, A.~Madhukumar, and R.~Swaminathan, ``Switching-based cooperative
  decode-and-forward relaying for hybrid {FSO/RF} networks,'' \emph{IEEE/OSA J.
  Optical Commun. Netw.}, vol.~11, no.~6, pp. 267--281, Jun. 2019.

\bibitem{swaminathan2021haps}
R.~Swaminathan, S.~Sharma, N.~Vishwakarma, and A.~Madhukumar, ``{HAPS}-based
  relaying for integrated space-air-ground networks with hybrid {FSO/RF}
  communication: {A} performance analysis,'' \emph{IEEE Trans. Aero. Electron.
  Syst.}, vol.~57, no.~3, pp. 1581--1599, Jan. 2021.

\bibitem{nath2019impact}
S.~Nath, S.~Sengar, S.~K. Shrivastava, and S.~P. Singh, ``Impact of atmospheric
  turbulence, pointing error, and traffic pattern on the performance of
  cognitive hybrid {FSO/RF} system,'' \emph{IEEE Trans. Cognitive Commun.
  Netw.}, vol.~5, no.~4, pp. 1194--1207, Nov. 2019.

\bibitem{gupta2019hard}
A.~Gupta, P.~Garg, and N.~Sharma, ``Hard switching-based hybrid {RF/VLC} system
  and its performance evaluation,'' \emph{Wiley Trans. Emerg. Telecommun.
  Technol.}, vol.~30, no.~2, p. e3515, Feb. 2019.

\bibitem{althunibat2020secure}
S.~Althunibat, R.~Mesleh, and K.~Qaraqe, ``Secure index-modulation based hybrid
  free space optical and millimeter wave links,'' \emph{IEEE Trans.
  Veh.Technol.}, vol.~69, no.~6, pp. 6325--6332, Apr. 2020.

\bibitem{shah2021adaptive}
S.~Shah, M.~Siddharth, N.~Vishwakarma, R.~Swaminathan, and A.~Madhukumar,
  ``Adaptive-combining-based hybrid {FSO/RF} satellite communication with and
  without {HAPS},'' \emph{IEEE Access}, vol.~9, pp. 81\,492--81\,511, Jun.
  2021.

\bibitem{rakia2015outage}
T.~Rakia, H.-C. Yang, M.-S. Alouini, and F.~Gebali, ``Outage analysis of
  practical {FSO/RF} hybrid system with adaptive combining,'' \emph{IEEE
  Commun. Lett.}, vol.~19, no.~8, pp. 1366--1369, Jun. 2015.

\bibitem{makki2016performance}
B.~Makki, T.~Svensson, T.~Eriksson, and M.-S. Alouini, ``On the performance of
  {RF-FSO} links with and without hybrid {ARQ},'' \emph{IEEE Trans. Wireless
  Commun.}, vol.~15, no.~7, pp. 4928--4943, Apr. 2016.

\bibitem{he2009bit}
B.~He and R.~Schober, ``Bit-interleaved coded modulation for hybrid {RF/FSO}
  systems,'' \emph{IEEE Trans. Commun.}, vol.~57, no.~12, pp. 3753--3763, Dec.
  2009.

\bibitem{zhang2009soft}
W.~Zhang, S.~Hranilovic, and C.~Shi, ``Soft-switching hybrid {FSO/RF} links
  using short-length raptor codes: design and implementation,'' \emph{IEEE J.
  Sel. Areas Commun.}, vol.~27, no.~9, pp. 1698--1708, Dec. 2009.

\bibitem{abdulhussein2010rateless}
A.~AbdulHussein, A.~Oka, T.~T. Nguyen, and L.~Lampe, ``Rateless coding for
  hybrid free-space optical and radio-frequency communication,'' \emph{IEEE
  Trans. Wireless Commun.}, vol.~9, no.~3, pp. 907--913, Mar. 2010.

\bibitem{trichili2021retrofitting}
A.~Trichili, A.~Ragheb, D.~Briantcev, M.~A. Esmail, M.~Altamimi, I.~Ashry,
  B.~S. Ooi, S.~Alshebeili, and M.-S. Alouini, ``Retrofitting {FSO} systems in
  existing {RF} infrastructure: A non-zero sum game technology,'' \emph{IEEE
  Open J. Commun. Soc.}, vol.~2, pp. 2597--2615, Nov. 2021.

\bibitem{Moradi2010}
H.~Moradi, M.~Falahpour, H.~H. Refai, P.~G. LoPresti, and M.~Atiquzzaman, ``On
  the capacity of hybrid {FSO/RF} links,'' in \emph{Proc. IEEE GLOBECOM'2010},
  Miami, FL, USA, Dec. 2010, pp. 1--5.

\bibitem{usman2014practical}
M.~Usman, H.-C. Yang, and M.-S. Alouini, ``Practical switching-based hybrid
  {FSO/RF} transmission and its performance analysis,'' \emph{IEEE Photon. J.},
  vol.~6, no.~5, pp. 1--13, Aug. 2014.

\bibitem{bag2018performance}
B.~Bag, A.~Das, I.~S. Ansari, A.~Proke{\v{s}}, C.~Bose, and A.~Chandra,
  ``Performance analysis of hybrid {FSO} systems using {FSO/RF-FSO} link
  adaptation,'' \emph{IEEE Photon. J.}, vol.~10, no.~3, pp. 1--17, May 2018.

\bibitem{boulogeorgos2019error}
A.-A.~A. Boulogeorgos and A.~Alexiou, ``Error analysis of mixed {THz-RF}
  wireless systems,'' \emph{IEEE Commun. Lett.}, vol.~24, no.~2, pp. 277--281,
  Dec. 2019.

\bibitem{bhardwaj2021performance}
P.~Bhardwaj and S.~Zafaruddin, ``Performance of dual-hop relaying for {THz-RF}
  wireless link over asymmetrical $\alpha-\mu$ fading,'' \emph{IEEE Trans. Veh.
  Technol.}, vol.~70, no.~10, pp. 10\,031--10\,047, Jul. 2021.

\bibitem{li2021performancecon}
S.~Li, L.~Yang, and Y.~Hu, ``Performance analysis of mixed {THz/FSO} relaying
  systems,'' in \emph{Proc. IEEE WCSP}.\hskip 1em plus 0.5em minus 0.4em\relax
  IEEE, Changsha, China, 2021, pp. 1--5.

\bibitem{li2022mixed}
S.~Li, L.~Yang, J.~Zhang, P.~S. Bithas, T.~A. Tsiftsis, and M.-S. Alouini,
  ``Mixed {THz/FSO} relaying systems: {S}tatistical analysis and performance
  evaluation,'' \emph{IEEE Trans. Wireless Commun.}, 2022.

\bibitem{henniger2010introduction}
H.~Henniger and O.~Wilfert, ``An introduction to free-space optical
  communications.'' \emph{Radioengineering}, vol.~19, no.~2, Jun. 2010.

\bibitem{farid2007outage}
A.~A. Farid and S.~Hranilovic, ``Outage capacity optimization for free-space
  optical links with pointing errors,'' \emph{J. Lightwave Technol.}, vol.~25,
  no.~7, pp. 1702--1710, Jul. 2007.

\bibitem{jung2020unified}
K.-J. Jung, S.~S. Nam, M.-S. Alouini, and Y.-C. Ko, ``Unified finite series
  approximation of {FSO} performance over strong turbulence combined with
  various pointing error conditions,'' \emph{IEEE Trans. Commun.}, vol.~68,
  no.~10, pp. 6413--6425, Jul. 2020.

\bibitem{abramowitz1988handbook}
M.~Abramowitz and I.~A. Stegun, ``{Handbook of Mathematical Functions with
  Formulas, Graphs, and Mathematical Tables},'' New York, NY, USA: Dover, 1972.

\bibitem{wolframe}
\BIBentryALTinterwordspacing
\emph{{The Wolfram Function Site}}. [Online]. Available:
  \url{http://functions.wolfram.com/}
\BIBentrySTDinterwordspacing

\bibitem{papasotiriou2021experimentally}
E.~N. Papasotiriou, A.-A.~A. Boulogeorgos, K.~Haneda, M.~F. de~Guzman, and
  A.~Alexiou, ``An experimentally validated fading model for {THz} wireless
  systems,'' \emph{Nature Scientific Reports}, vol.~11, no.~1, pp. 1--14, Sep.
  2021.

\bibitem{kokkoniemi2021line}
J.~Kokkoniemi, J.~Lehtom{\"a}ki, and M.~Juntti, ``A line-of-sight channel model
  for the 100--450 gigahertz frequency band,'' \emph{EURASIP J. Wireless
  Commun. Netw.}, vol. 2021, no.~1, pp. 1--15, Dec. 2021.

\bibitem{gradshteyn2000table}
I.~Gradshteyn and I.~Ryzhik, \emph{Table of Integrals, Series and
  Products}.\hskip 1em plus 0.5em minus 0.4em\relax 6th ed. {N}ew {Y}ork, {NY},
  {USA}: Academic, 2000.

\end{thebibliography}

	\end{document}